\documentclass[11pt]{article}

\usepackage[T1]{fontenc}
\usepackage[utf8]{inputenc}
\usepackage{xcolor}
\usepackage{tikz-feynman}
\usepackage[italian,english]{babel}
\usepackage{url}
\usepackage{siunitx}
\usepackage{amsmath}
\usepackage{amssymb}
\usepackage{graphicx}
\usepackage{graphicx,caption}
\usepackage{siunitx}
\usepackage{bm}
\usepackage{amsfonts}
\usepackage[left=2cm,right=2cm,top=2.5cm,bottom=2.5cm]{geometry}
\usepackage{amsmath}
\usepackage{graphicx}
\usepackage{subfigure}
\usepackage{slashed}
\usepackage{verbatim}
\usepackage[hidelinks]{hyperref}

\let\originaleqref\eqref 
\makeatletter
\renewcommand{\eqref}[1]{%
	\let\ref\@refstar%
	\hyperref[#1]{%%
		~\originaleqref{#1}%
	}%
}
\makeatother

\newcommand{\la}{\lambda}
\numberwithin{equation}{section}

\usepackage{tikz}
\usepackage{tikz-feynman}
\usepackage{pgf}
\usetikzlibrary{arrows}
\usetikzlibrary{snakes}
\usetikzlibrary{decorations,decorations.pathmorphing,decorations.markings}

\def\tor{\leftrightarrow}

\def\kbf{\mathbf k}
\def\pbf{\mathbf p}

\def\Tr{\text {Tr}}
\def\ra{\rangle}
\def\be{\begin{equation}}
\def\ee{\end{equation}}

\usepackage{adjustbox}
\usepackage{standalone}
\usepackage{amsmath}
\newcommand{\beq}{\begin{equation}}
\newcommand{\eeq}{\end{equation}}
\newcommand{\beqa}{\begin{eqnarray}}
\newcommand{\eeqa}{\end{eqnarray}}
\newcommand{\veps}{\varepsilon}
\newcommand{\nn}{\nonumber}

\def\id{\leavevmode\hbox{\small1\kern-3.3pt\normalsize1}}
\def\sigbf{\boldsymbol \sigma}

\def\kbf{\mathbf k}
\def\pbf{\mathbf p}

\def\Tr{\text {Tr}}
\def\la{\lambda}
\newcommand{\matr}[1]{ \begin{pmatrix} #1 \end{pmatrix}}

\usepackage{adjustbox}
\usepackage{standalone}
\usepackage{amsmath}

\def\bbx#1\ebx{\begin{empheq}[box={\tcbhighmath[colframe=blue!20!white,colback=blue!10!white]}]{align} #1 \end{empheq}}

\def\bbxd#1\ebxd{\begin{identity} \vskip -.4cm #1 \end{identity}\vskip-.2cm}

\def\btbox#1\etbox{\begin{titlebox} \vskip -.0cm #1 \end{titlebox}\vskip-.0cm}
\usepackage{tcolorbox}

\begin{document}
\vspace{1.5cm}
\begin{center}
{\bf  \Large Axion-like Quasiparticles and Topological States of Matter: \\ } 
\vspace{0.2cm}
{\bf \Large  Finite Density Corrections of the Chiral Anomaly Vertex \\}
\vspace{0.2cm}
{\bf  \Large }

{\Large \bf  }
\vspace{0.1cm}
{\Large \bf }

 \vspace{0.3cm}
\vspace{1cm}
{\bf Claudio Corian\`o,$^{1,2,3}$ Mario Cret\`i,$^{1,2,4}$ Stefano Lionetti,$^{1,2}$ Riccardo Tommasi$^{1,2}$ \\}

\vspace{1cm}
{\it  $^{1}$Dipartimento di Matematica e Fisica, Universit\`{a} del Salento 
	and\\  INFN Sezione di Lecce, Via Arnesano 73100 Lecce, Italy\\}
\vspace{0.3cm}
{\it  $^{2}$
	National Center for HPC, Big Data and Quantum Computing,\\ Via Magnanelli 2, 40033 Casalecchio di Reno, Italy\\}
\vspace{0.3cm}
{\it  $^{3}$  Institute of Nanotechnology, \\ National Research Council (CNR-NANOTEC), Lecce 73100\\}
\vspace{0.3cm}

{\it $^{4}$Center for Biomolecular Nanotechnologies,\\ Istituto Italiano di Tecnologia, Via Barsanti 14,
	73010 Arnesano, Lecce, Italy\\}
\vspace{0.5cm}

\begin{abstract}
We investigate the general structure of the chiral anomaly $AVV/AAA$  and $(LLL, RRR)$ vertices, 
in the presence of chemical potentials in perturbation theory. The study finds application in anomalous transport, whenever chirally unbalanced matter is present, with propagating external currents that are classically conserved. Examples are topological materials and the chiral magnetic effect in the plasma state of matter of the early universe. We classify the minimal number of form factors of the $AVV$ parametrization, by a complete analysis of the Schouten identities in the presence of a heat-bath.
We show that the longitudinal (anomaly) sector in the axial-vector channel, for on-shell and off-shell photons, is protected against corrections coming from the insertion of a chemical potential  in the fermion loop. 
When the photons are on-shell, we prove that the transverse sector, in the same channel, is also $\mu$-independent and vanishes. The related effective action is shown to be always described by the exchange of a massless anomaly pole, as in the case of vanishing chemical potentials.  
The pole is interpreted as an interpolating axion-like quasiparticle generated by the anomaly.  In each axial-vector channel,  it is predicted to be a correlated fermion/antifermion pseudoscalar (axion-like) quasiparticle appearing in the response function, once the material is subjected to an external chiral perturbation. The cancellation of the $\mu$ dependence extends to any chiral current within the Standard Model, including examples such as $B$ (baryon), $L$ (lepton), and $B-L$. This holds true irrespective of whether these currents exhibit anomalies.

\end{abstract}

\end{center} 
\newpage
\tableofcontents

\section{Introduction} 
The interest in exploring anomalies in chiral matter, specifically the Adler-Bell-Jackiw (ABJ) anomaly \cite{Adler:1969gk,Bell:1969ts} (see \cite{Bonora:2023soh,Scrucca:2004jn} for overviews), has witnessed substantial growth over the past two decades. This surge extends across both condensed matter theory \cite{Qi:2008ew,Qi:2010qag,Fruchart:2013tza,Hasan:2010xy,Chernodub:2021nff,Landsteiner:2013sja,Frohlich:2023uqc,Landsteiner:2011tf,Landsteiner:2011cp,Chernodub:2020yaf,Tutschku:2020rjq,Grushin:2019uuu,Arouca:2022psl,Wehling:2014cla,Burkov:2015hba, Nissinen:2019mkw,Nissinen:2018dnq}, and high-energy physics, particularly in the theoretical and experimental study of matter under high densities, and in transport \cite{Son:2004tq,Ferrer:2020ulz,Tarasov:2021yll,Tarasov:2020cwl,Rebhan:2009vc,Gorbar:2009bm,Metlitski:2005pr,Glorioso:2017lcn,Kanazawa:2015xna}. Experiments involving heavy ion collisions have tested and confirmed  the anomalous behavior of matter in the presence of finite density chiral asymmetric backgrounds and strong fields \cite{Huang:2015oca}.\\
The chiral interaction implies that, for massless fermions coupled to electric and magnetic fields ($\vec{E},\vec{B}$) the chiral fermion number $N_5$, if nonzero, is not conserved,  but is modified  by  the anomaly 
\beq
\frac{d N_5}{dt}=\frac{e^2}{2\pi^2} {\bf E}\cdot {\bf B}.
\label{prima}
\eeq
A similar effect was pointed out to be possible in a crystal subjected to the same external fields, more than a decade later, after the discovery of the ABJ anomaly, by Nielsen and Ninomiya 
\cite{Nielsen:1983rb}.\\
In condensed matter physics, the more recent identification of Dirac and Weyl semimetals has paved the way for the exploration of analogous anomaly-related phenomena \cite{Xiong:2015nna,Li:2014bha,Zyuzin:2012tv}. These materials exhibit distinctive features, notably the presence of "Dirac-points" where the bands come into contact. In Weyl metals, these Dirac points are concealed within the Fermi surface \cite{Son:2012bg}.\\
In the case of topological materials, their relativistic description is categorized as an analog, as the relativistic dynamics are not associated with a genuine relativistic dispersion relation in the propagation of the fundamental (real or virtual) states. Instead, it manifests through the Fermi velocity ($v_F$), which remains relativistic-like. This is how the Weyl and Dirac equations manifest in these systems, providing a table-top environment for testing exotic phenomena in particle physics.\\
Eq. \eqref{prima} clearly shows that if chiral asymmetries are present in the background in the form of propagating external currents, with a net nonzero chiral charge, the response of the matter system may be characterized by a dynamical evolution 
that, as it has been pointed out, can even lead to instabilities.  \\
The general feature of the phenomenon is such that it can be incorporated into the magnetohydrodynamic (MHD) equations of the early universe plasma as well \cite{Brandenburg:2017rcb}, generating a chiral magnetic cascade.  Indeed, a notable example of a similar phenomenon is the chiral magnetic effect (CME) in the quark gluon plasma \cite{Kharzeev:2013ffa,Fukushima:2008xe}.
 The CME, for example, is characterized by the generation of an electric charge separation along an external magnetic field, due to a chirality unbalance that is driven by the anomaly. \\
In this case one introduces the chiral current 
 \beq
 \vec{J_5}=\mu_5 \vec{B},
 \eeq
 with $\mu_5=\mu_L-\mu_R$ being the chiral chemical potential of the left $(L)$ and right $(R)$ Weyl fermions present in the initial state, and $\vec{B}$ a magnetic field.  $\vec{J_5}$ is the external source triggering the anomalous response at quantum level. The result of the chiral anomaly interaction is the generation of an electric field $\vec{E}$ parallel to the magnetic field.\\
 This macroscopic phenomenon gives rise to a collective motion within the Dirac sea. The topological nature of 
the unbalance $(\mu_5\neq 0)$ imparts a unique property to the CME current, ensuring a non-dissipative behavior even with the inclusion of radiative corrections. 
Therefore, in matter under extreme conditions, as found, for example, at neutron star densities \cite{Ferrer:2018kpd} or in the primordial plasma \cite{Brandenburg:2017rcb,Boyarsky:2015faa}, chiral anomalies induce a non-equilibrium phase that deserves a close attention. Given the different values of the external chiral asymmetries and of the related chemical potentials, the evaluation of the finite density corrections is expected to be of remarkable phenomenological relevance.\\ 
These interactions may leave an imprint in the stochastic background(s) of gravitational waves via the gravitational chiral anomaly \cite{Kamada:2021kxi,delRio:2021bnl} as well as contributing to the generation of primordial magnetic fields \cite{Joyce:1997uy,Kamada:2022nyt}. This may clearly occur at any phase of the early universe, and impact the production of gravitational waves in strongly first order phase transitions. Similar spectral asymmetries for spin-1 particles in the anomaly loop can be generated by  Chern-Simons currents (see the discussion and references in \cite{Coriano:2023gxa}). \\
Both chiral and conformal anomalies have been associated with the presence of interpolating massless states. In the chiral case, they are clearly recognized in the on-shell effective action, since the anomaly diagram is entirely represented by a pole. However, they appear in combination with transverse sectors as soon as in axial-vector/vector/vector ($AVV$) interactions one moves off-shell in the vector lines. \\   
The possibility that future experiments in Weyl semimetals may shed light on this phenomenon is the motivation of our analysis.
 Our goal in this work is to show how the entire anomaly interaction is dominated also at finite density by the exchange of a massless pole, that can be interpreted by an interpolating, on-shell fermion/antifermion pair in the axial vector channel. This intermediate state can be classified as an axion-like quasiparticle excitation of the medium.  \\
 We demonstrate that the pole remains unaffected by any correction and reaches saturation in its interaction when external electromagnetic fields are on-shell. Systems with these characteristics  exhibit a topological response, and conducting a table-top experiment provides a viable means to investigate this interpolating state.
 
\subsection{Topological protection in the Ward identity }
The Ward identity associated with the chiral interaction has been shown to be protected by the quantum corrections in a specific way. For example, at zero (massless) fermion density and temperature, the anomaly 
coefficient $e^2/(2 \pi)$ in \eqref{prima} is not modified by radiative corrections  \cite{Adler:1969er}. This is the content of the Adler-Bardeen theorem. On the other end, finite fermion mass corrections are shown to modify \eqref{prima} in the form
\beq
\partial\cdot \langle J_5\rangle = \frac{e^2}{2 \pi^2} F\tilde{F} + 2 m \langle \bar\psi\gamma^5\psi \rangle,
\label{seconda}
\eeq
\begin{figure}[t]
\centering
\subfigure[]{\includegraphics[scale=0.8]{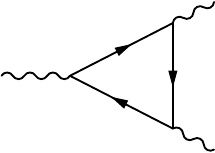}}  \hspace{2cm}
\subfigure[]{\includegraphics[scale=0.8]{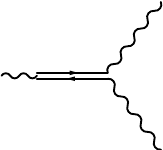}} \hspace{2cm}
\subfigure[]{\includegraphics[scale=0.8]{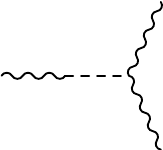}}
\caption{The fermion loop (a); the collinear region in the loop integration (b); the effective pseudoscalar exchange as an effective axion (c). }
\label{ddiag}
\end{figure}
where the mass term appears together with a pseudoscalar interaction inserted in the anomaly loop. Such second contribution in \eqref{seconda}, in axion physics, is responsible for the finite mass corrections of the axion-anomaly interaction. The corrections are obtained by coupling the axion field, 
$\varphi$, treated as an asymptotic state, to \eqref{seconda} in the standard form
\beq
\frac{\varphi}{f_a} \,\partial\cdot \langle J_5 \rangle
\label{eff}
\eeq
with the inclusion of a scale $f_a$ in order to preserve the quartic mass dimensions of the interaction. The $\varphi\to \gamma \gamma$ decay of an axion is then obtained by differentiating twice \eqref{seconda} with respect to the background electromagnetic gauge field.  The appearance of a scale $f_a$ in \eqref{eff} identifies the most complex part of axion physics within the local formulation, which in the context of strongly interacting theories such as QCD can be justified by introducing the $\theta$ vacuum. The question of the generation of an axion mass at the hadron phase transition is then answered by the inclusion of two very different scales. The first (large) scale identifies the breaking of a chiral $U(1)$ symmetry $(f_a)$ - for example the Peccei-Quinn symmetry -  while the second is far smaller and typical of the strong interactions. \\
In the absence of strong interactions, such as in QED, the link between the two descriptions, the first essentially based on the  description offered by the one-particle irreducible (1PI) effective action, and the local one, leaves open the way for further analysis. We refer to  \cite{Mottola:2023emy} for a discussion of a modified variational approach that attempts to bring together the two formulations.\\
Our analysis will be entirely based on the 1PI approach, which is quite direct and allows to identify conclusively the structure of the interaction and not just its Ward identity, which is the content of \eqref{eff}.\\
Concerning the generalization of the chiral Ward identity at finite density, previous studies have shown that \eqref{seconda} is preserved in this case \cite{Itoyama:1982up,GomezNicola:1994vq,Qian:1994pp}.\\
The goal of the current analysis will be to identify the structure of the  3-point functions containing both one and three insertions of $J_5$ at finite density. These are correlators that we will be denoting equivalently as $\langle J_5 JJ\rangle $ and $\langle J_5 J_5 J_5\rangle$ or $AVV$ and $AAA$, whose structure we will be investigating at finite density in perturbation theory. We will also consider correlators with chiral fermions $\langle J_LJ_LJ_L\rangle $ and $\langle J_RJ_RJ_R\rangle $ defined by the inclusion of left and right chirality projectors. \\
The analysis covers any chiral current associated with global symmetries of the Standard Model as well, and can be easily generalized to the nonabelian case by invoking the gauge invariance of our results.  Examples are baryon $(B)$ and lepton $(L)$ current associated with the conservation of the baryon and lepton numbers of the Standard Model. Both are conserved at classical level, but are anomalous at quantum level. The $B-L$ chiral current cancels the mixed $(B-L)\,Y^2$ anomaly, where $Y$ is the hypercharge, but requires a singlet right-handed neutrino in order to erase the $(B-L)^3 $ anomaly. Our results hold for all these cases as well, since any chiral diagram, in the absence of a spontaneously broken phase, are not affected by density corrections on-shell. As we are going to clarify, this result can be viewed as a consequence of conformal symmetry, since the conformal Ward identities of the $AVV$ or $AAA$ interactions are trivially satisfied in the on-shell photon case. 

\subsection{Content and organization of this work}
The $AVV$ vertex provides the simplest (free field) realization of the correlators mentioned above, 
that can be studied, at least in principle, even at finite density, using modified conformal Ward identities.  Notice that the structure of such correlators is determined in parity-odd CFT's at zero density by the inclusion of the anomaly contribution 
 in the solutions of the conformal Ward identities (CWIs), as shown in  \cite{Coriano:2023hts}, without resorting to perturbation theory. This result is a consequence of the conformal symmetry of the interaction at $\mu=0$, that is preserved, as we are going to see, even at finite $\mu$, for on-shell photons. \\
 We will proceed with an explicit identification of all the sectors of a chiral anomaly interaction, classifying all the tensor structures and form factors that are part of it. \\
 Our representation of the entire $AVV$ vertex differs in the number of form factors and tensor structures presented in \cite{Itoyama:1982up}, and is worked out in full detail. Our goal is to illustrate explicitly 
how the reduction of the general parametrization of the vertex proceeds. The original parametrization, valid at finite $T$ and $\mu$,  involves a large number of tensor structure - sixty -  due to the presence 
of three independent 4-vectors, and gets reduced by requiring Bose symmetry and conservation of the vector currents together with the use of the Schouten relations. \\
In \cite{Hou:2011ze} the authors analyze various aspects of the $\langle AVV \rangle$ correlator, both with and without a density background. The main form factors decomposition is carried out for a simplified version of the diagram, where the index associated with the axial leg is fixed to the Wick-rotated time coordinate. This significantly alters the number of Schouten identities and the final count of form factors. Additionally, the authors focus particularly on the infrared limit. 
In contrast, our paper examines the general full open indices amplitude in its form factors decomposition, paying close attention to details regarding transverse and longitudinal projections and their susceptibility to the density background. We work in a covariant manner, which is crucial for extracting an effective action that describes the general features of the anomaly phenomenon.\\
We will work in the simplest case of an abelian theory, such as QED coupled to external fields, vector and axial-vector. 
The generalization of our results to the nonabelian case is straightforward, since it can be obtained by imposing gauge invariance on the abelian result. There are some interesting implications  that emerge once one works, for example, with the full spectrum of the particles of the Standard Model rather than in QED, as we are going to do in the current work. These, specialized to the fermion families and gauge currents of the Standard Model,  are part of  an analysis that will be discussed elsewhere. They impact axion physics in a dense astrophysical environment.

\section{The nonlocal action}

\begin{figure}[t]
{\centering \resizebox*{10.8cm}{!}{\rotatebox{0}
{\includegraphics{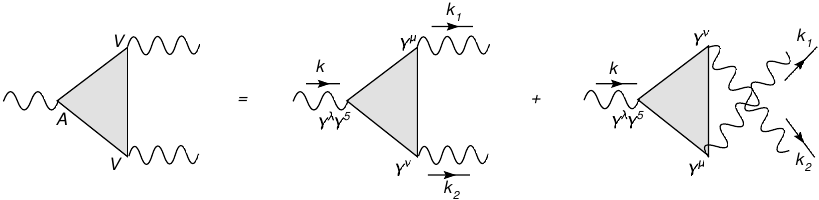}}}\par}
\caption{The ${\bf AVV}$ diagrams (direct and exchanged) in the expansion of the anomaly action $\mathcal{S}_{eff}$ }
\label{VAA1}
\end{figure}

The structure of the 1PI  effective action $\mathcal{S}_{eff}$ for an $AVV$ diagram is pictorially described in Fig 1. 
 The expansion variable in these actions is a dimensionless combination of fields and massless propagators 
 of the form $\partial\cdot B \Box^{-1}$ (bilinear mixing) for the chiral anomaly effective action, and $R  \Box^{-1}$ for the conformal anomaly action. Here, $B_{\mu}$ is an axial-vector source and $R$ is the Ricci scalar from an external gravitational metric. The expansion does not involve any scale and captures the scale-independent part of the anomaly interaction. On the other end, 
 a scale dependence is expected to come from explicit chiral symmetry breaking terms, in the form of mass corrections or finite density and thermal corrections, as mentioned above.\\
 In the context of the conformal anomaly, a similar nonlocal structure appears starting with the $TJJ$ vertex at the first order and progresses to the $TTT$ vertex at the third order. Here, $T$ denotes the stress-energy tensor, and $J$ represents a vector current.\\
The $1/\Box$ contribution present in the interaction, symbolized by the dashed line in Fig. \ref{ddiag} is the common tract of these interactions in flat space. The corresponding chiral anomaly effective action can then be simply described by the nonlocal term 
\beq
\mathcal{S}_{eff}\sim\frac{e^2}{2\pi}\int d^4 x\ d^4 y\ \partial\cdot B\ \Box^{-1}(x,y) \ F\tilde{F}(y). 
\label{nonl}
\eeq
Extensive studies of chiral and conformal correlators in perturbation theory as well as nonperturbatively, using conformal Ward identities (CWIs) prove conclusively that anomaly vertices are characterized by contributions similar to the one shown in Fig. 1. These bilinear insertions couple both to the longitudinal component of the external axial-vector source and to the anomaly. In the case of the conformal anomaly, a similar (dilaton) interaction couples to the scalar curvature $R$ and to the Gauss-Bonnet $(E)$ and Weyl tensor squared term ($C^2$).     \\ 
Without much ado, the interaction \eqref{nonl} is generally traded for a local one 
\beq
\mathcal{S}_{eff}\sim\frac{e^2}{2\pi}\int d^4 y\varphi(y) \ F\tilde{F}(y),
\label{nonl}
\eeq
by a field redefinition that involves the external background 
\beq
\label{naive}
\varphi(y)=\int d^4 y  \partial\cdot B \Box^{-1}(x,y). 
\eeq
Bilinear mixings are typical of spontaneously broken phases in gauge theories, and are usually removed by gauge-fixing conditions. In the case of external sources, not associated with gauge symmetries, this redefinition is not strictly necessary,  and the description provided by the nonlocal action \eqref{nonl} is indeed genuine. It is summarized by the statement that an anomaly-driven interaction is characterized by the exchange of an anomaly pole.\\
In topological materials it is natural to interpret such interactions as quasiparticle excitations generated by an external chiral source or - for conformal anomalies - by a thermal gradient. In the case of the conformal anomaly, Luttinger's equivalence relation between thermal gradients and the gravitational potential  \cite{Luttinger:1964zz}, similar nonlocal interactions are expected to appear as interpolating states between a thermal source and other (final state) currents, that can be both electromagnetic or thermal (i.e. correlators with three stress energy tensors). These materials indeed offer a natural realization of chiral and conformal anomalies in a surprising manner. The possibility of testing the essential tracts of axion physics in a simplified experimental setting is the motivation for  recent and less recent studies.  The reformulation of the nonlocal exchange as a local action with a $\varphi F \tilde{F}$ direct coupling, is a common tract of the literature on axions and anomalies, even in the absence of a strong interaction that may allow to classify the axion as a composite state rather than an elementary one. \\
Our analysis, however, will not deal with the issue of how to relate the local and the nonlocal structure of the chiral anomaly interaction. A discussion of this point can be found in \cite{Mottola:2023emy}.  
\subsection{The on-shell effective action}
 By requiring that the anomaly interaction takes place on a null surface, then the only surviving contribution in the chiral vertex is the pole. From the perturbative picture this simply means that when the two vector currents in the $AVV$ are on-shell, then the entire interaction is simply described, in the $m=0$ limit, simply by the exchange of the pole.  \\
As already mentioned, similar nonlocal actions are predicted by conformal symmetry, as shown in several studies of the conformal Ward identities, both in  free field theory realizations and non-perturbatively \cite{Coriano:2017mux,Coriano:2018bsy} using methods of reconstruction of conformal correlators in momentum space \cite{Bzowski:2013sza,Bzowski:2015pba,Bzowski:2018fql,Coriano:2013jba,Coriano:2018bsy,Coriano:2018zdo,Jain:2021wyn,Marotta:2022jrp}  in several dimensions.  \\
 This protection has significant ramifications, particularly in influencing the hydrodynamical and transport properties of systems containing chiral fermions. The entire chiral anomaly interaction can be built from the pole, as shown in \cite{Coriano:2023hts}, being constrained by conformal symmetry. \\
How this picture is modified by the presence of density effects in some materials and in the quark gluon plasma is only partially known. 
Conformal symmetry is expected to be broken by the new scale $(\mu)$. Previous analyses have shown, however, that the anomalous Ward identity is not modified at finite density \cite{Itoyama:1982up,GomezNicola:1994vq,Qian:1994pp}, except for extra contributions that originate from the transverse sector - with respect to the momentum of the axial vector current -  of this correlator.\\
The interaction is generally viewed as the response of the Dirac sea to an external chiral perturbation, as a result of the generation of a correlated fermion-antifermion interpolating state between the external chiral perturbation and the gauge fields of the final state.

\section{The action at finite density and the solutions}
The covariant structure of a chiral anomaly interaction at finite density and temperature is very involved. 
As in the case of the ordinary $AVV$ (i.e. $JJJ_5$) at $\mu=0$, we are forced to perform an analysis of the Schouten relations in the presence of a 
heat-bath in order to identify a suitable basis of tensor structures and form factors in the expansion of  
this correlator. At the same time, we discuss their perturbative realization, which is absent in the previous literature on the topic. The infrared properties of the form factors are investigated from the perturbative side and the proof of their infrared finiteness is shown explicitly by a direct computation  performed in a special reference frame.  \\
The Lagrangian is defined as
	\begin{equation}
	\label{dx}
		\mathcal{L} = \bar \psi i \slashed \partial \psi - i e J^\mu A_\mu -i g_B J_A^\mu B_\mu  +\mathcal{L}_C
\eeq
where		
\beq
\label{chiral}
\mathcal{L}_C=	-  \mu  \psi^\dagger \gamma^5 \psi
	\end{equation}
is the contribution of a chiral chemical potential. In this case $\mu_L=-\mu$ and $\mu_R=\mu$ and
 \beq
J^\mu=\bar{\psi}\gamma^\mu \psi \qquad 	J_A^\mu=\bar{\psi}\gamma^\mu\gamma^5 \psi 
\eeq
being the vector and axial-vector currents coupled to external fields $A_\mu$ and $B_\mu$. 
$g_B$ is the coupling of the axial-vector source to the Dirac fermion $\psi$, that we will split into its left-handed and right-handed chiral components. The chemical potential $\mu$ accounts for the presence of a nonvanishing background density of particles ($\mu>0$), of total charge $Q$ by the inclusion of a term 
$-\mu Q$  in the action.  \\
The action has a $U(1)_L\times U(1)_R\times U(1)_V$ global symmetry. The $U(1)_V$ symmetry can be gauged by the coupling to the (photon) field $A_\mu$ and the integral of the zeroth component of the current $J^\mu$ is the total charge density  $Q$.
Both the chiral charges of the left and right components, 

\beq
Q_{L/R}=\int d^3 x\ n_{L/R}^0    \qquad \qquad    Q= - e(Q_L + Q_R),\qquad  e>0
\eeq
are conserved, where 
\beq
n_{R}=\mu_R\bar\psi_R \gamma^0\psi_R,  \qquad \qquad n_{L}=\mu_L \bar\psi_L \gamma^0\psi_L 
\eeq
are the density of the $L$ and $R$ fermionic modes. Charge conjugation $(C)$ symmetry is broken by the external 
conditions. $L$ and $R$ modes are intertwined by $C$ but the finite density vacuum is not $C$ invariant. A more comprehensive examination of the character of the physical vacuum, particularly under external constraints associated with the specific background choice, will be thoroughly explored in the upcoming sections. This detailed analysis will extend to the description of the diverse contributions embedded within the thermal propagator.\\
 The chiral anomaly vertex will render $J_A$ not conserved at quantum level with the anomalous Ward identity
\beq
\partial_\mu J^\mu_A= a_n F \tilde F= a_n {\bf E}\cdot {\bf B}
\eeq
where $a_n$ is the anomaly, and the chiral charge 
\beq
Q_5\equiv Q_L-Q_R, \qquad    
\eeq
will acquire a nonzero time dependence 
\beq
\dot{Q}_5=\dot{N}_L -\dot{N}_R= \int d^3 x \ {\bf E}\cdot{\bf  B}. 
\eeq
We use the Dirac matrices in the chiral base, given by
\begin{equation}
	\gamma^\mu = \matr{ 0 & \sigma^\mu  \\  \bar \sigma^\mu & 0 }
\end{equation}
	where $\sigma^\mu = ( \mathbb{I}, \sigbf)$, $\bar \sigma^\mu = ( \mathbb{I}, -\sigbf)$ and $\sigbf=(\sigma_1, \sigma_2, \sigma_3)$ are the Pauli matrices. The chiral matrix $\gamma^5$ in this basis is defined as
\begin{equation}
		\gamma^5 = \matr{ - \mathbb I & 0  \\  0 & \mathbb I }\,\, .
\end{equation}
	Left- and right-handed spinors are defined by the projectors
	\begin{equation}
		P_R = \frac{1 +\gamma^5}{2}  \qquad 	P_L = \frac{1 -\gamma^5}{2}.
	\end{equation}
	We define Dirac spinors as sum of two 4 components chiral spinors $\psi_L$ and $\psi_R$	
	\begin{equation}
		\psi= \psi_L  + \psi_R \qquad \psi_L= \matr{\chi_L \\ 0}  \, \, \,  \psi_R= \matr{0 \\ \chi_R} \,\, .
	\end{equation}
\subsection{Vector chemical potential}
For a vector chemical potential $\mu_L=\mu_R=\mu$ the finite density Lagrangian is modified with 
$\mathcal{L}_C\to \mathcal{L}_V$ in \eqref{dx}
\begin{equation}
\mathcal L_V=-\mu \bar\psi \gamma^0\psi.
\end{equation}
The equations of motions are
\begin{equation}
i\gamma^\mu\partial_\mu \psi-\mu\gamma^0\psi=0,
\end{equation}
that we investigate with the ansatz
\begin{equation}
\label{ans}
\psi\sim e^{-ipx}u(p)=e^{-ipx}\begin{pmatrix} \chi_L\\ \chi_R \end{pmatrix}.
\end{equation} 
For the metric we use the signature $(1,-1,-1,-1)$.
Using \eqref{ans} we obtain
\begin{equation}
\begin{pmatrix}
0 & \sigma\cdot p\\
\bar\sigma\cdot p & 0
\end{pmatrix} \begin{pmatrix} \chi_L \\ \chi_R \end{pmatrix}-\mu 
\begin{pmatrix} 0 & \mathbb I \\ \mathbb I &0 \end{pmatrix} \begin{pmatrix} \chi_L \\ \chi_R \end{pmatrix}=0.\end{equation}

\begin{eqnarray} 
\label{cc}
\sigma \cdot \pbf \, \chi_R &=& (E-\mu) \, \chi_R\\
\sigma \cdot \pbf \, \chi_L &=& - (E-\mu) \, \chi_L. 
\end{eqnarray}
Iterating these equations we get the constraint
\begin{eqnarray} \label{dispersion}
\label{cc}
|\pbf|^2 \, \chi_{L/R} &=& (E - \mu )^2  \chi_{L/R} \,  ,
\end{eqnarray}
giving 
\begin{eqnarray}
\label{oneE}
E^{(\pm)}\equiv E_{L/R}^{(\pm)} &=& \mu \pm |\pbf| .
\end{eqnarray} 
Both equations in \eqref{cc} are of the form

\beq
\sigma\cdot\mathbf n \chi=\lambda\chi
\eeq
with
\beq
\mathbf n=\frac{1}{|\mathbf p|}(p_1,p_2,p_3)=(\sin\theta \cos\varphi, \sin\theta\sin\varphi,\cos\theta)
\eeq
with $\lambda=\pm 1$. 
The two Weyl spinors that solve the equations are classified by their helicities 
\beq
\chi^{(\pm)}=\begin{pmatrix} (\frac{1\pm n_3}{2})^{1/2}\\ \\ \pm(\frac{1\pm n_3}{2})^{1/2}\frac{1\mp n_3}{n_1-in_2}  \end{pmatrix}\,\,.
\eeq
The equation for $\chi_R$ becomes 
\beq
 {\mathbf \sigma} \cdot \mathbf n\chi_R=\frac{E-\mu}{|\mathbf p|}\chi_R \,\, .
\eeq
If we define 
\beq
E_1=\mu+|\mathbf p|\qquad E_2=\mu -|\mathbf p |,  
\eeq
then
\beq 
\mathbf\sigma \cdot \mathbf n\,\chi_R^{(+)}=+\chi_R^{(+)}, \,\,\, \textrm{if}\qquad  E=E_1
\label{typ}
\eeq
with
\beq
\chi_{R,1}^{(+)}=\begin{pmatrix} (\frac{1+n_3}{2})^{1/2}\\ \\ (\frac{1+n_3}{2})^{1/2}\frac{1-n_3}{n_1-i\,n_2}  \end{pmatrix},
\eeq
of helicity $+1$. If $E=E_2$, then 
\beq 
\sigma \cdot \mathbf n\chi_R^{(-)}=-\chi_R^{(-)}
\eeq
with
\beq
\chi_{R,2}^{(-)}=\begin{pmatrix} (\frac{1-n_3}{2})^{1/2}\\ \\- (\frac{1-n_3}{2})^{1/2}\frac{1+n_3}{n_1-i\, n_2}  \end{pmatrix}.
\eeq
One proceeds similarly for $\chi_L$. 
In this way we derive, for $E=E_1$, the solution  $\chi_{L,1}^-$, which has helicity $-1$. Analogously, for $E=E_2$, we derive the solutions $\chi_L=\chi_{L,2}^+$ of helicity $+1$.

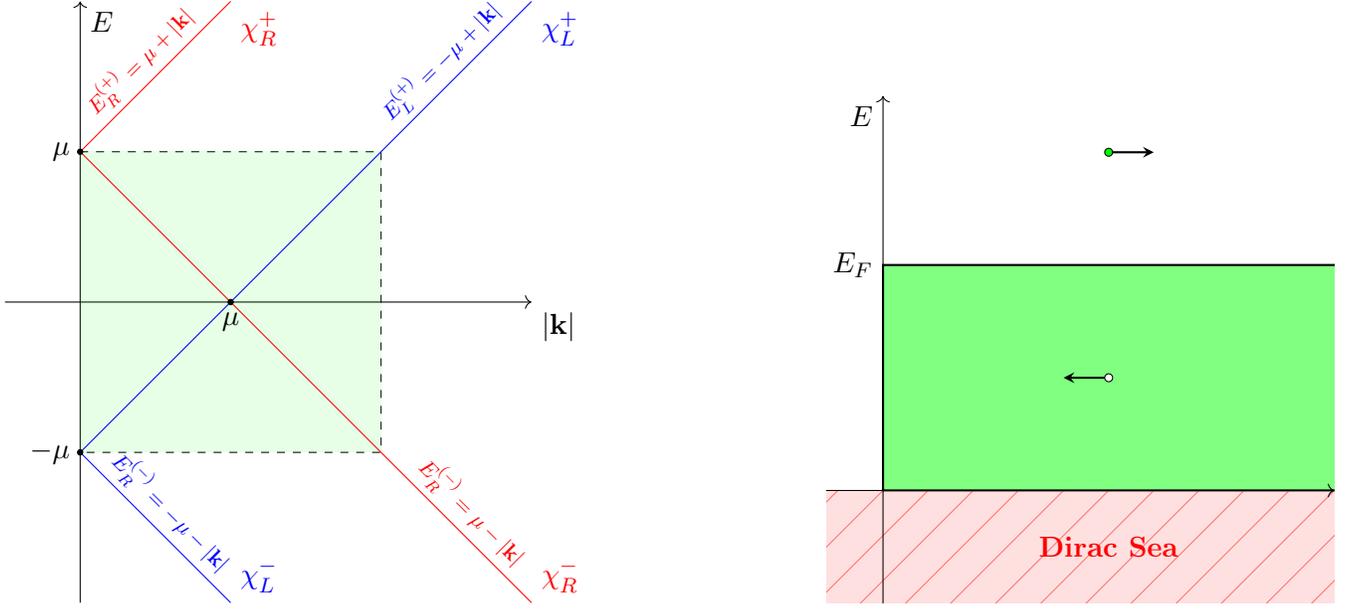
\begin{figure}
\begin{tikzpicture}
		\draw[->] (-1,0) -- (6,0) node[below right] {$|\mathbf k|$} ;
		\draw[->] (0,-4) -- (0,4) node[below right] {$E$} ;
		\draw[dashed] (0,2) -- (4,2) -- (4,-2) -- (0,-2);
		\fill[green, opacity=.1] (0,-2) rectangle (4,2);
		\draw[red] (2,4) node[below right] {$\chi_R^+$} -- node[sloped, above, scale=.75] {$E_R^{(+)} = \mu + |\mathbf k|$} (0,2) -- (4,-2) -- node[sloped, above, scale=.75] {$E_R^{(-)} = \mu - |\mathbf k|$} +(2,-2) node[above right] {$\chi_R^-$};
		\draw[blue] (6,4)  node[below right] {$\chi_L^+$} -- node[sloped, above, scale=.75] {$E_L^{(+)} = -\mu + |\mathbf k|$} (4,2) -- (0,-2) -- node[sloped, above, scale=.75] {$E_R^{(-)} = - \mu - |\mathbf k|$} +(2,-2)  node[above right] {$\chi_L^-$};
		\filldraw[black] (0,-2) circle (1pt) node[left] {$-\mu$} ;
		\filldraw[black] (2,0) circle (1pt) node[below] {$\mu$} ;
		\filldraw[black] (0,2) circle (1pt) node[left] {$\mu$} ;
\label{fourfold}
\end{tikzpicture} \hspace*{3cm}
\begin{tikzpicture}[scale=3]
	\clip (-.25,-.5) rectangle (2,1.75);
	\filldraw[fill=green!50!white, draw=black, thick] (0,0) rectangle (3,1);
	\draw (0,1) node[left] {$E_F$};
	\fill[red!40!white, opacity=0.3] (-3,-3) rectangle (3,0);
	\foreach \h in {-5,-4.8,...,3}{ \draw[red, opacity=0.5, shift={(\h,0)}] (-3,-3) -- (0,0);}
	\draw[red, thick] (1,-0.25) node {\textbf{Dirac Sea}};
	\draw[->] (-3,0) -- (2,0);
	\draw[->] (0,-3) -- (0,1.75) node[below left] {$E$};
	\draw[-{stealth},thick] (1,.5) -- +(-.2,0); 
	\draw[-{stealth},thick] (1,1.5) -- +(.2,0); 
	\filldraw[fill=white] (1,.5) circle (.5pt);
	\filldraw[fill=green] (1,1.5) circle (.5pt);
\end{tikzpicture}

\caption{The four folds of the dispersion relation in the  $E$, {$|\pbf|$} plane in the case of a chiral chemical potential. For a vector chemical potential the two folds at $E=-\mu$ are superimposed with the other two at $E=\mu$, giving a two-folds solution.}

\end{figure}

\subsection{Chiral chemical potential}
Now let us consider a chiral chemical potential, with the finite density Lagrangian term $\mathcal{L}_C$ as in \eqref{chiral}.
The equation of motion for the fermion is 
\beq
i\gamma^\mu\partial_\mu\psi-\mu\bar \psi \gamma^0\gamma^5\psi=0,
\eeq
or equivalently 
\beq
\sigma\cdot \mathbf p\chi_R=(E-\mu)\chi_R,
\eeq
and
\beq
\sigma\cdot \mathbf p\chi_L=-(E+\mu)\chi_L.
\eeq
Iterating we get
\beq
\mathbf p^2\chi_R=(E-\mu)^2\chi_R,
\eeq
giving two dispersion relations  $E_{1,2}=\mu\pm |\mathbf p|$ for the $R$ component. For the $L$ component 
we get other two, separate dispersion relations $E_3=-\mu+|\mathbf p|$ and $E_4=-\mu-|\mathbf p|$.\\
 Also in this case we encounter the same equations of the vector case. For $E=E_1=\mu+|\mathbf p|$
 we have $\chi_{R,1}^{(+)}$ with helicity $+1$, while for $E=E_2=\mu -|\mathbf p|$, the chiral solution   $\chi_{R,2}^{(-)}$ has helicity $-1$. \\
 Similar results hold for $E_3$ and $E_4$, with $\chi_{L3}^{(-)}$ of energy $E_3$ and  $\chi_{L 4}^{(+)}$ of energy $E_4$. The four folds are shown in Fig. 3. 
\subsection{The massive fermion case and the propagator} 
The generalization to the massive case can be worked out in a  similar way. We choose a vector chemical potential. 
The equation of motion 
\beq\label{eqref1}
i\gamma^\mu\partial_\mu\psi-m\psi-\mu\gamma^0\psi=0,
\eeq
can be rewritten in the Hamiltonian form 
\beq
H\psi=E\psi.
\eeq
We have the Hamiltonian equations
\begin{align}&E\chi_L=-\sigma\cdot\mathbf p\chi_L+\mu\chi_L+m\chi_R\notag\\&E\chi_R= \sigma\cdot\mathbf p\chi_R+\mu\chi_R+m\chi_L, \end{align}
that we rewrite as before in the form
\begin{align}&\sigma\cdot\mathbf p\chi_L=-E'\chi_L+m\chi_R\\\notag& \sigma\cdot\mathbf p\chi_R=E'\chi_R-m\chi_L, \end{align}
with $E'=E-\mu$. 
The eigenvalues can be easily found by iterating the equations, yielding the eigenvalues
\beq
E_1=\mu- E(m),\ \ \ \ E_2=\mu+E(m),
\eeq
where 
\beq
E(m)=\sqrt{\mathbf p^2+m^2}.
\eeq 
The equations for the $L$ and $R$ modes are obviously coupled. One can check that the solutions can be obtained in each case by selecting for $\chi_R$ the two options of spin up $\chi_R^\uparrow=(1,0)$ and down $\chi_R^\downarrow=(0,1)$ and then solving the coupled equations. For the eigenvalue $E_1=\mu-E(m)$ we derive the two degenerate eigenfunctions
\begin{align}&\psi_{11}=\biggl [-\frac{-p_1+ip_2}{m}, -\frac{-p_3+E(m)}{m},0,1   \biggl]\\\notag& \psi_{12}=\biggl [-\frac{-p_3+E(m)}{m}, -\frac{-p_1+ip_2}{m},1,0   \biggl],\end{align}
while, for eigenvalue $E_2=\mu+E(m)$, they are given by 
\begin{align}&\psi_{21}=\biggl [-\frac{-p_1+ip_2}{m}, -\frac{p_3-E(m)}{m},0,1   \biggl]\\\notag& \psi_{22}=\biggl [-\frac{-p_3-E(m)}{m}, -\frac{-p_1-ip_2}{m},1,0   \biggl].\end{align} \\
Our analysis, in the next sections, will be limited to the massless case. In the massive case, as already mentioned in the Introduction, the corrections to the anomaly vertex coming from nonzero $\mu$ and $m$, generate an axial-vector Ward identity that contains, besides the anomaly, also an additional contribution proportional to the product of $m$ and $\mu$.  \\
Coming back to the massless case, the four chiral solutions of fixed helicities can be mapped into the usual $u$ and $v$ particle/antiparticle solutions  
\beq
u_1^T=(\chi_L^{(-)},\chi_L^{(-)}), u_2^T=(\chi_R^{(+)},-\chi_R^{(+)}), \qquad v_1^T=(\chi_L^{(+)},\chi_L^{(+)})
\qquad v_2^T=(\chi_R^{(-)},-\chi_R^{(-)})
\eeq
and can be used in order to derive the propagator. 
\subsection{Anatomy of the propagator at finite $T_{emp}$ and $\mu$ and the $T_{emp}\to 0$ limit}
In this section we proceed with an analysis of the propagator at nonzero $\mu$.\\
\begin{figure}
\begin{center}
	\begin{tikzpicture}[scale=1.5]
	\draw (-2,0) -- (4,0) node[below right] {Re $\tilde k_0$} ;
	\draw[red, line width=1.5] (-2,0) --  node[above] {$E<0$} (0,0) ;
	\draw[blue,  line width=1.5 , dashed] (0,0) --  node[above] {$0<E<\mu$} (2,0) ;
	\draw[->, line width=1.5, red] (2,0) --  node[above] {$E>0$}  (4,0) ;
	\filldraw[black] (0,0) circle (.7pt) node[below] {$0$} ;
	\filldraw[black] (2,0) circle (.7pt) node[below] {$\mu$} ;
	\draw[->] (3,-.05) -- (3,-.43) ;
	\filldraw[black] (3,-.5) circle (.7pt) node[below] {$|\mathbf k| + \mu$} ;
	\draw[rounded corners=10, postaction={decorate, decoration={
			markings,
			mark=between positions 0.1 and 2 step .4 with {\arrow{stealth}}}}] (-2,-.4) -- (1.5,-.4) -- (2.5, .4) -- (4,.4);

\end{tikzpicture}
\end{center}
\caption{The path in the integration regions in the finite density propagator. } 
\end{figure}
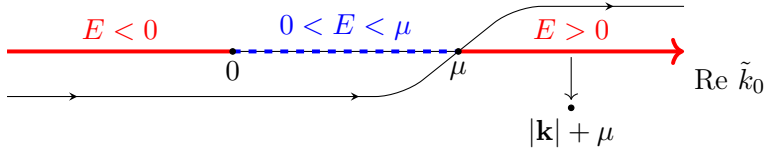
\noindent Computations at finite temperature have been traditionally performed both in the real and in the imaginary time formalisms \cite{Laine:2016hma,Gross:1980br,Niemi:1983ea}. While the imaginary time formalism is very efficient in the computation of vacuum diagrams, such as in the derivation of the equations of state of QED \cite{Coriano:1994re,Parwani:1994xi} and QCD \cite{Arnold:1994eb} to rather large orders, and in the resummation program of hard thermal loops \cite{Braaten:1989mz,Parwani:1991gq},
 the real time formalism has the advantage of providing direct access to time-dependent quantities directly \cite{Kobes:1984vb}. \\
The derivation of the propagator is rather subtle in the relativistic case, due to the negative energy solutions that appear in its retarded part due to the finite temperature background. These parts describe antiparticles which can be present in  the thermal bath due to thermal excitations, as we are going to illustrate below.  At zero density, the propagator can be derived more simply by sending the positive energy solutions of momentum $|\textbf k|> \mu$ in the future, as part of the retarded propagator. The negative energy solutions together with the positive ones with   $|k|< \mu$ become part of the advanced propagator. A discussion of this point in the Furry picture can be found in \cite{Elmfors:1993bm} that we are going to specialize to the zero temperature case.\\
 One introduces the expansion of the ordinary fermionic field operator with creation and annihilation operators of fermions $(b^\dagger, b)$ and antifermions $(d^\dagger, d)$
 \beq
  \Psi({\bf x},t) = \sum _{\lambda,k}
 b_{\lambda k}\psi^{(+)}_{\lambda k }({\bf x},t) +
  d_{\lambda k }^{\dagger}\psi^{(-)}_{\lambda k}({\bf x},t),
\eeq
where $\lambda$ runs over the spin states, $k $ denotes the energy
 and momentum 

 \be
\{d_{\lambda' k '},d_{\lambda k}^{\dagger}\} =
\delta_{\lambda'\lambda}\delta_{k' k} = \{b_{\lambda' k
'},b_{\lambda k}^{\dagger}\},
\ee
while other anti-commutators are zero. The statistical averages are obtained from the expressions
 \beqa
\label{fermidirac}
      \langle b_{\lambda k}^{\dagger}b_{\lambda' k'}\rangle
      &=& f_{F}^{+}(E_{ k})\delta_{\lambda\lambda'}
       \delta_{k k'},\nonumber\\
      \langle d_{\lambda k}^{\dagger}d_{\lambda' k'}\rangle
      &=& f_{F}^{-}(E_{ k})\delta_{\lambda\lambda'}
      \delta_{k k'}.
\eeqa
with the statistical averages given by the Fermi distributions  
 
\beq
f^+(E)=\frac{1}{e^{\beta(E-\mu)} +1}, \qquad  f^{-}(E) = 1 - f^{+}(-E)
\eeq
for the positive and negative energy states. Notice that the dispersion relation defining the fermionic excitation modes in this formulation is the ordinary one, that for massless fermions is derived from the usual equations of motion 
\beq
\slashed k u(k)=0\qquad \slashed{k} v(k)=0,
\label{simple}
\eeq
with $k^2=0$. \\
In the expressions above, we have set $E\equiv +|\textbf k|$ and identified the negative energy branch simply as $-E$. Then, the expansion of the propagator in the vacuum into its retarded 
($\sim \theta (t'-t)$)  and advanced ($\sim \theta (t'-t)$) parts

\beqa
iS_{F}(x' -x)~=~\langle 0|{\bf T}\left( \Psi({\bf x'},t') \overline{\Psi}({\bf
x},t)\right)|0\ra  =~~~~~~~~~~~~~~~~~~~~~~&&\nonumber \\
 \theta(t'-t)
\sum_{\lambda k}\psi^{(+)}_{\lambda k}({\bf x'},t')
\overline{\psi}^{(+)}_{\lambda k}({\bf x},t) -  \theta(t-t')
\sum_{\lambda k}\psi^{(-)}_{\lambda k}({\bf x'},t')
\overline{\psi}^{(-)}_{\lambda k}({\bf x},t)~~~,
\label{prop}
 \eeqa
at finite $T$ and $\mu$ is given by  
\beqa
 iS(x'-x)
& = &\nonumber \\
\sum_{\lambda,k} 
  \left[\theta(t'-t)
\left([1-f^{+}(E)]\psi^{(+)}_{\lambda\kappa}({\bf x'},t')
\overline{\psi}^{(+)}_{\lambda k}({\bf x},t) \right.\right.& +
&\left.\left.
[1-f^{+}( -E)]\psi^{(-)}_{\lambda k}({\bf x'},t')
\overline{\psi}^{(-)}_{\lambda k}({\bf x},t) \right)
 \right.  \nonumber \\
- \theta(t-t') \left.
\left(f^{+}(-E)\psi^{(-)}_{\lambda k}({\bf x'},t')
\overline{\psi}^{(-)}_{\lambda k}({\bf x},t) \right.\right.& +
&\left.\left.
f^{+}(E)\psi^{(+)}_{\lambda k}({\bf x'},t')
\overline{\psi}^{(+)}_{\lambda k}({\bf x},t) \right)
 \right].
\label{comb}
\eeqa
The sum over  $k$, in the continuum limit, turns into an ordinary  momentum integration over the on-shell states of energy, performed over the various kinematical regions.  \\
Both the retarded and advanced parts in \eqref{comb} are characterized by two contributions, proportional both to positive and negative energy modes.  For example, the retarded part, in the thermal vacuum, is generated by the combination
\beq
 i S_{ret}\equiv  \theta(t'-t)
\sum_{\lambda k}\left( \langle bb^\dagger \rangle\psi^{(+)}_{\lambda k}({\bf x'},t')
\overline{\psi}^{(+)}_{\lambda k}({\bf x},t) +\langle d^\dagger d\rangle \psi^{(-)}_{\lambda k}({\bf x'},t')
\overline{\psi}^{(-)}_{\lambda k}({\bf x},t)\right)
 \eeq
where the products $b b^\dagger$ and $d^\dagger d$ are statistically averaged.\\
At finite temperature and density, as one can see from \eqref{comb}, the retarded part contains both a term 
\beq
\langle b b^\dagger \rangle =1-\langle b^\dagger b\rangle=1-f^{+}(E),
\eeq
and a second term proportional to 
\beq
\langle d^\dagger d\rangle =f^-(E)=1 - f^+(-E),
\eeq
generated by finite temperature fluctuations, that vanishes in the zero temperature limit. This means that at finite temperature, thermal fluctuations can deplete the Dirac sea of negative energy states that propagate into the future. At temperature $T_{emp}=0$, these states are simply not present, and the physical picture of propagation is the one described above. Indeed we have, in the retarded part

\beq
1-f^+(E)\to 1- \theta(\mu-|\textbf k |) =\theta (|\textbf k|-\mu), \qquad f^-(E)\to \theta(-\mu-|\textbf k |)=0 \qquad (\mu>0),
\eeq
with only  the states with $E>\mu$ propagating into the future. \\
On the other end, the advanced part, 

\beqa
i S_{ret}=\theta(t-t') 
\left(\langle d d^\dagger\rangle\psi^{(-)}_{\lambda k}({\bf x'},t')
\overline{\psi}^{(-)}_{\lambda k}({\bf x},t)  +
\langle b^\dagger b\rangle\psi^{(+)}_{\lambda k}({\bf x'},t')
\overline{\psi}^{(+)}_{\lambda k}({\bf x},t) \right)
\eeqa
in the same $T_{emp}\to 0$ limit, will contain all the negative energy states since 
\beq
 \langle d d^\dagger\rangle= f^+(-E) =\frac{1}{ e^{\beta(-|\textbf k|-\mu)} +1}\to 1 
  \eeq
   and in the same limit, the positive energy states with $0<E<\mu$
   \beq
 \langle b^\dagger b\rangle= f^+(E)\to  \theta(\mu - |\textbf k|)
\eeq
are sent to the past. Eq. \eqref{comb} can be separated into a vacuum part $S_0$, which is an ordinary Feynman propagator, and a finite density/temperature  part $S_1$
\beq
       iS(x'-x) =
      iS_{0}(x'-x) + iS_{1}(x'-x).
\label{eq:thermalexp}
\eeq
The thermal part given by
\beqa
&& S_1(x'- x)~~~=  \nonumber \\
&&i \sum_{\lambda,k}
\left(f^{+}(E_k)\psi^{(+)}_{\lambda k}({\bf x'},t')
\overline{\psi}^{(+)}_{\lambda k}({\bf x},t)  -
f^{-}(E_k)\psi^{(-)}_{\lambda k}({\bf x'},t')
\overline{\psi}^{(-)}_{\lambda k}({\bf x},t) \right).
\label{thermal}
\eeqa
Notice that of this part, only the first term survives in the $T_{emp}\to 0$ limit, due to $f^-(E)\to 0$, and the integration over the momentum is restricted to the $0<E< \mu$ region.  
\subsection{The covariant formulation}
In order to derive a covariant expression of the propagator at finite density, one needs to be specific about the form of the expansion and effectively rewrite the integration as a four-dimensional one. \\\
We illustrate how to do this for the retarded part of the propagator, the other terms being similar.  \\
We rewrite the statistical average as 
\beqa
i S_{ret}(x'-x)&=&\theta(x'_0- x_0)\langle \Psi(x')\overline{\Psi(x)}\rangle\nonumber \\
&=&i\int \frac{d^3\vec{ k }d \tilde{k_0}}{(2 \pi)^4}e^{-i (\mu+|\textbf k|)(x'_0-x_0) +i \vec{k}\cdot (\vec{x'}-\vec{x})}\frac{(\gamma_0| \textbf k| - \vec{\gamma}\cdot \vec{k})}{2 |\textbf k|}
\frac{e^{-i \tilde{k_0}(x'_0-x_0)}}{\tilde{k_0}+ i\epsilon}
\eeqa
where the expansion covers the positive energy branch $E(k)= \mu + |k|$ of the Hamiltonian. We have used the contour representation of the step function $\theta(x'_0-x_0)$ using the integration variable $\tilde{k_0}$, and performed the usual average over the spin  states 
\beq
\sum_{\lambda} u_\lambda(k)\overline{u_\lambda(k)}=\frac{\slashed{k}}{2 |\textbf k|}
\eeq
taken on shell.  Notice that in this expansion, formulated over plane waves 
$\sim e^{i{(\mu +|\textbf k|)}x_0 + i \vec{k}\cdot \vec{x}}$, the Dirac operator at finite density density requires that the spinors satisfy the ordinary massless on-shell relations \eqref{simple}. At this stage one performs a change of variables by redefining the total phase $k_0=\tilde{k_0}+\mu +|k|$ in order to rewrite the expression above as 
\beqa
i S_{ret}(x'-x) &=& i \int \frac{d^4 k}{(2 \pi)^4} e^{-i k(x'_0-x_0)}\theta(k_0-\mu)\frac{\slashed{\bar{k}}}
{\left(k_0-(\mu + |\textbf k|- i \epsilon)\right)\left(k_0-(\mu - |\textbf k| + i \epsilon)\right)}\nonumber \\
&=& i \int \frac{d^4 k}{(2 \pi)^4} e^{-i k(x'_0-x_0)}\theta(k_0-\mu)\frac{\bar{\slashed{k}}}{\bar{k}^2 + i \epsilon}
\eeqa
with $\bar{k}=(k_0-\mu,\vec{k})$. Notice that the poles, in general,  are chosen in such a way to extract one of the two residues present in the covariant formulation. Notice also that, the integration over the the momentum variable $k_0$, once the covariant expression is obtained, leaves a phase $e^{i \mu ((x'_0-x_0)}$ in the propagator. This phase cancels in any closed loop and can be omitted.

\subsection{Simplifications in momentum space}
It is widely acknowledged that a naive extension of zero-temperature perturbation theory,  proves to be inconsistent in any Feynman diagram featuring two lines with overlapping momenta, as the square of distributions cannot be defined as another distribution. The remedy for this challenge involves a meticulous derivation of the Feynman rules for real-time perturbation theory, extensively discussed in the former literature \cite{Landsman:1986uw}. However, this situation is not encountered in the computation of the form factors in our case, and we can simply use the fermionic version of the real time propagator (at $T_{emp}=0$ and finite density) in the form

\begin{eqnarray}
S_F(k) &=&S _0(k) + S_1(k)  \nonumber\\
S_0(k)&=&\frac{\slashed{k}}{k^2} \qquad S_1(k) = 2\pi i \slashed{k}\theta(k_0)\theta(\mu-k_0),
\label{above}
\end{eqnarray}
with only $S_1$ related to the chemical potential. 
More generally, this result can be formulated covariantly in the form 
\begin{equation}
	S_F (k) = {1 \over \slashed k} + 2 \pi i \slashed{k} \, \delta (k^2) \, \theta(\eta \cdot k) \, \theta(\mu - \eta \cdot k),
	\label{agr}
\end{equation}
where the final expression has been written in terms of a 4-vector $\eta^\mu$, which is the  velocity of the heat-bath. This simplified expression can be obtained from \eqref{thermal}, that in general takes the form ($\beta=1/T_{emp}$)

\begin{equation}\label{eq:propmassivtemp}
	\begin{aligned}
		S_F(k,\beta,\mu)&\equiv (\not k+m) G_F(k,\beta,\mu)=\\&(\not k+m)\Bigg\{\frac{1}{ k^2-m^2}+2 \pi i  \delta\left(k^2-m^2\right)\left[\frac{\theta\left(k_0\right)}{e^{\beta(E-\mu)}+1}+\frac{\theta\left(-k_0\right)}{e^{\beta(E+\mu)}+1}\right]\Bigg\}\, .
	\end{aligned}
	\end{equation}
Notice that in the $T_{emp}\to 0$ and $m\to 0$ limits this result reduces to \eqref{above}.\\
The computations that we will present in the next sections will be performed in a special frame where $\eta=(1,\textbf {0})$.  The propagator for chiral fermions can be obtained 
from \eqref{above} by the inclusion of the appropriate chiral projectors $P_L$ and $P_R$ and corresponding chemical potential $\mu\to \mu_{L/R}$ in \eqref{agr}, as in previous analyses performed within the Standard Model \cite{Notzold:1987ik}.
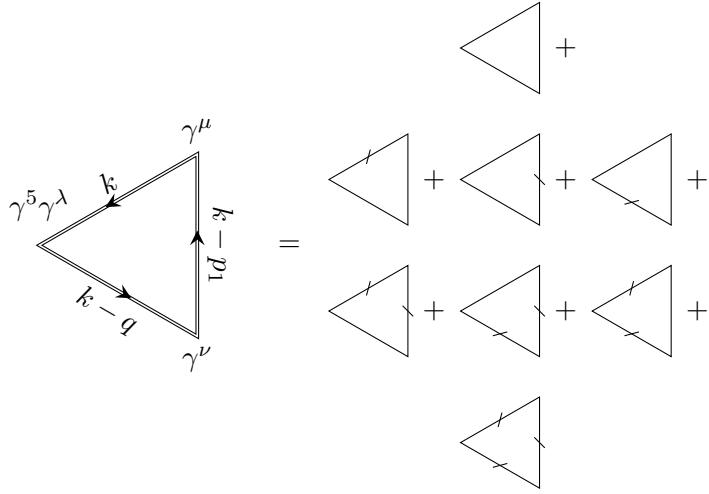
\begin{figure}
\begin{center}
\begin{tikzpicture}[scale=0.7]
	\draw[scale=2, shift={(-2.25,-1.875)} , double , postaction={decorate, decoration={
			markings,
			mark=between positions 1/6 and 1 step 1/3 with {\arrow{stealth[reversed]} } } } ]
	(180:1) node[above=.2] {$\gamma^5 \gamma^\lambda$} -- node[sloped, above] {$k$} 
	(60:1) node[above] {$\gamma^\mu$} -- node[sloped, above] {$k-p_1$} 
	(-60:1) node[below] {$\gamma^\nu$} -- node[sloped, below] {$k-q$}  cycle; 
	
		\draw[shift={(2.5,0)}]
	(180:1) -- 	(60:1) -- (-60:1) -- cycle; 
	
		\draw[shift={(0,-2.5)} , postaction={decorate, decoration={
			markings,
			mark=at position 1/6 with {\draw (-2pt, -2pt) -- (2pt, 2pt); } } } ]
	(180:1) -- 	(60:1) -- (-60:1) -- cycle; 
	
		\draw[shift={(2.5,.-2.5)} , postaction={decorate, decoration={
		markings,
		mark=at position 3/6 with {\draw (-2pt, -2pt) -- (2pt, 2pt); } } } ]
	(180:1) -- 	(60:1) -- (-60:1) -- cycle;

		\draw[shift={(5,-2.5)} , postaction={decorate, decoration={
		markings,
		mark=at position 5/6 with {\draw (-2pt, -2pt) -- (2pt, 2pt); } } } ]
	(180:1) -- 	(60:1) -- (-60:1) -- cycle; 

		\draw[shift={(0,-5)} , postaction={decorate, decoration={
		markings,
		mark=between positions 1/6 and 3/6 step 1/3 with {\draw (-2pt, -2pt) -- (2pt, 2pt); } } } ]
	(180:1) -- 	(60:1) -- (-60:1) -- cycle; 

	\draw[shift={(2.5,.-5)} , postaction={decorate, decoration={
		markings,
		mark=between positions 3/6 and 5/6 step 1/3 with {\draw (-2pt, -2pt) -- (2pt, 2pt); } } } ]
	(180:1) -- 	(60:1) -- (-60:1) -- cycle;

	\draw[shift={(5,-5)}, rotate=120 , postaction={decorate, decoration={
		markings,
		mark=between positions 1/6 and 3/6 step 1/3 with {\draw (-2pt, -2pt) -- (2pt, 2pt); } } } ]
	(180:1) -- 	(60:1) -- (-60:1) -- cycle; 
	
		\draw[shift={(2.5,-7.5)} , postaction={decorate, decoration={
			markings,
			mark=between positions 1/6 and 5/6 step 1/3 with {\draw (-2pt, -2pt) -- (2pt, 2pt); } } } ]
	(180:1) -- 	(60:1) -- (-60:1) -- cycle; 
	
	\draw (-1.75,-3.75) node {$=$};
	\draw  (3.5, 0) node {$+$} (3.5, -2.5) node {$+$} (3.5, -5) node {$+$} ;
	\draw  (1, -2.5) node {$+$} (1, -5) node {$+$} ;
	\draw  (6, -2.5) node {$+$} (6, -5) node {$+$} ;
\end{tikzpicture}
\end{center}
\caption{List of the vacuum and finite density contributions in the expansion of the $AVV$  3-point function. The finite density insertions are indicated with cut lines.}  
\end{figure}

\section{General parametrization of the correlator}
The correlation function 
\begin{equation}
\Gamma^{\lambda\mu\nu}\equiv\langle J_A^\lambda(q) J^\mu(p_1) J^\nu(p_2) \rangle_\mu
\label{corr}
\end{equation} 
is realized in free field theory by the simple $AVV$ Feynman diagram. In \eqref{corr} the suffix $\mu$ indicates that the quantum average is performed at finite density. \\
The analysis of this correlator will be performed both on general grounds, by the classification of the minimal number of form factors appearing in its expression and, at the same time, perturbatively, by using the expression of the propagator at finite density.  
 The general dependence of $\Gamma$ on the external momenta  of the incoming axial vector line $(q)$ and outgoing vextor lines ($p_1$ and $p_2$),  $(q=p_1 +p_2)$, is extended by the inclusion of a four-vector $\eta^\mu$ characterizing the velocity of the heat-bath in a covariant formulation.\\
As already mentioned, the analysis of these interactions has been performed in the previous literature by focusing the attention on the Ward identities satisfied by the correlator at perturbative level, showing that the anomaly is protected against these corrections \cite{Qian:1994pp}. Our goal is to extend such previous  studies and proceed with a complete analysis of the vertex, proving the emergence of a $1/q^2$ interaction in its anomaly form factor, which is the signature of the new effective degree of freedom associated with the chiral anomaly. \\
When the two vector lines (photons) are on-shell, we are going to show explicitly that the interaction reduces just to the anomaly pole, as in the vacuum case. For off-shell photons,  the transverse sector of the correlator will be shown to be $\mu$-dependent, but the longitudinal contribution will be also shown to exhibit the same massless pole structure, deprived of any correction. 
The minimal decomposition of $\Gamma$ differs from the standard $AVV$ vertex at zero $\mu$ (the vacuum contribution), which is rather elementary, since it requires only 6 form factors. In that case two separate formulations are the most useful ones in the perturbative analysis: the Rosenberg decomposition and the longitudinal/transverse (L/T) ones. The latter has been used in \cite{Coriano:2023hts} in the derivation of the expression of the interaction from conformal field theory (CFT) in momentum space.  \\
The expansion is complicated by the presence of a four-vector characterizing the heat-bath $(\eta)$, and the number of allowed tensor structures increases quite significantly. 
The complete list of 60 tensor structures $\tau^i$ and form factors $B_i$ formally written as 
\begin{equation}
\label{fdecom}
\Gamma^{\lambda\mu\nu} =\sum_{i=1}^{60} B_i(q^2,p_1^2,p_2^2,p_1\cdot\eta,p_2\cdot\eta)\tau_i^{\lambda\mu\nu}
\end{equation}
is reported in Appendix \ref{structures}.\\
We have denoted with $q$ the momentum of the axial-vector line and with $p_1$ and $p_2$ the momenta of the two photons.
We pause for a moment and define our conventions for the vacuum contributions and the finite density parts. 
We are going to denote with $\Gamma$ the entire contribution to the vertex, that we split into a vacuum part $(\Gamma^{(0)})$ and a finite density part $\Gamma^{(1)}$. This second part is $\mu$-dependent  
   \beq
  \Gamma=\Gamma^{(0)} + \Gamma^{(1)}.
   \eeq
The action of the longitudinal and transverse projectors defined in the axial-vector channel allows us to separate the entire vertex $\Gamma$ as
\beq
\pi^{\lambda}_{\rho} \Gamma^{\rho\mu\nu}\equiv\Gamma^{\rho\mu\nu}_{T}, \qquad  \Sigma^\lambda_\rho
\Gamma^{\rho\mu\nu}\equiv \Gamma^{\rho\mu\nu}_L, \qquad 
\pi^{\lambda\rho}=g^{\lambda \rho} -  \frac{q^\lambda q^\rho}{q^2}, \qquad \Sigma^{\lambda\rho}=\frac{q^\lambda q^\rho}{q^2}.
\eeq
Similar projections can be introduced on both $\Gamma^{(0)}$ and $\Gamma^{(1)}$.
This decomposition gets reduced, once we impose the Ward identity on the free indices $\mu,\nu$. \\
The general dependence on the external momenta in each of the form factors $B_i$, can be expressed in terms of the scalar products of external momenta ($p_1^2, \, p_2^2, \, q^2,\,  p_1 \cdot \eta, \, p_2 \cdot \eta$). Using the Schouten identities, we can significantly lower the number of form factors. We will consider their reduction, first in the general case and then, at a second stage, assuming a simplifying kinematical condition, in which both vector lines (i.e. the two photons) have equal projection on the 4-vector of the heat-bath $(p_1\cdot \eta=p_2\cdot \eta)$. 
\subsection{Schouten relations}
The analysis of the Schouten relation is rather involved. In $d=4$ spacetime dimensions, tensors created by the complete antisymmetrization over five indices, must vanish. We need to consider all the tensors of such type contracted with the two momenta $p_1, p_2$ and $\eta$ in order to obtain the corresponding identities.\\
We start with tensors with three free indices ($\lambda \mu \nu$) and three contracted indices ($\alpha \beta \gamma$). The possible tensors are
\begin{equation}
	\epsilon^{[\lambda\mu\nu\alpha}\delta^{\beta]\gamma}, \ \ \ \epsilon^{ [\mu\nu\alpha\beta} \delta^{\gamma]\lambda}, \ \ \ \epsilon^{[\lambda\mu\alpha\beta}\delta^{\gamma]\nu},\ \ \ \epsilon^{[\lambda\nu\alpha\beta}\delta^{\gamma]\mu}.
\end{equation}
Contracting these with $p_1, p_2,$ and  $\eta$ yields 12 equations, with 11 of them independent, that allow us to solve for the structures below. We use the simplified notation $\epsilon^{\mu \nu {p_1}{p_2}}\equiv \epsilon^{\mu \nu {\alpha}{\beta}}p_{1\alpha}p_{2\beta}$, and similarly in the other cases, to obtain the relations
{\allowdisplaybreaks
\begin{eqnarray} \label{schouten1}
\epsilon^{\mu \nu {p_1}{p_2}} {p_1}^{\lambda }& = & \epsilon^{\lambda \nu {p_1}{p_2}} {p_1}^{\mu }-{\epsilon}^{\lambda \mu {p_1}{p_2}} {p_1}^{\nu}+\epsilon^{\lambda \mu \nu {p_2}}{p_1}^{2}-\epsilon^{\lambda \mu \nu {p_1}} \left({p_1}\cdot {p_2}\right) \notag \\ 
\epsilon^{\mu \nu	{p_1}{\eta}} {p_1}^{\lambda } & = & {\epsilon}^{\lambda \nu {p_1}{\eta}} {p_1}^{\mu}-\epsilon^{\lambda \mu {p_1}{\eta}}{p_1}^{\nu }+\epsilon^{\lambda \mu \nu {\eta}}{p_1}^{2}-\epsilon^{\lambda \mu \nu {p_1}}\left({p_1}\cdot {\eta}\right) \notag \\ \epsilon^{\mu \nu	{p_1}{p_2}} {p_2}^{\lambda } & = & \epsilon^{\lambda \nu {p_1}{p_2}}{p_2}^{\mu }-\epsilon^{\lambda \mu	{p_1}{p_2}} {p_2}^{\nu }+{\epsilon}^{\lambda \mu \nu {p_2}} \left({p_1}\cdot{p_2}\right)-\epsilon^{\lambda \mu \nu {p_1}}{p_2}^{2} \notag \\ 
\epsilon^{\mu \nu {p_1}{p_2}} {\eta}^{\lambda } & = & {\epsilon}^{\lambda \nu {p_1}{p_2}} {\eta}^{\mu}-\epsilon^{\lambda \mu {p_1}{p_2}} {\eta}^{\nu }+\epsilon^{\lambda \mu \nu {p_2}}\left({p_1}\cdot {\eta}\right)-\epsilon^{\lambda \mu \nu	{p_1}} \left({p_2}\cdot {\eta}\right) \notag \\ 
\epsilon^{\mu \nu {p_1}{\eta}}{p_2}^{\lambda } & = & \epsilon^{\lambda \nu	{p_1}{\eta}} {p_2}^{\mu }-{\epsilon}^{\lambda \mu {p_1}{\eta}} {p_2}^{\nu}+\epsilon^{\lambda \mu \nu {\eta}} \left({p_1}\cdot{p_2}\right)-\epsilon^{\lambda \mu \nu {p_1}}\left({p_2}\cdot {\eta}\right) \notag \\ \epsilon^{\mu \nu{p_2}{\eta}} {p_1}^{\lambda } & = & {\epsilon}^{\lambda \nu {p_2}{\eta}} {p_1}^{\mu}-\epsilon^{\lambda \mu {p_2}{\eta}}{p_1}^{\nu }+\epsilon^{\lambda \mu \nu {\eta}}\left({p_1}\cdot {p_2}\right)-\epsilon^{\lambda \mu	\nu {p_2}} \left({p_1}\cdot {\eta}\right) \notag \\
\epsilon^{\mu \nu {p_1}{\eta}} {\eta}^{\lambda } & = & \epsilon^{\lambda \nu {p_1}{\eta}}{\eta}^{\mu }-\epsilon^{\lambda \mu {p_1}{\eta}} {\eta}^{\nu }+\epsilon^{\lambda \mu \nu {\eta}}\left({p_1}\cdot {\eta}\right)-\epsilon^{\lambda \mu \nu {p_1}} {\eta}^{2} \notag \\ 
\epsilon^{\mu \nu {p_2}{\eta}} {p_2}^{\lambda } & = & {\epsilon }^{\lambda \nu {p_2}{\eta}} {p_2}^{\mu }-\epsilon^{\lambda \mu {p_2}{\eta}} {p_2}^{\nu }+\epsilon^{\lambda \mu \nu {\eta}} {p_2}^{2}-\epsilon^{\lambda \mu \nu {p_2}}
\left({p_2}\cdot {\eta}\right) \notag \\
\epsilon^{\mu \nu{p_2}{\eta}} {\eta}^{\lambda } & = & {\epsilon}^{\lambda \nu {p_2}{\eta}} {\eta}^{\mu}-\epsilon^{\lambda \mu {p_2}{\eta}} {\eta}^{\nu }+\epsilon^{\lambda \mu \nu {\eta}}\left({p_2}\cdot {\eta}\right)-\epsilon^{\lambda \mu \nu{p_2}} {\eta}^{2} \notag \\ 
\epsilon^{\mu	{p_1}{p_2}{\eta}}\delta^{\lambda \nu } & = &\epsilon^{\lambda {p_1}{p_2}{\eta}}
\delta^{\mu \nu }-\epsilon^{\lambda \mu {p_2}{\eta}}{p_1}^{\nu }+\epsilon^{\lambda \mu{p_1}{\eta}} {p_2}^{\nu }-{\epsilon}^{\lambda \mu {p_1}{p_2}} {\eta}^{\nu} \notag \\
\epsilon^{\nu {p_1}{p_2}{\eta}}\delta^{\lambda \mu } & = & \epsilon^{\lambda{p_1}{p_2}{\eta}} \delta^{\mu \nu}-\epsilon^{\lambda \nu {p_2}{\eta}}{p_1}^{\mu }+\epsilon^{\lambda \nu{p_1}{\eta}} {p_2}^{\mu }-{\epsilon}^{\lambda \nu {p_1}{p_2}} {\eta}^{\mu }.
\end{eqnarray}
}    
A second set of Schouten identities can be obtained by considering tensors with two free indices and four contracted ones
\begin{equation}
	\epsilon^{[\mu\nu\alpha\beta}\delta^{\gamma]\rho}, \ \ \ \epsilon^{[\lambda\mu\alpha\beta}\delta^{\gamma]\rho}, \ \ \ \epsilon^{[\lambda\nu\alpha\beta}\delta^{\gamma]\rho}.
\end{equation}
After contracting with the three available momenta, we complete the rank-3 structure using $p_1, p_2, \eta$ and the remaining free index. Using \eqref{schouten1} to eliminate redundancies, we get 21 new independent relations, that we can use to eliminate the following structures
\begin{align} \label{schouten2}
&{p_1}^{\lambda } {p_1}^{\mu }\epsilon^{\nu{p_1}{p_2}\eta}  &   
& {p_1}^{\mu }{p_2}^{\lambda }\epsilon^{\nu{p_1}{p_2}\eta}  & 
& \eta^{\lambda }{p_1}^{\mu }\epsilon^{\nu{p_1}{p_2}\eta}  &  
&  {p_1}^{\lambda} {p_2}^{\mu }\epsilon^{\nu {p_1}{p_2}\eta}  & 
&  {p_2}^{\lambda} {p_2}^{\mu }\epsilon^{\nu{p_1}{p_2}\eta}  &  
&  \eta^{\lambda }{p_2}^{\mu }\epsilon^{\nu{p_1}{p_2}\eta}   &   
&   \eta^{\mu } {p_1}^{\lambda }\epsilon^{\nu {p_1}{p_2}\eta}   \notag\\    
&\eta^{\mu }{p_2}^{\lambda }\epsilon^{\nu {p_1}{p_2}\eta}  &  
& \eta^{\lambda } \eta^{\mu }\epsilon^{\nu {p_1}{p_2}\eta}  &  
& {p_1}^{\lambda} {p_1}^{\nu }\epsilon^{\mu {p_1}{p_2}\eta}  &   
& {p_1}^{\lambda} {p_2}^{\nu }\epsilon^{\mu {p_1}{p_2}\eta}  &    
&  \eta^{\nu } {p_1}^{\lambda }\epsilon^{\mu  {p_1}{p_2}\eta}  &   
&   {p_1}^{\nu } {p_2}^{\lambda }\epsilon^{\mu	{p_1}{p_2}\eta}  &  
&    {p_2}^{\lambda} {p_2}^{\nu }\epsilon^{\mu	{p_1}{p_2}\eta}  \notag\\  
&    \eta^{\nu }{p_2}^{\lambda }\epsilon^{\mu	{p_1}{p_2}\eta}  & 
&      \eta^{\lambda }{p_1}^{\nu }\epsilon^{\mu	{p_1}{p_2}\eta}   &  
&       \eta^{\lambda }{p_2}^{\nu }\epsilon^{\mu	{p_1}{p_2}\eta}  &  
& \eta^{\lambda }\eta^{\nu }\epsilon^{\mu	{p_1}{p_2}\eta}    &  
& \eta^{\mu }{p_1}^{\nu }\epsilon^{\lambda	{p_1}{p_2}\eta}  &    
&  {p_1}^{\nu }{p_2}^{\mu }\epsilon^{\lambda	{p_1}{p_2}\eta}  & 
&   \eta^{\nu }{p_2}^{\mu }\epsilon^{\lambda	{p_1}{p_2}\eta} .
\end{align}
All the possible relations obtained from other types of antisymmetric tensors have been verified to be automatically satisfied using Eqs. \eqref{schouten1} and \eqref{schouten2}. There are altogether 32 Schouten identities, that lower the number of independent tensors to 28 from the original 60
\begin{align}\label{tensorfinali}
		&{\epsilon }^{\lambda \mu \nu {p_1}} &
		& {\epsilon }^{\lambda \mu \nu {p_2}} &
		 & {\epsilon }^{\lambda \mu \nu {\eta }} &
		  &{\delta}^{\mu \nu } {\epsilon }^{\lambda{p_1}{p_2}{\eta }} & 
		  &{p_1}^{\mu }{\epsilon }^{\lambda \nu {p_1}{p_2}} &
		  &{p_1}^{\mu } {\epsilon }^{\lambda \nu{p_1}{\eta }} &
		 &{p_1}^{\mu } {\epsilon}^{\lambda \nu {p_2}{\eta }} \notag \\
		&{p_2}^{\mu } {\epsilon }^{\lambda \nu{p_1}{p_2}} & 
		&{p_2}^{\mu } {\epsilon}^{\lambda \nu {p_1}{\eta }} & 
		&{p_2}^{\mu }{\epsilon }^{\lambda \nu {p_2}{\eta }} & 
		&{\eta}^{\mu } {\epsilon }^{\lambda \nu {p_1}{p_2}} &
		&{\eta }^{\mu } {\epsilon }^{\lambda \nu {p_1}{\eta}} & 
		&{\eta }^{\mu } {\epsilon }^{\lambda \nu{p_2}{\eta }} & 
		&{p_1}^{\nu } {\epsilon	}^{\lambda \mu {p_1}{p_2}} \notag \\
		&{p_1}^{\nu } {\epsilon }^{\lambda \mu{p_1}{\eta }} & 
		&{p_1}^{\nu } {\epsilon}^{\lambda \mu {p_2}{\eta }} & 
		&{p_1}^{\mu }{p_1}^{\nu } {\epsilon }^{\lambda {p_1}{p_2}{\eta }} & 
		&{p_2}^{\nu }{\epsilon }^{\lambda \mu {p_1}{p_2}} &
		&{p_2}^{\nu } {\epsilon }^{\lambda \mu{p_1}{\eta }} & 
		&{p_2}^{\nu } {\epsilon}^{\lambda \mu {p_2}{\eta }} &
		 &{p_1}^{\mu }{p_2}^{\nu } {\epsilon }^{\lambda{p_1}{p_2}{\eta }}\notag \\
		&{p_2}^{\mu } {p_2}^{\nu } {\epsilon }^{\lambda {p_1}{p_2}{\eta }} & 
		&{\eta }^{\mu }{p_2}^{\nu } {\epsilon }^{\lambda{p_1}{p_2}{\eta }} &
		 &{\eta }^{\nu }{\epsilon }^{\lambda \mu {p_1}{p_2}} &
		 &{\eta}^{\nu } {\epsilon }^{\lambda \mu {p_1}{\eta }} &
		&{\eta }^{\nu } {\epsilon }^{\lambda \mu {p_2}{\eta}} &
		 &{\eta }^{\nu } {p_1}^{\mu } {\epsilon }^{\lambda {p_1}{p_2}{\eta }} & 
		 &{\eta }^{\mu }{\eta }^{\nu } {\epsilon }^{\lambda{p_1}{p_2}{\eta }}.
\end{align}
At this stage, the vertex is simplified into the form 
\begin{align}\label{le28}
\Gamma^{\lambda\mu\nu}(p_1,p_2,\eta)&=
	 B_1({p_1},{p_2},\eta ) {\epsilon }^{\lambda \mu \nu {{p_1}}}
	+B_2({p_1},{p_2},\eta) {\epsilon }^{\lambda \mu \nu {{p_2}}}
	+B_3({p_1},{p_2},\eta ) {\epsilon }^{\lambda \mu \nu {\eta}} \notag \\ &
	+{{p_1}}^{\mu } B_4({p_1},{p_2},\eta ) {\epsilon }^{\lambda \nu {{p_1}}{{p_2}}}
   +{{p_1}}^{\mu } B_5({p_1},{p_2},\eta ) {\epsilon }^{\lambda \nu {{p_1}}{\eta }}
   +{{p_1}}^{\mu } B_6({p_1},{p_2},\eta) {\epsilon }^{\lambda \nu{{p_2}}{\eta}} \notag \\ &
   +{{p_2}}^{\mu } B_7({p_1},{p_2},\eta) {\epsilon }^{\lambda \nu{{p_1}}{{p_2}}}
   +{{p_2}}^{\mu } B_8({p_1},{p_2},\eta ) {\epsilon}^{\lambda \nu {{p_1}}{\eta}}
   +{{p_2}}^{\mu } B_9({p_1},{p_2},\eta  ) {\epsilon }^{\lambda \nu   {{p_2}}{\eta }}  \notag \\ &
   +{\eta }^{\mu   } B_{10}({p_1},{p_2},\eta ) {\epsilon   }^{\lambda \nu   {{p_1}}{{p_2}}}
   +{\eta   }^{\mu } B_{11}({p_1},{p_2},\eta ) {\epsilon   }^{\lambda \nu {{p_1}}{\eta   }}
   +{\eta }^{\mu } B_{12}({p_1},{p_2},\eta   ) {\epsilon }^{\lambda \nu   {{p_2}}{\eta   }} \notag \\ &
   +{{p_1}}^{\nu }  B_{13}({p_1},{p_2},\eta ) {\epsilon }^{\lambda   \mu   {{p_1}}{{p_2}}}
   +{{p_1}}^{\nu } B_{14}({p_1},{p_2},\eta ) {\epsilon   }^{\lambda \mu {{p_1}}{\eta   }}
   +{{p_1}}^{\nu }   B_{15}({p_1},{p_2},\eta ) {\epsilon }^{\lambda   \mu {{p_2}}{\eta   }}  \notag \\ &
   +{{p_2}}^{\nu }   B_{16}({p_1},{p_2},\eta ) {\epsilon }^{\lambda   \mu   {{p_1}}{{p_2}}}
   +{{p_2}}^{\nu } B_{17}({p_1},{p_2},\eta ) {\epsilon}^{\lambda \mu {{p_1}}{\eta   }}
   +{{p_2}}^{\nu }   B_{18}({p_1},{p_2},\eta ) {\epsilon }^{\lambda   \mu {{p_2}}{\eta }}  \notag \\ &
   +{\eta   }^{\nu } B_{19}({p_1},{p_2},\eta ) {\epsilon   }^{\lambda \mu   {{p_1}}{{p_2}}}
   +{\eta   }^{\nu } B_{20}({p_1},{p_2},\eta ) {\epsilon   }^{\lambda \mu {{p_1}}{\eta   }}
   +{\eta }^{\nu } B_{21}({p_1},{p_2},\eta   ) {\epsilon }^{\lambda \mu   {{p_2}}{\eta   }}  \notag \\ &
   +{{p_1}}^{\mu } {{p_1}}^{\nu }   B_{22}({p_1},{p_2},\eta ) {\epsilon }^{\lambda   {{p_1}}{{p_2}}{\eta   }}
   +{{p_1}}^{\mu } {{p_2}}^{\nu }   B_{23}({p_1},{p_2},\eta ) {\epsilon }^{\lambda   {{p_1}}{{p_2}}{\eta   }}  \notag \\ &
   +{\eta }^{\nu } {{p_1}}^{\mu }   B_{24}({p_1},{p_2},\eta ) {\epsilon }^{\lambda   {{p_1}}{{p_2}}{\eta   }}
   +{{p_2}}^{\mu } {{p_2}}^{\nu }   B_{25}({p_1},{p_2},\eta ) {\epsilon }^{\lambda   {{p_1}}{{p_2}}{\eta   }}  \notag \\ &
   +{\eta }^{\mu } {{p_2}}^{\nu }   B_{26}({p_1},{p_2},\eta ) {\epsilon }^{\lambda   {{p_1}}{{p_2}}{\eta   }}
   +{\eta }^{\mu } {\eta }^{\nu }   B_{27}({p_1},{p_2},\eta ) {\epsilon }^{\lambda   {{p_1}}{{p_2}}{\eta }}  \notag \\ &
   +{\delta}^{\mu \nu } B_{28}({p_1},{p_2},\eta )   {\epsilon }^{\lambda   {{p_1}}{{p_2}}{\eta   }}
\end{align}
(reduction ${\bf 60\to 28})$ 
that we are going to simplify even further by imposing the Bose symmetry of the two photon lines. 
The relations among the form factors can be found in Appendix \ref{bose}, where we report the structure of the vertex after imposing this symmetry. 
\subsection{Bose symmetry} 
The relations among the form factors obtained by imposing the Bose symmetry are presented in the Appendix \ref{bose}. Using those relations the expression of the amplitude takes the form 
	\begin{align}
	\Gamma^{\lambda\mu\nu}(p_1,p_2,\eta) & =
B_1({p_1},{p_2},\eta ) {\epsilon }^{\lambda    \mu \nu {{p_1}}}
-B_1({p_2},{p_1},\eta   ) {\epsilon }^{\lambda \mu \nu   {{p_2}}}
+B_3({p_1},{p_2},\eta )   {\epsilon }^{\lambda \mu \nu {\eta   }}  \notag \\ &
+{{p_1}}^{\mu } B_4({p_1},{p_2},\eta   ) {\epsilon }^{\lambda \nu   {{p_1}}{{p_2}}}
-{{p_2}}^{\nu } B_4({p_2},{p_1},\eta ) {\epsilon   }^{\lambda \mu   {{p_1}}{{p_2}}}
+{{p_1}}^{\mu } B_5({p_1},{p_2},\eta ) {\epsilon   }^{\lambda \nu {{p_1}}{\eta   }}   \notag \\ 
\end{align}
\begin{align}
& +{{p_2}}^{\nu } B_5({p_2},{p_1},\eta   ) {\epsilon }^{\lambda \mu   {{p_2}}{\eta   }}
+{{p_1}}^{\mu } B_6({p_1},{p_2},\eta   ) {\epsilon }^{\lambda \nu   {{p_2}}{\eta   }}   
	+{p_2}^\nu B_6(p_2,p_1,\eta   ) \epsilon^{\lambda \mu  p_1 \eta }   \notag \\ &
	+{{p_2}}^{\mu } B_7({p_1},{p_2},\eta   ) {\epsilon }^{\lambda \nu   {{p_1}}{{p_2}}}
	-{{p_1}}^{\nu } B_7({p_2},{p_1},\eta ) {\epsilon   }^{\lambda \mu   {{p_1}}{{p_2}}}+{{p_2}}^{\mu } B_8({p_1},{p_2},\eta ) {\epsilon   }^{\lambda \nu {{p_1}}{\eta   }}   \notag \\ &
	+{{p_1}}^{\nu } B_8({p_2},{p_1},\eta   ) {\epsilon }^{\lambda \mu   {{p_2}}{\eta   }}
	+{{p_2}}^{\mu } B_9({p_1},{p_2},\eta   ) {\epsilon }^{\lambda \nu   {{p_2}}{\eta   }}
	+{{p_1}}^{\nu } B_9({p_2},{p_1},\eta   ) {\epsilon }^{\lambda \mu   {{p_1}}{\eta }}   \notag \\ &
	+{\eta }^{\mu   } B_{10}({p_1},{p_2},\eta ) {\epsilon   }^{\lambda \nu   {{p_1}}{{p_2}}}
	-{\eta   }^{\nu } B_{10}({p_2},{p_1},\eta ) {\epsilon   }^{\lambda \mu   {{p_1}}{{p_2}}}
	+{\eta   }^{\mu } B_{11}({p_1},{p_2},\eta ) {\epsilon   }^{\lambda \nu {{p_1}}{\eta   }}   \notag \\ &
	+{\eta }^{\nu } B_{11}({p_2},{p_1},\eta   ) {\epsilon }^{\lambda \mu   {{p_2}}{\eta }}
	+{\eta }^{\mu   } B_{12}({p_1},{p_2},\eta ) {\epsilon   }^{\lambda \nu {{p_2}}{\eta   }}
	+{\eta }^{\nu } B_{12}({p_2},{p_1},\eta   ) {\epsilon }^{\lambda \mu   {{p_1}}{\eta   }}   \notag \\ &
	+{{p_1}}^{\mu } {{p_1}}^{\nu }   B_{22}({p_1},{p_2},\eta ) {\epsilon }^{\lambda   {{p_1}}{{p_2}}{\eta   }}
	-{{p_2}}^{\mu } {{p_2}}^{\nu }   B_{22}({p_2},{p_1},\eta ) {\epsilon }^{\lambda   {{p_1}}{{p_2}}{\eta   }}  \notag \\ &
	+{{p_1}}^{\mu } {{p_2}}^{\nu }   B_{23}({p_1},{p_2},\eta ) {\epsilon }^{\lambda   {{p_1}}{{p_2}}{\eta   }} 
	+{\eta }^{\nu } {{p_1}}^{\mu }   B_{24}({p_1},{p_2},\eta ) {\epsilon }^{\lambda   {{p_1}}{{p_2}}{\eta   }}   \notag \\ &
	-{\eta }^{\mu } {{p_2}}^{\nu }   B_{24}({p_2},{p_1},\eta ) {\epsilon }^{\lambda   {{p_1}}{{p_2}}{\eta   }}
	+{\eta }^{\mu } {\eta }^{\nu }   B_{27}({p_1},{p_2},\eta ) {\epsilon }^{\lambda   {{p_1}}{{p_2}}{\eta }}     \notag \\ &
	+{\delta}^{\mu \nu } B_{28}({p_1},{p_2},\eta )   {\epsilon }^{\lambda   {{p_1}}{{p_2}}{\eta   }},
\end{align}
which, at this stage, is expressed in terms of 16 independent form factors. Therefore, we have reduced the structures according to the sequence  ${\bf 60\to 28\to 16}$ from the original (${\bf 60}$) ones.
\subsection{Vector Ward identities}
The last step involves the implementation of the Ward identities on the vector lines. 
We proceed with the first requirement, and impose the conservation Ward identities on the two vector indices $\mu,\nu$
\begin{equation}
{p_1}_\mu \Gamma^{\lambda\mu\nu}(p_1,p_2,\eta)={p_2}_\nu\Gamma^{\lambda\mu\nu}(p_1,p_2,\eta)=0.
\end{equation}
By imposing these conditions, we get the additional relations
\begin{equation}\label{wicasononsimm1}
B_8({p_2},{p_1},\eta )= -{{p_1}}^{2}
   B_{22}({p_1},{p_2},\eta )-{{p_2}}^{2} B_{22}({p_2},{p_1},\eta
   )-B_8({p_1},{p_2},\eta )
\end{equation}
\begin{equation}\label{wicasononsimm2}
B_{11}({p_2},{p_1},\eta )=
   \frac{-\left({{p_1}}\cdot {\eta }\right)
   B_{11}({p_1},{p_2},\eta )-{{p_1}}^{2} \left({{p_2}}\cdot
   {\eta }\right) B_{24}({p_1},{p_2},\eta )-{{p_2}}^{2}
   \left({{p_1}}\cdot {\eta }\right) B_{24}({p_2},{p_1},\eta
   )}{{{p_2}}\cdot {\eta }}
\end{equation}
that will be essential for further simplifications. \\
Substituting \eqref{VecWIoff} into $\Gamma^{\lambda\mu\nu}(p_1,p_2,\eta)$ we recover the final expression of the amplitude 
{\allowdisplaybreaks
\begin{align}
 \label{gen}
\Gamma^{\lambda\mu \nu } (p_1, p_2,\eta)&= \chi_1^{\lambda\mu\nu}\left(p_1,p_2,\eta \right)\bar B_1\left(p_1,p_2,\eta\right) + \chi_1^{\lambda\mu\nu}\left(p_2,p_1,\eta \right)\bar B_1\left(p_2,p_1,\eta\right)   \notag \\ &
+ \chi_2^{\lambda\mu\nu}\left(p_1,p_2,\eta \right)\bar B_2\left(p_1,p_2,\eta \right)  + \chi_2^{\lambda\mu\nu}\left(p_2,p_1,\eta \right)\bar B_2\left(p_2,p_1,\eta\right)   \notag \\ &
+ \chi_3^{\lambda\mu\nu}\left(p_1,p_2,\eta \right)\bar B_3\left(p_1,p_2,\eta \right)  + \chi_3^{\lambda\mu\nu}\left(p_2,p_1,\eta \right)\bar B_3\left(p_2,p_1,\eta\right)   \notag \\ &
+ \chi_4^{\lambda\mu\nu}\left(p_1,p_2,\eta \right)\bar  B_4\left(p_1,p_2,\eta \right)+ \chi_4^{\lambda\mu\nu}\left(p_2,p_1,\eta \right)\bar B_4\left(p_2,p_1,\eta\right)   \notag \\ &
+ \chi_5^{\lambda\mu\nu}\left(p_1,p_2,\eta \right)\bar  B_5\left(p_1,p_2,\eta \right)+ \chi_5^{\lambda\mu\nu}\left(p_2,p_1,\eta \right)\bar B_5\left(p_2,p_1,\eta\right)   \notag \\ &
+ \chi_6^{\lambda\mu\nu}\left(p_1,p_2,\eta \right)\bar B_{6}\left(p_1,p_2,\eta \right) + \chi_6^{\lambda\mu\nu}\left(p_2,p_1,\eta \right)\bar B_6\left(p_2,p_1,\eta\right)   \notag \\ &
+ \chi_7^{\lambda\mu\nu}\left(p_1,p_2,\eta \right)\bar B_{7}\left(p_1,p_2,\eta \right) + \chi_7^{\lambda\mu\nu}\left(p_2,p_1,\eta \right)\bar B_7\left(p_2,p_1,\eta\right)   \notag \\ &
+ \chi_8^{\lambda\mu\nu}\left(p_1,p_2,\eta \right)\bar B_{8}\left(p_1,p_2,\eta \right) + \chi_8^{\lambda\mu\nu}\left(p_2,p_1,\eta \right)\bar B_8\left(p_2,p_1,\eta\right)   \notag \\ &
+ \chi_9^{\lambda\mu\nu}\left(p_1,p_2,\eta \right)\bar B_{9}\left(p_1,p_2,\eta \right) +  \chi_{10}^{\lambda\mu\nu}\left(p_1,p_2,\eta \right)\bar B_{10}\left(p_1,p_2,\eta \right) ,
\end{align} 

(reduction {${\bf 28\to 10}$ generic),
where we have introduced the new tensor structures
{ \allowdisplaybreaks
\begin{align}
\label{tens}
&\chi_{1}^{\lambda\mu\nu}(p_1,p_2,\eta)=\left({\epsilon }^{\lambda \mu \nu {p}_2} p_1^2+{\epsilon }^{\lambda \nu {p}_1{p}_2} {p}_1{}^{\mu }\right)\notag\\
&
\chi_{2}^{\lambda\mu\nu}(p_1,p_2,\eta)=\left({\epsilon }^{\lambda \nu	{p}_2{\eta }} {p}_1{}^{\mu }- \frac{{\epsilon }^{\lambda \nu {p}_2{\eta }} {\eta }^{\mu } p_1^2}{{\eta }\cdot 	{p}_1}\right)\notag\\&
\chi_{3}^{\lambda\mu\nu}(p_1,p_2,\eta)=\left({\epsilon }^{\lambda \nu 	{p}_1{p}_2} {p}_2{}^{\mu }+{\epsilon }^{\lambda \mu \nu  {p}_2} \left({p}_1\cdot {p}_2\right)\right)\notag\\&
\chi_{4}^{\lambda\mu\nu}(p_1,p_2,\eta)=\left({\epsilon}^{\lambda \nu {{p_2}}{\eta }} {{p_2}}^{\mu }-\frac{{\epsilon }^{\lambda \nu {{p_2}}{\eta }} {\eta }^{\mu } \left({{p_1}}\cdot {{p_2}}\right)}{{{p_1}}\cdot {\eta }}\right)\notag\\&     
\chi_{5}^{\lambda\mu\nu}(p_1,p_2,\eta)=\left({\epsilon }^{\lambda \nu {{p_1}}{{p_2}}} {\eta }^{\mu }+{\epsilon }^{\lambda \mu \nu
   {{p_2}}} \left({{p_1}}\cdot {\eta }\right)\right) \notag\\&    
\chi_{6}^{\lambda\mu\nu}(p_1,p_2,\eta)=\biggl({\epsilon }^{\lambda {{p_1}}{{p_2}}{\eta }} {{p_1}}^{\mu } {{p_1}}^{\nu }-\frac{1}{2} {\epsilon }^{\lambda \mu {{p_2}}{\eta }} {{p_1}}^{2} {{p_1}}^{\nu }-\frac{1}{2} {\epsilon }^{\lambda {{p_1}}{{p_2}}{\eta }} {g}^{\mu \nu } {{p_1}}^{2}-\frac{1}{2}{\epsilon }^{\lambda \nu {{p_1}}{\eta }} {{p_2}}^{\mu } {{p_1}}^{2}\notag\\&+{\epsilon }^{\lambda \nu {{p_1}}{\eta }} {{p_1}}^{\mu } \left({{p_1}}\cdot {{p_2}}\right)+\frac{1}{2} {\epsilon }^{\lambda \mu \nu {\eta }} {{p_1}}^{2} \left({{p_1}}\cdot {{p_2}}\right)\biggl)\notag\\&    
\chi_{7}^{\lambda\mu\nu}(p_1,p_2,\eta)=\left(\frac{1}{2} {\epsilon }^{\lambda {{p_1}}{{p_2}}{\eta }} {{p_1}}^{\mu } {{p_2}}^{\nu }-\frac{1}{2} {\epsilon }^{\lambda \mu {{p_2}}{\eta }}
   {{p_1}}^{2} {{p_2}}^{\nu }+\frac{1}{2} {\epsilon }^{\lambda \nu {{p_1}}{\eta }} {{p_1}}^{\mu } {{p_2}}^{2}+\frac{1}{2} {\epsilon }^{\lambda \mu \nu {\eta }} {{p_1}}^{2} {{p_2}}^{2}\right) \notag\\&     
\chi_{8}^{\lambda\mu\nu}(p_1,p_2,\eta)=\biggl({\epsilon }^{\lambda {{p_1}}{{p_2}}{\eta }} {{p_1}}^{\mu } {\eta }^{\nu }-\frac{1}{2} {\epsilon }^{\lambda \mu {{p_2}}{\eta }} {{p_1}}^{2} {\eta }^{\nu }-\frac{{\epsilon }^{\lambda {{p_1}}{{p_2}}{\eta }} {\eta }^{\mu } {{p_1}}^{2} {\eta }^{\nu }}{2 \left({{p_1}}\cdot {\eta }\right)}\notag\\&+{\epsilon }^{\lambda \nu {{p_1}}{\eta }} {{p_1}}^{\mu } \left({{p_2}}\cdot {\eta }\right)+\frac{1}{2} {\epsilon }^{\lambda \mu \nu {\eta
   }} {{p_1}}^{2} \left({{p_2}}\cdot {\eta }\right)-\frac{{\epsilon }^{\lambda \nu {{p_1}}{\eta }} {\eta }^{\mu } {{p_1}}^{2} \left({{p_2}}\cdot {\eta }\right)}{2 \left({{p_1}}\cdot {\eta }\right)}\biggl) \notag\\&        
\chi_{9}^{\lambda\mu\nu}(p_1,p_2,\eta)=\left({\epsilon }^{\lambda {{p_1}}{{p_2}}{\eta }} {g}^{\mu \nu }+{\epsilon }^{\lambda \nu {{p_1}}{\eta }}
   {{p_2}}^{\mu }-{\epsilon }^{\lambda \mu {{p_2}}{\eta }} {{p_1}}^{\nu }+{\epsilon }^{\lambda \mu \nu {\eta }} \left({{p_1}}\cdot {{p_2}}\right)\right) \notag\\&   
\chi_{10}^{\lambda\mu\nu}(p_1,p_2,\eta)=\left({\epsilon }^{\lambda \mu \nu {\eta }}+\frac{{\epsilon }^{\lambda \nu {{p_1}}{\eta }} {\eta }^{\mu }}{{{p_1}}\cdot {\eta }}-\frac{{\epsilon }^{\lambda \mu {{p_2}}{\eta }} {\eta }^{\nu }}{{{p_2}}\cdot {\eta }}+\frac{{\epsilon }^{\lambda {{p_1}}{{p_2}}{\eta }} {\eta }^{\mu } {\eta }^{\nu }}{\left({{p_1}}\cdot {\eta}\right) \left({{p_2}}\cdot {\eta }\right)}\right).       
\end{align}}
Therefore, using the Schouten relations and imposing the symmetry of the interaction, we have simplified the off-shell vertex from its original expression in terms of 60 form factors in \eqref{fdecom}, down to 10. 
The procedure can be summarized according to this sequence: {\bf 60 } initial form factors $\to$ {\bf 28} form factors using the 32 Schouten relations $\to$ {\bf 16} using the Bose symmetry $\to$ {\bf 10} using the vector WIs.

\section{Symmetric decomposition with off-shell photons}
A simplified decomposition follows if we impose some kinematical constraints. We consider the following additional conditions
\begin{align} \label{symmetricconditions}
	p_1^2 &= p_2^2 = p^2 \notag \\
	p_1 \cdot \eta &= p_2 \cdot \eta = p \cdot \eta,  
\end{align}
in which the two photons have equal invariant mass as well as equal projections on the four-vector of the heat-bath. 
These relations constrain the form factors to depend  only on three scalar products, namely $q^2$, $p^2$ and $p\cdot \eta$ and are automatically invariant under the exchange ($p_1\leftrightarrow p_2$).\\ 
The Schouten identities are left almost untouched, the main difference being merely a relabeling of their scalar products.
From the 60 structures included in \eqref{60tensors}, using the Schouten relations, also in this case we 
end up with 28 form factors and tensor structures, on which we apply the constraints from Bose symmetry. 
These are more constraining, leaving us with only 12 form factors instead of 16. They take the form 

\begin{align} \label{BosSybSubON}
		B_2 \, &= -B_1 &
		B_3 \,& = -B_3 \notag \\
		B_{13}&= -B_7 &
		B_{14}&= B_9 \notag \\
		B_{15}&= B_8&
		B_{16}&= -B_4 \notag \\
		B_{17}&= B_6  &
		B_{18}&= B_5  \notag \\
		B_{19}&= -B_{10} &
		B_{20}&= B_{12} \notag \\
		B_{21}&= B_{11} &
		B_{23}&= -B_{23} \notag \\
		B_{25}&= -B_{22} &
		B_{26}&= -B_{24} \notag \\
		B_{27}&= -B_{27} &
		B_{28}&= -B_{28}  
\end{align}
that differ from \eqref{BosSybSub}. 
It is clear that the relations regarding $B_3$, $B_{23}$, $B_{27}$ and $B_{28}$ now force such form factors to be zero. Therefore their number is reduced to 12 instead of 16 and their decomposition is given by
\begin{align} \label{FFDecompB1}
	\Gamma^{\lambda\mu\nu}(p_1,p_2,\eta)&=
	B_5 \left({{p_1}}^{\mu } {\epsilon }^{\lambda \nu {{p_1}}{\eta }}+{{p_2}}^{\nu } {\epsilon }^{\lambda  \mu {{p_2}}{\eta }}\right)
	+B_6 \left({{p_1}}^{\mu } {\epsilon }^{\lambda \nu {{p_2}}{\eta }}+{{p_2}}^{\nu } {\epsilon }^{\lambda \mu {{p_1}}{\eta }}\right)
	+B_8 \left({{p_2}}^{\mu } {\epsilon }^{\lambda \nu {{p_1}}{\eta }}+{{p_1}}^{\nu } {\epsilon }^{\lambda \mu  {{p_2}}{\eta }}\right)  \notag \\ &
	+B_9 \left({{p_1}}^{\nu } {\epsilon }^{\lambda \mu {{p_1}}{\eta}}+{{p_2}}^{\mu } {\epsilon }^{\lambda \nu {{p_2}}{\eta }}\right) 
	+B_{10} \left({\eta }^{\mu } {\epsilon	}^{\lambda \nu {{p_1}}{{p_2}}}-{\eta }^{\nu } {\epsilon }^{\lambda \mu	{{p_1}}{{p_2}}}\right) 
	+B_{11} \left({\eta }^{\mu } {\epsilon }^{\lambda \nu {{p_1}}{\eta	}}+{\eta }^{\nu } {\epsilon }^{\lambda \mu {{p_2}}{\eta }}\right)  \notag \\ &
	+B_{12} \left({\eta }^{\nu } {\epsilon}^{\lambda \mu {{p_1}}{\eta }}+{\eta }^{\mu } {\epsilon }^{\lambda \nu {{p_2}}{\eta	}}\right)  
	+B_{22} \left({{p_1}}^{\mu } {{p_1}}^{\nu } {\epsilon }^{\lambda {{p_1}}{{p_2}}{\eta }}-{{p_2}}^{\mu } {{p_2}}^{\nu } {\epsilon }^{\lambda	{{p_1}}{{p_2}}{\eta }}\right)  \notag \\ &
	+B_4 \left({{p_1}}^{\mu } {\epsilon }^{\lambda \nu	{{p_1}}{{p_2}}}-{{p_2}}^{\nu } {\epsilon }^{\lambda \mu {{p_1}}{{p_2}}}\right)
	+B_7\left({{p_2}}^{\mu } {\epsilon }^{\lambda \nu {{p_1}}{{p_2}}}-{{p_1}}^{\nu } {\epsilon	}^{\lambda \mu {{p_1}}{{p_2}}}\right) \notag \\ &
	+B_{24} \left({\eta }^{\nu } {{p_1}}^{\mu } {\epsilon }^{\lambda {{p_1}}{{p_2}}{\eta }}-{\eta }^{\mu } {{p_2}}^{\nu } {\epsilon }^{\lambda 	{{p_1}}{{p_2}}{\eta }}\right) 
	+B_1 \left({\epsilon }^{\lambda \mu \nu {{p_1}}}-{\epsilon }^{\lambda	\mu \nu {{p_2}}}\right).
\end{align}
At this stage, we finally need to impose the conservation Ward identity
\begin{align}
	{p_2}_{\nu} \Gamma^{\lambda\mu\nu} (p_1, p_2 , \eta)&={\epsilon }^{\lambda \mu {{p_1}}{{p_2}}} \Big(-B_{10} \left({p}\cdot {\eta }\right)-B_7 \left({{p_1}}\cdot{{p_2}}\right)-B_1-B_4 \, p^2\Big)  \notag\\
	&+{\epsilon }^{\lambda \mu {{p_2}}{\eta }} \Big(B_{11} \left({p}\cdot {\eta}\right)+B_8 \left({{p_1}}\cdot {{p_2}}\right)  + B_5 \, p^2 \Big) +{\epsilon }^{\lambda \mu {{p_1}}{\eta }} \Big(B_{12}	\left({p}\cdot {\eta }\right)+B_9 \left({{p_1}}\cdot {{p_2}}\right)  + B_6 \, p^2\Big)  \notag\\
	&+ {\epsilon }^{\lambda {{p_1}}{{p_2}}{\eta }} \Big(- (B_{11} + B_{24} \, p^2) {\eta }^{\mu }+{{p_1}}^{\mu } \left[ B_{24} \left({p}\cdot	{\eta }\right)+B_{22} \left({{p_1}}\cdot {{p_2}}\right)-B_5\right]-(B_8 + B_{22} \, p^2) {p_2}^{\mu } \Big)=0,
\end{align}
obtaining the following set of relations
\begin{eqnarray}\label{eqreffff}
	B_1 &=& -B_7 \left( p_1 \cdot p_2 \right)-B_{10} (p\cdot \eta )-B_4 \, p^2  \notag\\
	B_5  &=& B_{22} \left(p_1 \cdot p_2  \right)+B_{24} (p\cdot \eta )  \notag\\
	B_8  &=& -B_{22} \, p^2  \notag\\
	B_{11}  &=& -B_{24} \, p^2  \notag\\
	B_{12}  &=& -\frac{B_9 \left(p_1 \cdot {p}_2\right)+B_6 \, p^2}{p\cdot	\eta }. 
\end{eqnarray}
The decomposition thus can be expressed in terms of a reduced number of 7 form factors, compared to the 10 of the general decomposition 

\begin{align} 
\label{gen1}
&\Gamma^{\lambda\mu\nu}(p_1,p_2,\eta) =
 \bar B_1\chi_1^{\lambda\mu\nu} +\bar B_2\chi_2^{\lambda\mu\nu} + \bar B_{3} \chi_3^{\lambda\mu\nu}+\bar B_{4}\chi_4^{\lambda\mu\nu}+ \bar B_{5}\chi_5^{\lambda\mu\nu} + \bar B_6\chi_6^{\lambda\mu\nu}+\bar B_7\chi_7^{\lambda\mu\nu},
   \end{align}
   (reduction ${\bf 10\to 7}$ for off-shell photons of equal invariant mass and $p_1\cdot \eta=p_2\cdot \eta$)\\
 where  
 \begin{align}
&\chi_1^{\lambda\mu\nu}=\left(-\frac{{p}^{2} {\eta }^{\nu } \bar{\epsilon }^{\lambda \mu {{p_1}}{\eta }}}{p\cdot {\eta }}-\frac{{p}^{2} {\eta }^{\mu } \bar{\epsilon }^{\lambda \nu {{p_2}}{\eta }}}{p\cdot {\eta }}+{{p_1}}^{\mu } \bar{\epsilon }^{\lambda \nu {{p_2}}{\eta }}+{{p_2}}^{\nu } \bar{\epsilon }^{\lambda \mu {{p_1}}{\eta }}\right)\notag\\&
\chi_2^{\lambda\mu\nu}=\left(-\frac{{\eta }^{\mu } \left({{p_1}}\cdot {{p_2}}\right) \bar{\epsilon }^{\lambda \nu {{p_2}}{\eta }}}{p\cdot {\eta }}-\frac{{\eta }^{\nu } \left({{p_1}}\cdot {{p_2}}\right) \bar{\epsilon }^{\lambda \mu {{p_1}}{\eta }}}{p\cdot {\eta }}+{{p_1}}^{\nu } \bar{\epsilon }^{\lambda \mu {{p_1}}{\eta }}+{{p_2}}^{\mu } \bar{\epsilon }^{\lambda \nu {{p_2}}{\eta }}\right)\notag\\&
\chi_3^{\lambda\mu\nu}=\left(-\left(p\cdot {\eta }\right) \bar{\epsilon }^{\lambda \mu \nu {{p_1}}}+\left(p\cdot {\eta }\right) \bar{\epsilon }^{\lambda \mu \nu {{p_2}}}+{\eta }^{\mu } \bar{\epsilon }^{\lambda \nu{{p_1}}{{p_2}}}-{\eta }^{\nu } \bar{\epsilon }^{\lambda \mu {{p_1}}{{p_2}}}\right)\notag\\&
 \chi_4^{\lambda\mu\nu}=\left(-{p}^{2} {{p_1}}^{\nu } \bar{\epsilon }^{\lambda \mu {{p_2}}{\eta }}-{p}^{2} {{p_2}}^{\mu } \bar{\epsilon }^{\lambda \nu {{p_1}}{\eta }}+{{p_1}}^{\mu } {{p_1}}^{\nu } \bar{\epsilon }^{\lambda {{p_1}}{{p_2}}{\eta }}-{{p_2}}^{\mu } {{p_2}}^{\nu } \bar{\epsilon }^{\lambda {{p_1}}{{p_2}}{\eta }}+{{p_1}}^{\mu } \left({{p_1}}\cdot {{p_2}}\right) \bar{\epsilon }^{\lambda \nu {{p_1}}{\eta }}+{{p_2}}^{\nu } \left({{p_1}}\cdot {{p_2}}\right) \bar{\epsilon }^{\lambda \mu {{p_2}}{\eta }}\right)\notag\\&
 \chi_5^{\lambda\mu\nu}=\left(-{p}^{2} {\eta }^{\mu } \bar{\epsilon }^{\lambda \nu {{p_1}}{\eta }}+{{p_1}}^{\mu } \left(p\cdot {\eta }\right) \bar{\epsilon }^{\lambda \nu {{p_1}}{\eta }}-{p}^{2} {\eta }^{\nu } \bar{\epsilon }^{\lambda \mu {{p_2}}{\eta }}+{{p_2}}^{\nu }\left(p\cdot {\eta }\right) \bar{\epsilon }^{\lambda \mu {{p_2}}{\eta }}+{\eta }^{\mu } {{p_2}}^{\nu } \left(-\bar{\epsilon }^{\lambda {{p_1}}{{p_2}}{\eta }}\right)+{\eta }^{\nu } {{p_1}}^{\mu } \bar{\epsilon }^{\lambda {{p_1}}{{p_2}}{\eta }}\right)\notag\\&
\chi_6^{\lambda\mu\nu}=\left(-{p}^{2} \bar{\epsilon }^{\lambda \mu \nu {{p_1}}}+{p}^{2} \bar{\epsilon }^{\lambda \mu \nu {{p_2}}}+{{p_1}}^{\mu } \bar{\epsilon }^{\lambda \nu {{p_1}}{{p_2}}}-{{p_2}}^{\nu } \bar{\epsilon}^{\lambda \mu {{p_1}}{{p_2}}}\right)\notag\\&
\chi_7^{\lambda\mu\nu}= \left({{p_2}}^{\mu } \bar{\epsilon }^{\lambda \nu {{p_1}}{{p_2}}}-{{p_1}}^{\nu } \bar{\epsilon }^{\lambda \mu {{p_1}}{{p_2}}}-\left({{p_1}}\cdot {{p_2}}\right) \bar{\epsilon }^{\lambda \mu \nu {{p_1}}}+\left({{p_1}}\cdot {{p_2}}\right) \bar{\epsilon }^{\lambda \mu \nu {{p_2}}}\right).
\end{align}

\section{The perturbative realization } 
Let us now analyze the perturbative realization of the vertex. The perturbative contribution is expanded into a zero density and a finite density part 
\begin{equation}
\label{eqq}
\Gamma^{\lambda\mu\nu}_{pert}=\Gamma^{\lambda\mu\nu},
\end{equation} 
on the (left-hand) side of  \eqref{eqq}, while the right-hand side of the same equation is identified by the expansion into form factors presented in \eqref{gen} in the general case, and in \eqref{gen1} in the symmetric $p^2\neq 0$ case. 
The perturbative expansion into form factors present in $\Gamma^{\lambda\mu\nu}$,  will be reduced to various sets of scalar equations by suitable projections. \\ We will act with several types of projections on both sides of \eqref{eqq} in order to eliminate the dependence of the antisymmetric $\epsilon$ tensor present on both sides of this equation. \\
 The list of projections is rather long and the equations that are obtained need to be carefully investigated in order to extract the independent ones.  At a final step, the inversion of the linear system of equations derived by the procedure, allows to derive the perturbative expressions of the form factors on the rhs of \eqref{eqq}.\\
While the general description of this procedure will be discussed in a following section, one can check whether the approach matches known results on the perturbative $AVV$ at zero density. For example, it is well-known, in this specific case, that the pertubative vertex reduces only to its longitudinal sector, if the two vector lines are set on-shell. In our case, this check is performed by taking, for example, the $\eta\to 0$ limit of $\eqref{gen1}$, combined with the 
$p^2\to 0$ condition or the transversality of the polarization of the two photons ($\varepsilon_1\cdot p_1=0, 
\varepsilon_2\cdot p_2=0$) on the rhs of \eqref{eqq}. We illustrate the method in some detail.\\
The two contributions to the  $AVV$, given by the direct and the exchange triangle diagrams, are modified at finite $\mu$ as
\begin{equation}
\label{trace}
		\Gamma^{\lambda\mu \nu }_{pert} = \Gamma^{(0)}_{pert} + \Gamma^{(1)}_{pert} =\int {d^4 k \over (2\pi)^4} \Tr[ (\slashed{k}-\slashed{q} )\, \gamma^\nu \, (\slashed{k}-\slashed{p_1}) \, \gamma^\mu \, \slashed{k} \, \gamma^\lambda \, \gamma^5]\, G_F(k) \, G_F(k-p_1) \, G_F(k-q)+ 
		{p_1 \tor p_2 \choose \mu \tor \nu} 
	\end{equation}
where we have separated the scalar and tensor components of the propagator $S_F(k)$ in \eqref{above}, together with their 
	finite density and vacuum parts as
	\begin{eqnarray}
	  S_F(k) &=& \slashed{k} \, G_F(k) \notag \\
		G_F(k) &=& G_0(k) +  G_1(k) \notag \\
		G_0(k) &=& {1 \over k^2}  \notag \\
		 G_1(k) &=& 2 \pi i \, \delta (k^2) \, \theta(\eta \cdot k) \, \theta(\mu - \eta \cdot k).
	\end{eqnarray}
	Notice that the trace part factorizes from the scalar part, as shown in \eqref{trace}. We expand the trace in the following form 
	\begin{equation}
		\Tr[ (\slashed{k}-\slashed{q} )\, \gamma^\nu \, (\slashed{k}-\slashed{p_1}) \, \gamma^\mu \, \slashed{k} \, \gamma^\lambda \, \gamma^5] = i
		D_{1\ \alpha}{}^{\lambda\mu\nu}(p_1, p_2) \, k^\alpha + iD_{2\ \alpha\beta}{}^{\lambda\mu\nu} (p_1, p_2)\, k^\alpha k^\beta + iD_{3\ \alpha\beta\gamma}{}^{\lambda\mu\nu} (p_1, p_2) \, k^\alpha k^\beta k^\gamma
	\end{equation}
	where we have defined the tensors
	\begin{eqnarray}
		D_{1\ \alpha}{}^{\lambda\mu\nu}(p_1,p_2) &=& 4  {g}^{\alpha \mu } {\epsilon }^{\lambda
			\nu {{p_1}}{q}}-4 
		{g}^{\lambda \nu } {\epsilon }^{\alpha
			\mu {{p_1}}{q}}+4 
		{{p_1}}^{\alpha } {\epsilon
		}^{\lambda \mu \nu {q}}-4 
		{q}^{\lambda } {\epsilon }^{\alpha
			\mu \nu {{p_1}}}+4 
		{{p_1}}^{\mu } {\epsilon
		}^{\alpha \lambda \nu {q}}+4 
		{q}^{\nu } {\epsilon }^{\alpha
			\lambda \mu {{p_1}}} \notag
		\\
		D_{2\ \alpha\beta}{}^{\lambda\mu\nu}(p_1,p_2) &=& 4 {g}^{\beta \lambda } {\epsilon
		}^{\alpha \mu \nu {{p_1}}}+4 
		{g}^{\beta \mu } {\epsilon }^{\alpha
			\lambda \nu {{p_1}}}-4 
		{g}^{\beta \nu } {\epsilon }^{\alpha
			\lambda \mu {{p_1}}}-4 
		{g}^{\alpha \beta } {\epsilon
		}^{\lambda \mu \nu {q}}-8 
		{g}^{\beta \mu } {\epsilon }^{\alpha
			\lambda \nu {q}}+4 
		{{p_1}}^{\beta } {\epsilon
		}^{\alpha \lambda \mu \nu } \notag \\
		D_{3\ \alpha\beta\gamma}{}^{\lambda\mu\nu}(p_1,p_2) &=&4  {g}^{\beta \gamma } {\epsilon
		}^{\lambda \mu \nu \alpha }.
	\end{eqnarray}
The perturbative integral can be written as
	\begin{eqnarray}\label{pert}
		\Gamma_{pert}^{\mu\nu\lambda}(p_1,p_2;\beta)
		&=& iD_{1\ \alpha}{}^{\lambda\mu\nu} (p_1 ,p_2) \int \frac{d^4k}{(2\pi)^4}\;k^\alpha \, G_F(k) \, G_F(k-q) \, G_F(k-p_1) \notag \\
		&& +iD_{2\ \alpha\beta}{}^{\lambda\mu\nu} (p_1 ,p_2)\int \frac{d^4k}{(2\pi)^4}\; k^\alpha \ k^\beta  \, G_F(k) \, G_F(k-q) \, G_F(k-p_1) \notag \\
		&& +iD_{3\ \alpha\beta\tau}{}^{\lambda\mu\nu} (p_1 ,p_2)\int \frac{d^4k}{(2\pi)^4k}\; k^\alpha \ k^\beta\ k^\tau  \, G_F(k) \, G_F(k-q) \, G_F(k-p_1) \notag \\
		&& + 
		{p_1 \tor p_2 \choose \mu \tor \nu}.
	\end{eqnarray}

	We will use the following notation to describe such integrals
	\begin{eqnarray}
		J [f(k,p_1,p_2,\eta)] &=& \int \frac{d^4k}{(2\pi)^4}\, f(k, p_1, p_2, \eta) \, G_F(k) \, G_F(k-p_1) \, G_F(k-q) \notag \\
		H [f(k,p_1,p_2,\eta)] &=& \int \frac{d^4k}{(2\pi)^4}\, f(k, p_1, p_2, \eta) \, G_F(k) \, G_F(k-p_2) \, G_F(k-q), \notag \\
	\end{eqnarray}
	where $f(k, p_1, p_2, \eta)$ indicates, generically, a contraction of $p_1, p_2, \eta$ with one or two four-momenta $k^\lambda$  in the numerators. These integrals turn into a sum of eight terms when we expand $G_F(k)$ as a sum of $G_0(k)$ and $ G_1(k)$. \\
We need to introduce some extra notation in order to illustrate the approach, by labeling every term of in $J[\dots]$ and $H[\dots]$ appropriately.\\
We introduce three upper indices, indicating for each $G_F(k), G_F(k-p_1), G_F(k-q)$ which  part of the propagator we are choosing, between $G_0$ (labeled as $f$), for the "free" or vacuum contribution, and $ G_1$ (labeled as $\delta$) for the $\mu$-dependent part. $J[\dots]$ and the $H[\dots]$ refer to the direct and the exchanged graph respectively. Here are some examples
	\begin{eqnarray*}
		J^{(\delta, f, f)}  [f(k,p_1,p_2,\eta)]  = &=& \int \frac{d^4k}{(2\pi)^4}\, f(k, p_1, p_2, \eta) \, G_1(k) \, G_0(k-p_1) \, G_0(k-q),  \\
		H^{(f, \delta, f)} [f(k,p_1,p_2,\eta)]  = &=& \int \frac{d^4k}{(2\pi)^4}\, f(k, p_1, p_2, \eta) \, G_0(k) \,  G_1(k-p_2) \, G_0(k-q),  \\
		J^{(\delta, f, \delta)} [f(k,p_1,p_2,\eta)] &=& \int \frac{d^4k}{(2\pi)^4}\,  [f(k,p_1,p_2,\eta)] \,  G_1(k) \, G_0(k-p_1) \,  G_1(k-q). 
	\end{eqnarray*}
	 In almost all of the finite $T_{emp}/ \mu$ cases, the integration cannot be performed covariantly. Indeed, such integrals can only be computed in a specific frame of reference. This defines the last step of the procedure, that takes place only after having totally contracted both sides of the amplitude in \eqref{eqq} with appropriate projectors in order to extract the scalar expressions  of the form factors. \\ 
	The only UV divergences contained in the $(f,f,f)$ part, that corresponds to the standard AVV diagram at zero temperature and density, are eliminated by the vector Ward identities, while other terms are automatically free of UV divergences due to the presence of a cutoff given by the chemical potential. There can be, however, other types of divergences, such as IR or collinear (the latter present in the on-shell photon case). We will deal with the latter using dimensional regularization (DR). We are going to show that while they are present in specific contributions, they cancel in the complete amplitude.

\subsection{$\Gamma^{(0)}_{pert}$, the pole and the on-shell limit }
\label{vacuum}
Before coming to study the perturbative contributions at finite density, we concentrate on $\Gamma^{(0)}_{pert}$, the $\mu=0$ part of the correlator, summarizing earlier findings. \\
As we have already mentioned, a chiral anomaly vertex can be completely identified by two conditions: the presence of an anomaly pole in the axial-vector channel and the conformal symmetry of the vertex. The method is entirely nonperturbative and has been presented in \cite{Coriano:2023hts}.\\
This is essentially one of the main reasons why it is conceivable to identify the {\em new degree of freedom} inherent to the anomaly, with the pole. The approach does not rely on a Lagrangian, and is therefore nonperturbative.  A similar result has been obtained in the case of the gravitational anomaly, in the $ TTJ_5$ vertex, where $T$ is the stress energy tensor \cite{Coriano:2023gxa}. In this case as well, the reconstruction of the vertex is performed just by the inclusion of a single pole in the axial-vector channel and by requiring conformal symmetry.
 
The first parametrization of the $AVV$ was presented in \cite{Rosenberg:1962pp} in the symmetric form 
(with $p_3^3\equiv q^2$ in this case)
\begin{align}
\langle J^{\mu_1}(p_1)J^{\mu_2}(p_2)J^{\mu_3}_5(p_3) \rangle &= \tilde{B}_1 (p_1, p_2) \varepsilon^{p_1\mu_1\mu_2\mu_3} + \tilde{B}_2 (p_1, p_2)\varepsilon^{p_2\mu_1\mu_2\mu_3} +
\tilde{B}_3 (p_1, p_2) \varepsilon^{p_1p_2\mu_1\mu_3}{p_1}^{\mu_2} \nonumber \\
&+  \tilde{B}_4 (p_1, p_2) \varepsilon^{p_1p_2\mu_1\mu_3}p_2^{\mu_2}
+ \tilde{B}_5 (p_1, p_2)\varepsilon^{p_1p_2\mu_2\mu_3}p_1^{\mu_1}
+ \tilde{B}_6 (p_1, p_2) \varepsilon^{p_1p_2\mu_2\mu_3}p_2^{\mu_1},\nonumber \\
\label{Ross}
\end{align}
with $\tilde{B}_1$ and $\tilde{B}_2$ divergent by power counting.
If we use a diagrammatic evaluation of the correlator, the four invariant amplitudes $\tilde{B}_i$ for $i\geq3$ are given by explicit parametric integrals 
\begin{align}
\label{ssym}
\tilde{B}_3(p_1, p_2) &= - \tilde{B}_6 (p_2, p_1) =   16 \pi^2 I_{11}(p_1, p_2), \nonumber \\
\tilde{B}_4(p_1,p_2) &= - \tilde B_5 (p_2, p_1) =- 16 \pi^2 \left[ I_{20}(p_1,p_2) - I_{10}(p_1,p_2) \right],
\end{align}
where the general $I_{st}$ integral is defined by
\begin{equation}
I_{st}(p_1,p_2) = \int_0^1 dw \int_0^{1-w} dz w^s z^t \left[ z(1-z) p_1^2 + w(1-w) p_2^2 + 2 w z (p_1\cdot p_2)  \right]^{-1}.
\end{equation}
Both $B_1$ and $B_2$ can be rendered finite by imposing the Ward identities on the two vector lines, giving
\begin{align}
\tilde{B}_1 (p_1,p_2) &= p_1 \cdot p_2 \, \tilde{B}_3 (p_1,p_2) + p_2^2 \, \tilde{B}_4 (p_1,p_2),
\label{WI1} \\
\tilde{B}_2 (p_1,p_2) &= p_1^2 \, \tilde{B}_5 (p_1,p_2) + p_1 \cdot p_2 \, \tilde{B}_6 (p_1,p_2),
\label{WI2}
\end{align}
which allow to re-express the formally divergent amplitudes in terms of the convergent ones. 	
This expansion will be particularly useful in order to check our ansatz for parametrization of the same vertex at nonzero density. 
We are going to extract the zero density part from the parametrization of the vertex \eqref{gen1}, taking the limit of zero four-vector $\eta$. Since the check will be performed for equal invariants $p_1^2=p_2^2$, we provide here the expression of \eqref{Ross} in the same kinematical limit. One obtains 
(we set $p_1^2=p_2^2\equiv M^2$)

{\allowdisplaybreaks
\beqa
\tilde{B}_1 (s, M^2, M^2) &=& -\frac{i}{4 \pi^2}\\
\tilde{B}_3 (s, M^2, M^2) &=&
- \frac{2\,  i \, M^4\,   } {\pi ^2 s^2 \left(s-4 M^2\right)^2} \, \,  \Phi_M   \, \, (s - M^2 ) \nn \\
&& -  \frac{i}   {2 \pi ^2 s \left(s-4 M^2\right)^2}
 \left[ s^2  - 6 s M^2+2  \left(2 M^2+s\right) \log  \left[\frac{M^2}{s}\right] M^2+8 M^4\right] \nn \\ \\
\tilde{B}_4 (s, M^2, M^2) &=&
\frac{i  M^2 }  {\pi ^2 s^2 \left(s - 4 M^2\right)^2}
\Phi_M \left(s^2 -3 s M^2 + 2 M^4 \right)  \nn \\
&& +\frac  {i}   {2 \pi ^2 s \left(s-4 M^2\right)^2}
\left[  2 s M^2  + \left(s^2-4 M^4\right) \log \left(\frac{M^2}{s} \right) -8 M^4  \right],
\eeqa
}

with the functions $\Phi(x,y)$ and $\lambda(x,y)$ defined in this  specific case by
\beqa
 \Phi_M \equiv \Phi (  \frac{M^2}{s} , \frac{M^2}{s}) &=& \frac{1}{\lambda_M}
\left[ \log ^2 \left(\frac{2 M^2} {s ( \lambda_M +1)  -2 M^2} \right) + 4 \textrm{Li}_2\left(\frac{2 M^2}
{- s ( \lambda_M +1) + 2 M^2 }\right)+ \frac{\pi ^2}{3}  \right], \nn \\ \\
\lambda_M &\equiv& \lambda(M^2/s, M^2/s) = \sqrt{1 - \frac{4 M^2}{s }},
\eeqa
as in Eqs.~(\ref{Phi},\ref{lambda}), with $x=y=M^2/s$. The symmetry relations on the external momenta give
\beqa
\tilde{B}_2 (s, M^2, M^2) &=& - \tilde{B}_1 (s, M^2, M^2), \\
\tilde{B}_5 (s, M^2, M^2) &=& - \tilde{B}_4 (s, M^2, M^2),  \\
\tilde{B}_6 (s, M^2, M^2) &=& - \tilde{B}_3 (s, M^2, M^2)
\eeqa
and in the total amplitude simplifies into the form
\beqa
\Gamma_{\mu=0}^{\la \mu \nu} (s, M^2, M^2) = \tilde{B}_3 (s, M^2, M^2) \, \eta_3 ^{\la \mu \nu}  (k_1, k_2) +
\tilde{B}_4 (s, M^2, M^2) \, \eta_4 ^{\la \mu \nu}  (k_1, k_2)  \nn \\
+ \tilde{B}_5 (s, M^2, M^2) \, \eta_5 ^{\la \mu \nu}  (k_1, k_2)
+ \tilde{B}_6 (s, M^2, M^2) \, \eta_6 ^{\la \mu \nu}  (k_1, k_2).
\eeqa
where the expressions of the tensor structures $\eta_i$  are given in table \eqref{table2}.  \\
\begin{table}[t]
\begin{center}
\begin{tabular}
{|c |c |}   \hline
$\eta_1$ & $\veps [p_1,p_2,\mu, \nu] \, p_1^{\la} $  \\ \hline
$\eta_2$ & $\veps [p_1,p_2,\mu, \nu] \, p_2^{\la}$ \\ \hline\hline
$\eta_3$         &           $k_1 \cdot p_2 \varepsilon[p_1,\la, \mu, \nu] + p_1^{\nu} \varepsilon[k_1,p_2,\mu, \la]$ \\ \hline
$\eta_4$ & $p_2 \cdot p_2 \varepsilon[p_1,\la, \mu, \nu] + p_2^{\nu} \varepsilon[p_1,p_2,\mu, \la] $ \\ \hline
$\eta_5$ & $ p_1 \cdot p_1 \varepsilon[p_2,\la, \mu, \nu] + p_1^{\mu} \varepsilon[p_1,p_2,\nu, \la] $  \\ \hline
$\eta_6$ & $p_1 \cdot p_2 \varepsilon[p_2,\la, \mu, \nu] + p_2^{\mu} \varepsilon[p_1,p_2,\nu, \la]  $  \\ \hline
\end{tabular}
\caption{ The six pseudotensors needed in the expansion of an amplitude $\Gamma^{\lambda \mu \nu}(k_1, k_2) $ satisfying the vector current conservation for $\mu=0$. \label{table2} }
\end{center}
\end{table}
An alternative parametrization of the $AVV$ correlator, using the longitudinal/transverse decomposition is given by \cite{Knecht:2003xy} (with $p_3\equiv q$ )
\beq
\label{deraf}
\langle J^{\mu_1}(p_1)J^{\mu_2}(p_2)J^{\mu_3}_5(p_3) \rangle=\frac{1}{8\pi^2} \left({W^{L}}^{\mu_1\mu_2\mu_3} - {  \mathcal \, W^{T}}^{\mu_1\mu_2\mu_3} \right)
\eeq
where the longitudinal component is specified in eq. \eqref{refe}, while
the transverse component is given by
\begin{align}
	{  \mathcal \, W^{T}}^{\mu_1\mu_2\mu_3}(p_1,p_2,p_3^2) &=
	w_T^{(+)}\left(p_1^2, p_2^2,p_3^2 \right)\,t^{(+)\,\mu_1\mu_2\mu_3}(p_1,p_2)
	+\,w_T^{(-)}\left(p_1^2,p_2^2,p_3^2\right)\,t^{(-)\,\mu_1\mu_2\mu_3}(p_1,p_2) \nonumber \\
	& +\,\, {\widetilde{w}}_T^{(-)}\left(p_1^2, p_2^2,p_3^2 \right)\,{\widetilde{t}}^{(-)\,\mu_1\mu_2\mu_3}(p_1,p_2).
\end{align}
This decomposition is in order with all the symmetries of the correlator. The transverse tensors are given by
\begin{equation}\label{tensors}
	\begin{aligned}
		t^{(+) \, \mu_1\mu_2\mu_3}(p_1,p_2) &=
		p_{1}^{\mu_2}\, \veps^{ \mu_1\mu_3 p_1p_2}  -
		p_{2}^{\mu_1}\,\veps^{\mu_2\mu_3 p_1 p_2}  - (p_{1} \cdot p_2)\,\veps^{\mu_1\mu_2\mu_3(p_1 - p_2)}\\ &\hspace{4cm}
		+  \frac{p_1^2 + p_2^2 - p_3^2}{p_3^2}\,  (p_1+p_2)^{\mu_3} \, 
		\veps^{\mu_1\mu_2 p_1 p_2}
		\nonumber  , \\
		t^{(-)\,\mu_1\mu_2\mu_3}(p_1,p_2) &= \left[ (p_1 - p_2)^{\mu_3} - \frac{p_1^2 - p_2^2}{p_3^2}\, (p_{1}+p_2)^{ \mu_3} \right] \,\veps^{\mu_1\mu_2 p_1 p_2}
		\nonumber\\
		{\widetilde{t}}^{(-)\, \mu_1\mu_2\mu_3}(p_1,p_2) &= p_{1}^{\mu_2}\,\veps^{ \mu_1\mu_3 p_1p_2} +
		p_{2}^{\mu_1}\,\veps^{\mu_2\mu_3 p_1 p_2} 
		- (p_{1}\cdot p_2)\,\veps^{ \mu_1 \mu_2 \mu_3 (p_1+p_2)}.
	\end{aligned}
\end{equation}
The map between the Rosenberg representation given in \eqref{Ross} and the current one is given by the relations 
\begin{align}
\tilde{B}_3 (p_1, p_2) &= \frac{1}{8 \pi^2} \left[ w_L - \tilde{w}_T^{(-)}
-\frac{p_1^2+p_2^2}{p_3^2}     w_T^{(+)}
- 2 \,  \frac{p_1 \cdot p_2 + p_2^2 }{p_3^2 }w_T^{(-)}  \right],  \\
\tilde{B}_4 (p_1, p_2) &= \frac{1}{8 \pi^2} \left[  w_L
+ 2 \, \frac{p_1 \cdot p_2}{p_3^2}       w_T^{(+)}
+ 2 \, \frac{p_1 \cdot p_2 + p_1^2}{p_3^2 }w_T^{(-)}  \right], 
\end{align}
that can be inverted in the form 
\begin{equation}
\label{ppo}
w_L (p_1^2, \, p_2^2,p_3^2) = \frac{8 \pi^2}{p_3^2} \left[\tilde{B}_1 - \tilde{B}_2 \right]
\end{equation}
or, after the imposition of the Ward identities in Eqs.(\ref{WI1},\ref{WI2}), as
\begin{align}
w_L ( p_1^2, \, p_2^2,p_3^2) &= \frac{8 \pi^2}{p_3^2}
\left[ (\tilde{B}_3-\tilde{B}_6) p_1 \cdot p_2 + \tilde B_4 \, p_2^2 - \tilde{B}_5 \, p_1^2 \right],
\label{wL}\\
w_T^{(+)} (p_1^2, \, p_2^2,p_3^2)  &= - 4 \pi^2 \left(\tilde{B}_3 - \tilde{B}_4 + \bar{B}_5 - \tilde{B}_6 \right),
\label{wTp}\\
w_T^{(-)} ( p_1^2, \, p_2^2,p_3^2)  &=  4 \pi^2 \left(\tilde{B}_4+ \tilde{B}_5 \right),
\label{wTm}\\
\tilde{w}_T^{(-)} ( p_1^2, \, p_2^2,p_3^2)  &= - 4\pi^2 \left( \tilde{B}_3 + \tilde{B}_4 + \tilde{B}_5 + \tilde{B}_6 \right),
\label{wTt}
\end{align}
where $\tilde{B}_i\equiv \tilde B_i(p_1,p_2)$. As already mentioned, \eqref{ppo} is a special relation, since it shows that the $1/p_3^2$ pole is not affected by Chern-Simons forms, telling us of the physical character of this part of the interaction. \\
Also in this case, the counting of the form factor is four, one for the longitudinal pole part and 3 for the transverse part. Notice that all of them are either symmetric or antisymmetric under the exchange of the momenta of the two photon lines by construction
\beqa
w_L(p_1^2,p_2^2,p_3^2)&=& w_L(p_2^2,p_1^2,p_3^2) \nn\\
w_T^{(+)}(p_1^2,p_2^2,p_3^2)&=& w_T^{(+)}(p_2^2,p_1^2,p_3^2) \nn\\
w_T^{(-)}(p_1^2,p_2^2,p_3^2)&=& - w_T^{(-)}(p_2^2,p_1^2,p_3^2) \nn \\
\tilde{w}_T^{(-)}(p_1^2,p_2^2,p_3^2)&=&- \tilde{w}_T^{(-)}(p_2^2,p_1^2,p_3^2). 
\eeqa
Notice that \beq
\label{refe}
W_L^{\mu_1\mu_2\mu_3}=w_L \, {p_3^{\mu_3}}\epsilon^{\mu_1\mu_2 \rho\sigma}{p_{1\rho}p_{2\sigma}}\equiv w_L\,
{p_3^{\mu_3}}\epsilon^{\mu_1\mu_2  p_1 p_2}
\eeq
where $w_L$ is the anomaly form factor, that 
in the massless photon limit has a ${1}/{p_3^2}$ pole. The entire structure emerges as a solution of the conformal Ward identities once we require that the longitudinal (anomaly) sector built around the axial-vector line has an anomaly. We are going to show that the same description holds, if the two photon lines are on-shell at finite density. In this case the $\mu$-dependence of the perturbative expansion vanishes. We will refer to this case as to the "on-shell $AVV$". \\
In other words, the on-shell $AVV$ at zero and at finite density are identical. This implies that at finite density this diagram can be reconstructed by on-shell conformal Ward identities in momentum space, just by the inclusion of a single pole in the axial-vector channel. The vertex has no transverse part and it is just given by the pole term. 

\subsubsection{Consistency of the expansion with the $\mu=0$ case for  off-shell photons in the symmetric configuration}	
In this subsection we perform a consistency check of our expansion 	\eqref{gen} (reduction ${\bf 60\to 10}$) derived before for the on-shell $AVV$.\\
The vacuum contribution to this amplitude is expected to agree with the expansions presented  in the previous section in the two forms \eqref{Ross} and \eqref{deraf}. The limit of zero density and with photons of equal virtuality can be performed from \eqref{gen1} (i.e. the reduction ${\bf 60\to 7}$) 
\beq
\label{conds}
p_1^2=p_2^2 \,\,\,\, p_1\cdot \eta=p_2\cdot\eta, 
\eeq
in which we will take the limit $\eta\to 0$. 
We start by checking that the decomposition \eqref{gen1}  reduces to the longitudinal sector of 
$W_L\equiv \Gamma^{(0)}_L$ of the $AVV$,  Eq. \eqref{Ross}, in the zero density limit. \\
In the $\eta \to 0$ limit of \eqref{gen1}, all the tensors in \eqref{FFDecompB2} vanish with the exception of $\chi_6$ and $\chi_7$. This limit leaves in the lhs of \eqref{eqq} 
only the zero density part. We obtain
\begin{equation}
\lim_{\eta\to 0} \Gamma^{\lambda \mu \nu} = \bar B_6\chi_6^{\lambda\mu\nu}+\bar B_7\chi_7^{\lambda\mu\nu}
\end{equation}
that we equate to the perturbative computation, subjected to the same limit and parametrized as in \eqref{Ross}.  They can be identified starting from the general expression of the zero density $AVV$, given in terms of the 4 form factors in $\Gamma^{(0)}_{pert}$, once the conditions \eqref{conds} are taken into account. 
The perturbative evaluation of the two form factors gives 
{\allowdisplaybreaks
\begin{eqnarray}
\label{int}
	\bar B_6 &=&\frac{i}{4 \pi ^2 \left(p^{2}-{p_1}\cdot {p_2}\right)^2 \left(p^{2}+{p_1}\cdot
   {p_2}\right)}\biggl\{ \biggl[({p_1}\cdot {p_2})^{2} \left(\log \left(\frac{2 \left(p^{2}+{p_1}\cdot {p_2}\right)}{p^{2}}\right)\right)\notag\\&&+p^{2} ({p_1}\cdot {p_2}) \left(2 \log \left(\frac{{2(p^{2}+{p_1}\cdot {p_2})}}{p^{2}}\right)-1\right)+p^{4}\biggl]\notag\\&&-{p^{2} ({p_1}\cdot {p_2}) \left(p^{2}+2 ({p_1}\cdot {p_2})\right) {C}_0\left(p^{2},p^{2},2 \left(p^{2}+{p_1}\cdot {p_2}\right),0,0,0\right)}\biggl\}, \notag \\
\bar B_7 &=& \frac{i}{4 \pi ^2 \left(p^{2}-{p_1}\cdot {p_2}\right)^2 \left(p^{2}+{p_1}\cdot{p_2}\right)} \biggl\{ p^{4} \left(p^{2}+2 ({p_1}\cdot {p_2})\right) {C}_0\left(p^{2},p^{2},2 \left(p^{2}+{p_1}\cdot {p_2}\right),0,0,0\right)\notag\\&&
	+({p_1}\cdot {p_2})^{2}-2 p^{4} \biggl(\log \left(\frac{2 \left(p^{2}+{p_1}\cdot {p_2}\right)}{p^{2}}\right)\biggl)-p^{2} ({p_1}\cdot {p_2}) \left(\log \left(\frac{2 \left(p^{2}+{p_1}\cdot {p_2}\right)}{p^{2}}\right)+1\right)\biggl\},\nonumber \\
\end{eqnarray}
 where we have defined the unique scalar three-point function with all the momenta off-shell and $q$ incoming, while $p_1$ and $p_2$ are outgoing
\begin{eqnarray}
C_0(q^2, p_1^2, p_2^2) &=&
\int d^d k  \frac{1}{(k-q)^2 \, (k-p_1)^2 \, k^2} = \frac{i \pi^2}{q^2} \Phi( x, y),
\end{eqnarray}
}
where the  $\Phi(x,y)$ function is defined as
\begin{equation}\label{Phi}
\Phi( x, y) = \frac{1}{\lambda} \biggl\{ 2 [Li_2(-\rho  x) + Li_2(- \rho y)]  +
\ln \frac{y}{ x}\ln \frac{1+ \rho y }{1 + \rho x}+ \ln (\rho x) \ln (\rho  y) + \frac{\pi^2}{3} \biggl\},
\end{equation}
with
\begin{align}\label{lambda} 
 &\lambda(x,y) = \sqrt {\Delta}, \qquad   \Delta=(1-  x- y)^2 - 4  x  y,\notag\\&
\rho( x,y) = 2 (1-  x-  y+\lambda)^{-1}, \qquad   x=y=\frac{p^2}{2(p^2+p_1\cdot p_2)},
\end{align}
Eq$.$ \eqref{int} is in agreement with the analysis in \cite{Coriano:2023hts} for the same kinematical case 
defined by \eqref{conds}.
 
\section{The form factors for on-shell photons in $\Gamma$ in the symmetric case}
The case of the decay of the axial-vector line into two on-shell photons ($p_1^2=p_2^2=p^2=0$,  $p_1\cdot\eta=p_2\cdot \eta$) is surely one of the most interesting, since the computations are slightly simplified and one discovers, as we have already anticipated, that the anomaly pole is not affected by any $\mu$ dependent
correction. We are going to illustrate this point by an explicit computation. \\
Notice that this kinematical constraint must be implemented from the very beginning of the expansion of the vertex into form factors, since the Schouten relations should be chosen in such a way to avoid possible $1/p^2$ singularities. 
For this reason the relations \eqref{schouten1} and the respective equations that are derived for the tensor structures listed in \eqref{schouten2} are modified. Eqs. \eqref{schouten1} become
\begin{eqnarray}
	{{p_1}}^{\lambda } {\epsilon }^{\mu \nu {{p_1}}{{p_2}}}& = &  {{p_1}}^{\mu } {\epsilon }^{\lambda
	\nu {{p_1}}{{p_2}}}-{{p_1}}^{\nu } {\epsilon }^{\lambda \mu
	{{p_1}}{{p_2}}}-\left({{p_1}}\cdot {{p_2}}\right) {\epsilon }^{\lambda \mu \nu
	{{p_1}}} \notag \\
{{p_1}}^{\lambda } {\epsilon }^{\mu \nu {{p_1}}{\eta }}& = &  -\left({p}\cdot {\eta }\right)
{\epsilon }^{\lambda \mu \nu {{p_1}}}+{{p_1}}^{\mu } {\epsilon }^{\lambda \nu {{p_1}}{\eta
}}-{{p_1}}^{\nu } {\epsilon }^{\lambda \mu {{p_1}}{\eta }} \notag \\
{{p_2}}^{\lambda } {\epsilon }^{\mu \nu {{p_1}}{{p_2}}}& = &  {{p_2}}^{\mu } {\epsilon }^{\lambda
	\nu {{p_1}}{{p_2}}}-{{p_2}}^{\nu } {\epsilon }^{\lambda \mu
	{{p_1}}{{p_2}}}+\left({{p_1}}\cdot {{p_2}}\right) {\epsilon }^{\lambda \mu \nu
	{{p_2}}} \notag \\
{\eta }^{\lambda } {\epsilon }^{\mu \nu {{p_1}}{{p_2}}}& = &  -\left({p}\cdot {\eta }\right)
{\epsilon }^{\lambda \mu \nu {{p_1}}}+\left({p}\cdot {\eta }\right) {\epsilon }^{\lambda \mu \nu
	{{p_2}}}+{\eta }^{\mu } {\epsilon }^{\lambda \nu {{p_1}}{{p_2}}}-{\eta }^{\nu } {\epsilon
}^{\lambda \mu {{p_1}}{{p_2}}} \notag \\
{{p_2}}^{\lambda } {\epsilon }^{\mu \nu {{p_1}}{\eta }}& = &  -\left({p}\cdot {\eta }\right)
{\epsilon }^{\lambda \mu \nu {{p_1}}}+{{p_2}}^{\mu } {\epsilon }^{\lambda \nu {{p_1}}{\eta
}}-{{p_2}}^{\nu } {\epsilon }^{\lambda \mu {{p_1}}{\eta }}+\left({{p_1}}\cdot
{{p_2}}\right) {\epsilon }^{\lambda \mu \nu {\eta }} \notag \\
{{p_1}}^{\lambda } {\epsilon }^{\mu \nu {{p_2}}{\eta }}& = &  -\left({p}\cdot {\eta }\right)
{\epsilon }^{\lambda \mu \nu {{p_2}}}+{{p_1}}^{\mu } {\epsilon }^{\lambda \nu {{p_2}}{\eta
}}-{{p_1}}^{\nu } {\epsilon }^{\lambda \mu {{p_2}}{\eta }}+\left({{p_1}}\cdot
{{p_2}}\right) {\epsilon }^{\lambda \mu \nu {\eta }} \notag \\
{\eta }^{\lambda } {\epsilon }^{\mu \nu {{p_1}}{\eta }}& = &  \left({p}\cdot {\eta }\right) {\epsilon
}^{\lambda \mu \nu {\eta }}+{\eta }^{\mu } {\epsilon }^{\lambda \nu {{p_1}}{\eta }}-{\eta }^{\nu }
{\epsilon }^{\lambda \mu {{p_1}}{\eta }}-{\eta }^{} {\epsilon }^{\lambda \mu \nu {{p_1}}} \notag \\
{{p_2}}^{\lambda } {\epsilon }^{\mu \nu {{p_2}}{\eta }}& = &  -\left({p}\cdot {\eta }\right)
{\epsilon }^{\lambda \mu \nu {{p_2}}}+{{p_2}}^{\mu } {\epsilon }^{\lambda \nu {{p_2}}{\eta
}}-{{p_2}}^{\nu } {\epsilon }^{\lambda \mu {{p_2}}{\eta }} \notag \\
{\eta }^{\lambda } {\epsilon }^{\mu \nu {{p_2}}{\eta }}& = &  \left({p}\cdot {\eta }\right) {\epsilon
}^{\lambda \mu \nu {\eta }}+{\eta }^{\mu } {\epsilon }^{\lambda \nu {{p_2}}{\eta }}-{\eta }^{\nu }
{\epsilon }^{\lambda \mu {{p_2}}{\eta }}-{\eta }^{} {\epsilon }^{\lambda \mu \nu {{p_2}}} \notag \\
g^{\lambda \nu } {\epsilon }^{\mu {{p_1}}{{p_2}}{\eta }}& = &  g^{\mu \nu } {\epsilon }^{\lambda
	{{p_1}}{{p_2}}{\eta }}-{{p_1}}^{\nu } {\epsilon }^{\lambda \mu {{p_2}}{\eta
}}+{{p_2}}^{\nu } {\epsilon }^{\lambda \mu {{p_1}}{\eta }}-{\eta }^{\nu } {\epsilon }^{\lambda \mu
	{{p_1}}{{p_2}}} \notag \\
g^{\lambda \mu } {\epsilon }^{\nu {{p_1}}{{p_2}}{\eta }}& = &  g^{\mu \nu } {\epsilon }^{\lambda
	{{p_1}}{{p_2}}{\eta }}-{{p_1}}^{\mu } {\epsilon }^{\lambda \nu {{p_2}}{\eta
}}+{{p_2}}^{\mu } {\epsilon }^{\lambda \nu {{p_1}}{\eta }}-{\eta }^{\mu } {\epsilon }^{\lambda \nu
	{{p_1}}{{p_2}} }
\end{eqnarray}
The overall picture is nonetheless the same, as we obtain a slightly modified version of the 32 relations we have derived in the off-shell-case, without any relation becoming redundant. Hence, under these conditions, we end up with the same 28 tensors and form factors of the off-shell case. It is thus clear that relations \eqref{BosSybSubON} are unmodified, hence the structure of the amplitude before the implementation of the conservation Ward identities for the vector currents is still given by \eqref{FFDecompB1}. \\
The first differences start to appear as we impose vectorial WIs. In this case the contraction of the on-shell decomposition with $p_2^\nu$ gives
\begin{align}
&{\epsilon }^{\lambda \mu {{p_1}}{{p_2}}} \Big(-B_{10} \left({p}\cdot {\eta }\right)-B_7 \left({{p_1}}\cdot
   {{p_2}}\right)-B_1\Big)+{\epsilon }^{\lambda \mu {{p_2}}{\eta }} \Big(B_{11} \left({p}\cdot {\eta
   }\right)+B_8 \left({{p_1}}\cdot {{p_2}}\right)\Big)\notag\\
   &+{\epsilon }^{\lambda \mu {{p_1}}{\eta }} \Big(B_{12}
   \left({p}\cdot {\eta }\right)+B_9 \left({{p_1}}\cdot {{p_2}}\right)\Big)+{\epsilon }^{\lambda
   {{p_1}}{{p_2}}{\eta }} \Big(-B_{11} {\eta }^{\mu }+{{p_1}}^{\mu } \left[ B_{24} \left({p}\cdot
   {\eta }\right)+B_{22} \left({{p_1}}\cdot {{p_2}}\right)-B_5\right]-B_8 {{p_2}}^{\mu } \Big)=0.
\end{align}
These equations allow to derive the constraints
\begin{eqnarray}
B_1&=& B_7 \left(-\left({{p_1}}\cdot
{{p_2}}\right)\right)-B_{10} \left({k}\cdot
{\eta }\right)\notag\\
B_5&=& B_{24} \left({k}\cdot
{\eta }\right)+B_{22} \left({{p_1}}\cdot
{{p_2}}\right)\notag\\
B_8&=& 0\notag\\
B_{11}&=& 0\notag\\
B_{12}&=& -\frac{B_9
	\left({{p_1}}\cdot
	{{p_2}}\right)}{{p}\cdot {\eta }},
   \end{eqnarray}
leading to the on-shell decomposition
   \begin{align} \label{FFDecompB2}
 \Gamma^{\lambda\mu\nu}(p_1,p_2,\eta)&=
\bar B_1\chi_1^{\lambda\mu\nu}+\bar B_{2}\chi_2^{\lambda\mu\nu}  +\bar B_{3} \chi_3^{\lambda\mu\nu} +\bar B_4 \chi_4^{\lambda\mu\nu}     +\bar B_5\chi_5^{\lambda\mu\nu}  +\bar B_{6}\chi_6^{\lambda\mu\nu} +\bar B_7  \chi_7^{\lambda\mu\nu},
   \end{align}
where
\begin{align}
&\chi_1^{\lambda\mu\nu}= \left(-\frac{{\eta }^{\mu } \left({{p_1}}\cdot {{p_2}}\right) {\epsilon }^{\lambda \nu  {{p_2}}{\eta }}}{{p}\cdot {\eta}}-\frac{{\eta }^{\nu } \left({{p_1}}\cdot {{p_2}}\right) {\epsilon }^{\lambda \mu {{p_1}}{\eta }}}{{p}\cdot {\eta}}+{{p_1}}^{\nu } {\epsilon }^{\lambda \mu  {{p_1}}{\eta 	}}+{{p_2}}^{\mu }    {\epsilon }^{\lambda \nu {{p_2}}{\eta   }}\right) \notag
\end{align}
\begin{align}
&\chi_2^{\lambda\mu\nu}=\left(-\left({p}\cdot {\eta }\right)   {\epsilon }^{\lambda \mu \nu   {{p_1}}}+\left({p}\cdot {\eta }\right)   {\epsilon }^{\lambda \mu \nu {{p_2}}}+{\eta   }^{\mu } {\epsilon }^{\lambda \nu   {{p_1}}{{p_2}}}-{\eta }^{\nu }  {\epsilon }^{\lambda \mu   {{p_1}}{{p_2}}}\right)\notag\\&
 \chi_3^{\lambda\mu\nu}=\left({{p_1}}^{\mu } \left({p}\cdot {\eta   }\right) {\epsilon }^{\lambda \nu {{p_1}}{\eta   }}+{{p_2}}^{\nu } \left({p}\cdot {\eta   }\right) {\epsilon }^{\lambda \mu {{p_2}}{\eta   }}-{\eta }^{\mu } {{p_2}}^{\nu }  {\epsilon }^{\lambda   {{p_1}}{{p_2}}{\eta   }}+{\eta }^{\nu } {{p_1}}^{\mu }   {\epsilon }^{\lambda   {{p_1}}{{p_2}}{\eta }}\right)\notag\\&
\chi_4^{\lambda\mu\nu}=\left({{p_1}}^{\mu } {\epsilon }^{\lambda \nu   {{p_2}}{\eta }}+{{p_2}}^{\nu }   {\epsilon }^{\lambda \mu {{p_1}}{\eta   }}\right) \notag\\&
 \chi_5^{\lambda\mu\nu}=\left({{p_1}}^{\mu } {\epsilon }^{\lambda \nu   {{p_1}}{{p_2}}}-{{p_2}}^{\nu }   {\epsilon }^{\lambda \mu   {{p_1}}{{p_2}}}\right)\notag\\&
\chi_6^{\lambda\mu\nu}= \left({{p_1}}^{\mu }  {{p_1}}^{\nu } {\epsilon }^{\lambda   {{p_1}}{{p_2}}{\eta   }}-{{p_2}}^{\mu } {{p_2}}^{\nu } {\epsilon   }^{\lambda {{p_1}}{{p_2}}{\eta   }}+{{p_1}}^{\mu } \left({{p_1}}\cdot   {{p_2}}\right) {\epsilon }^{\lambda \nu   {{p_1}}{\eta }}+{{p_2}}^{\nu }(p_1\cdot p_2)\epsilon^{\lambda\mu p_2\eta}\right)\notag\\&
\chi_7^{\lambda\mu\nu}= \left({{p_2}}^{\mu } {\epsilon }^{\lambda \nu   {{p_1}}{{p_2}}}-{{p_1}}^{\nu }   {\epsilon }^{\lambda \mu   {{p_1}}{{p_2}}}-\left({{p_1}}\cdot   {{p_2}}\right) {\epsilon }^{\lambda \mu \nu   {{p_1}}}+\left({{p_1}}\cdot   {{p_2}}\right) {\epsilon }^{\lambda \mu \nu   {{p_2}}}\right)
\end{align}
are the new tensor structures. Therefore both in the on-shell case and in the symmetric off-shell case - in this second case we require \eqref{conds} - we have 7 (but separately different) tensor structures and corresponding from factors  
instead of the 10 which have been identified in the most general off-shell case in \eqref{tens}.

\subsection{The longitudinal component}
In order to analyze the anomaly, we focus on the longitudinal component of the correlator. This component is defined with respect to the momentum of the axial vector line, connected with the anomaly contribution. Using the $\pi$ and $\Sigma$ projectors, we extract from the perturbative part its longitudinal components. We apply the longitudinal projectors on the amplitude, obtaining   
 \begin{align}\label{longPertFF}
\Sigma^\lambda_\alpha\, \Gamma^{\alpha\mu\nu}(p_1,p_2,\eta) = & ({{p_2}}^{\nu } {\epsilon }^{\lambda \mu {{p_1}}{{p_2}}} -{{p_1}}^{\mu } {\epsilon }^{\lambda \nu {{p_1}}{{p_2}}})\left(\frac{B_{24} (p\cdot {\eta })^{2}}{{{p_1}}\cdot {{p_2}}}+\frac{B_9 {\eta }^{2}}{2 \left(p\cdot {\eta }\right)}+\left(p\cdot {\eta }\right) \left(B_{22}-\frac{B_6+B_{10}}{{{p_1}}\cdot {{p_2}}}\right)-B_7\right)\notag\\&
+({\epsilon }^{\lambda \mu \nu {{p_2}}}-{\epsilon }^{\lambda \mu \nu {{p_1}}}) \left(\left({{p_1}}\cdot {{p_2}}\right) \left(\frac{B_7 - B_9 {\eta }^{2}}{2 \left(p\cdot {\eta }\right)}\right)+\left(B_9+B_{10}\right) \left(p\cdot {\eta }\right)\right)\notag\\&
   +\Big( {{p_1}}^{\nu } {\epsilon }^{\lambda \mu {{p_1}}{{p_2}}}
   -{{p_2}}^{\mu } {\epsilon }^{\lambda \nu {{p_1}}{{p_2}}} \Big) \left(\frac{B_9 {\eta }^{2}}{2 \left(p\cdot {\eta }\right)}-\frac{\left(B_9+B_{10}\right)
   \left(p\cdot {\eta }\right)}{{{p_1}}\cdot {{p_2}}}-B_7\right)\notag\\ &
   +\frac{1}{2} \Big[ p_1^\mu \epsilon^{\lambda \nu p_1 \eta } + p_2^\nu \epsilon^{\lambda \mu p_2 \eta }  - p_1^\mu \epsilon^{\lambda \nu p_2 \eta } - p_2^\nu \epsilon^{\lambda \mu p_1 \eta }  +(p_1^\mu p_1^\nu - p_2^\mu p_2^{\nu } ){ \epsilon^{\lambda {{p_1}}{{p_2}}{\eta }}  \over p_1\cdot p_2}
 \Big] \Big(B_{24} \left(p\cdot {\eta }\right) \notag \\ & \qquad \qquad + B_{22} \left({{p_1}}\cdot {{p_2}}\right)-B_6+B_9\Big)
\end{align}
having used \eqref{pert}. It is clear that in the expression above we have several tensorial structures multiplied by linear combinations of form factors. Now we introduce two new form factors that we call $C_1$ and $C_2$, so that we can write two of these linear combinations as
\begin{equation}
-B_9 \left(p\cdot {\eta }\right)-B_{10} \left(p\cdot {\eta }\right)+\frac{B_9  \left({{p_1}}\cdot {{p_2}}\right)}{2 \left(p\cdot {\eta }\right)}+B_7 \left(-\left({{p_1}}\cdot {{p_2}}\right)\right)=C_1,
\end{equation}
\begin{equation}
-B_{24} \left({k}\cdot {\eta }\right)-B_{22} \left({{p_1}}\cdot {{p_2}}\right)+B_6-B_9=C_2.
\end{equation}
Solving for $B_7$ and $B_9$, we get
\begin{align}
&B_7\to \frac{-2 B_9 (p\cdot {\eta })^{}-2 B_{10} (p\cdot {\eta })^{}+B_9  \left({{p_1}}\cdot {{p_2}}\right)/2\left(p\cdot {\eta }\right)-2 C_1 \left(p\cdot {\eta }\right)}{  \left({{p_1}}\cdot {{p_2}}\right)},\notag\\&
B_6\to B_{24} \left(p\cdot {\eta }\right)+B_{22} \left({{p_1}}\cdot {{p_2}}\right)+B_9+C_2.
\end{align}
Substituting these two solutions into \eqref{longPertFF}, we get
\begin{eqnarray}
\Sigma^\lambda_\alpha(q)\, \Gamma^{\alpha \mu\nu}(p_1,p_2,\eta)&=&C_2 \biggl(\frac{{{p_1}}^{\mu } \left(p\cdot {\eta }\right) {\epsilon }^{\lambda \nu {{p_1}}{{p_2}}}}{{{p_1}}\cdot {{p_2}}}-\frac{{{p_2}}^{\nu } \left(p\cdot {\eta }\right) {\epsilon }^{\lambda \mu {{p_1}}{{p_2}}}}{{{p_1}}\cdot {{p_2}}}+\frac{1}{2} {{p_1}}^{\mu } {\epsilon }^{\lambda \nu {{p_2}}{\eta }}+\frac{1}{2} {{p_2}}^{\nu } {\epsilon }^{\lambda \mu {{p_1}}{\eta }}\notag\\&&-\frac{{{p_1}}^{\mu } {{p_1}}^{\nu } {\epsilon }^{\lambda {{p_1}}{{p_2}}{\eta }}}{2
   \left({{p_1}}\cdot {{p_2}}\right)}+\frac{{{p_2}}^{\mu } {{p_2}}^{\nu } {\epsilon }^{\lambda {{p_1}}{{p_2}}{\eta }}}{2 \left({{p_1}}\cdot {{p_2}}\right)}-\frac{1}{2} {{p_1}}^{\mu } {\epsilon }^{\lambda \nu {{p_1}}{\eta }}-\frac{1}{2} {{p_2}}^{\nu } {\epsilon }^{\lambda \mu {{p_2}}{\eta }}\biggl)\notag\\&&-\frac{C_1 \left({{p_1}}^{\lambda }+{{p_2}}^{\lambda }\right) {\epsilon }^{\mu \nu {{p_1}}{{p_2}}}}{{{p_1}}\cdot {{p_2}}}.
\end{eqnarray}
Notice that $C_2$ does not contribute to the amplitude, since this form factor vanishes after the contraction with the physical polarization vectors of the two photons. Therefore we obtain in the on-shell photon case that

\begin{equation}\label{labellong}
\Sigma^\lambda_\alpha(q)\, \Gamma^{\alpha\mu\nu}(p_1,p_2,\eta)=\omega_{{L}}\epsilon^{\mu\nu p_1 p_2}\frac{q^\lambda}{q^2},
\end{equation}   
where $\omega_L\equiv C_1$ is the only longitudinal form factor that survives the limit. 
Contracting both sides of the equation with the antisymmetric tensor, we get a scalar equation for $\omega_L$ that we decompose in terms of a vacuum $(\mu=0)$ and of a finite density contribution
\begin{equation}
\label{ccd}
\omega_L=\omega_L^{\mu=0}+\omega_L^{\mu\neq0},
\end{equation}
a relation that we are going to investigate in detail.\\
The analysis of the zero temperature part in \eqref{ccd} is quite direct. Using the notations introduced above, and applying the same longitudinal projection of \eqref{labellong} on $\Gamma^{(0)}_{pert}$, the contribution to $\omega_L$ takes the form
\small
\begin{align}\label{blexpre}
\omega_L^{\mu=0}&=8i(p_1\cdot p_2)H^{(f,f,f)}[p_1\cdot k]+8i(p_1\cdot p_2)J^{(f,f,f)}[p_2\cdot k]-8iH^{(f,f,f)}[(p_1\cdot k) (p_2\cdot k)] +4i H^{(f,f,f)}[(k\cdot k ) (p_1\cdot k)]\notag\\
&-4iH^{(f,f,f)}[(k\cdot k)  (p_2\cdot k)]+8iH^{(f,f,f)}[(p_2\cdot k)(p_2\cdot k)]-8iJ^{(f,f,f)}[(p_1\cdot k)(p_2\cdot k)]-4iJ^{(f,f,f)}[(k\cdot k) (p_1\cdot k)]\notag\\
&+8iJ^{(f,f,f)}[(p_1\cdot k) (p_1\cdot k)]+4iJ^{(f,f,f)}[(k\cdot k) (p_2\cdot k)]-8iJ^{(f,f,f)}[(p_2\cdot k)(p_2\cdot k)]-8iH^{(f,f,f)}[(p_1\cdot k)(p_1\cdot k)].
\end{align}
\normalsize
A computation gives 
\begin{equation}
\omega_L^{\mu=0}=-\frac{i}{2\pi^2},
\end{equation}
as expected, corresponding to an anomaly pole term $-\frac{i}{2\pi^2}/q^2$. Indeed, when the two photons are on-shell, the $AVV$ coincides with the contribution of the anomaly pole and a vanishing transverse component. 
Instead, the finite density part is given by the contributions
{\begin{align}\label{blexpre}
\omega_L^{\mu\neq0}&=8i(p_1\cdot p_2)H^{(\delta,f,f)}[p_1\cdot k]+8i(p_1\cdot p_2)J^{(\delta,f,f)}[p_2\cdot k]-8iH^{(\delta,f,f)}[(p_1\cdot k) (p_2\cdot k)] \notag\\
&+4i H^{(\delta,f,f)}[(k\cdot k ) (p_1\cdot k)]-4iH^{(\delta,f,f)}[(k\cdot k)  (p_2\cdot k)]+8iH^{(\delta,f,f)}[(p_2\cdot k)(p_2\cdot k)]\notag\\
&-8iJ^{(\delta,f,f)}[(p_1\cdot k)(p_2\cdot k)]-4iJ^{(\delta,f,f)}[(k\cdot k) (p_1\cdot k)]+8iJ^{(\delta,f,f)}[(p_1\cdot k) (p_1\cdot k)]\notag\\&+4iJ^{(\delta,f,f)}[(k\cdot k) (p_2\cdot k)]-8iJ^{(\delta,f,f)}[(p_2\cdot k)(p_2\cdot k)]-8iH^{(\delta,f,f)}[(p_1\cdot k)(p_1\cdot k)]\notag\\&
+(f,\delta,f)+(f,f,\delta) + (\delta, f, \delta).
\end{align}
}
We are going to show by a direct computation that this contribution vanishes. The computation is performed in the $p^2=0$ case, i.e. for the on-shell $AVV$.  
\subsection{Evaluation of the integrals in a special frame}
{
	The explicit evaluation of \eqref{blexpre} is performed by choosing a frame of reference in which  the $J[\dots]$ and $H[\dots]$ integrals simplify. We choose a frame where the particle of momentum $q$ and invariant mass $4 p_0^2$ decays into two on-shell back-to-back photons.  The heat-bath is assumed to be at rest in this frame, with $\eta^\mu=(1,0,0,0)$. Together with \eqref{symmetricconditions} (and $p^2=0$), this choice translates into the following parametrizations 
	\begin{align}\label{frame}
	&p_1^\mu = (p_0, 0, 0, p_0) 	&&p_2^\mu=(p_0,0,0,-p_0) \notag \\
	&q^\mu = (2 p_0, 0 ,0 ,0)	 &&\eta^\mu=(1,0,0,0).
\end{align}
	As it will become apparent soon,  integrals in this frame are plagued by collinear divergences as well as IR ones, that will cancel out in the overall sum \eqref{blexpre}. As an explicit example, we will sketch the evaluation of the $H(p_1,k)$ integral, treated in DR.\\
 We start with ($1\delta$) that we evaluate in the special frame \eqref{frame}
\begin{align}
&H^{(\delta,f,f)}[p_1\cdot k]=i\int \frac{d^dk}{(2\pi)^{d-1}}k\cdot p_1 \biggl (\theta(\mu- \eta \cdot k ) \, \theta (\eta \cdot k ) \, \delta(k^2) \biggl )\frac{1}{(k-q)^2}\frac{1}{(k-p_2)^2}\notag\\
&=i\int \frac{d^dk}{(2\pi)^{d-1}}(k_0 \, p_0- |\kbf| p_0 \cos\theta) \biggl (\theta(\mu-k_0)\frac{\delta(k_0-|\mathbf k|)}{2|\mathbf k|} \biggl )\frac{1}{4p_0^2+k^2-4k_0 p_0}\frac{1}{k^2-2k_0p_0 + 2 |\kbf| p_0 \cos \theta},\notag\\
\end{align}
and perform the energy $dk^0$ integration 
\begin{equation}
H^{(\delta,f,f)}[p_1\cdot k]=\frac{i}{16 \, p_0}\int {d^{d-1}k \over (2\pi)^{d-3}} \, \frac{1}{{|\mathbf k|} \, (|\mathbf k|-p^0)}\frac{1-\cos\theta}{1+\cos\theta} \, \theta(\mu - |\kbf|). 
\end{equation}
We use spherical coordinates in $d-1$ dimensions
\begin{equation}
H^{(\delta,f,f)}[p_1\cdot k]=\frac{i}{16 \, p_0}\int {d |\kbf | \over (2\pi)^{d-3}} \ \theta(\mu-|\kbf|) \frac{ |\kbf |^{d-3}}{|\mathbf k|-p_0}\int d\theta_1d\theta_2...d\theta_{d-2}\  \frac{1-\cos\theta_1}{1+\cos\theta_1}\sin^{d-3}\theta_1 \sin^{d-4}\theta_2...\sin\theta_{d-3}.
\end{equation}
In the frame that we have chosen, there is a factorization of integrals, since the angular and radial integrations do not intertwine. This takes place in every integral computed in this special frame, and it is one of the main reasons why we are able to evaluate the result analytically. If we take the photons to be off-shell, this factorization is not present. \\
The radial integration is cut by the theta function. The angular part can be integrated for every integration variable but $\theta_1$, yielding a $\Omega(d-3)$ volume. The integral can be re-expressed in the form \begin{equation}
H^{(\delta,f,f)}[p_1\cdot k]= \frac{i\Omega(d-3)}{16\, p_0 \, (2\pi)^{d-1}}\int_0^\mu d|\kbf | \frac{ |\kbf |^{d-3}}{|\mathbf k|-p_0}\int_{-1}^{1} dt \, \frac{1-t}{1+t} \,(1-t^2)^{\frac{d}{2}-1}.
\label{vv}
\end{equation}
Using
\begin{eqnarray}
	\int_{-1}^{1} dt \, \frac{1-t}{1+t} \,(1-t^2)^{\frac{d}{2}-1} &=& { \sqrt{\pi} (d-2) \over 2} {\Gamma(d/2-1)\over \Gamma(d/2-1/2)} \notag \\
	\int_0^\mu d|\kbf | \frac{ |\kbf |^{d-3}}{|\mathbf k|-p^0} &=& {\mu^{d-2} \over (d-2) p_0} {}_2F_1(1,d-2,d-1,\mu/p_0) \notag \\
	\Omega(d-3) &=& \frac{\pi^{d/2-3/2}}{\Gamma(d/2-1/2)}
\end{eqnarray}
where $_2F_1$ is an Hypergeometric $_2 F_1$ function, we obtain
\begin{equation}
H^{(\delta,f,f)} [p_1\cdot k] = 
-\frac{2^{-d-4} \pi ^{-d/2} \mu ^{d-2} \Gamma \left(\frac{d}{2}-2\right) \, _2F_1\left(1,d-2;d-1;\frac{\mu }{p_0}\right)}{p_0^2 \,  \Gamma \left(\frac{d-1}{2}\right)^2}.
\end{equation} 
The limit $d \to 4$ is singular due to collinear divergences. Notice that we need to require that $p_0> \mu$ in order to guarantee the convergence of all the contributions. This requirement, in this case, is necessary only at the intermediate stage, since all the contributions, as already mentioned, add up to zero. \\
The integrals are reported in Appendix \ref{appIn}. One could argue, looking at the table of integrals, that the possibility of residual IR divergences cannot be ruled out a priori. We now demonstrate how in \eqref{blexpre} such divergences do not intervene in the overall sum. \\
From \eqref{radialintegrals}, it is apparent that the only cases in which IR divergences could be present are $J^{(f,\delta,f)}[p_2 \cdot k]$, 
$H^{(f,\delta,f)}[p_1 \cdot k]$, $J^{(f,\delta,f)}[(p_2 \cdot k)^2]$, $H^{(f,\delta,f)}[(p_1 \cdot k )^2]$. The IR divergent part of \eqref{blexpre} thus can be identified in the contributions
\begin{equation}
	\omega_{IR} = 8i \, (p_1\cdot p_2) (H_{IR}^{(f,\delta,f)}[p_1 \cdot k] + J_{IR}^{(f,\delta,f)}[p_2 \cdot k]) - 8i \, (H_{IR}^{(f,\delta,f)}[(p_1 \cdot k)^2] + J_{IR}^{(f,\delta,f)}[(p_2 \cdot k)^2]).
\end{equation}
In our frame, $p_1 \cdot p_2 = 2 p_0^2$ and $J^{(f,\delta,f)}[p_2 \cdot k] = H^{(f,\delta,f)}[p_1 \cdot k]$, $J^{(f,\delta,f)}[(p_2 \cdot k)^2] = H^{(f,\delta,f)}[(p_1 \cdot k )^2]$. In these integrals the IR divergent part takes the form
\begin{equation}
	\omega_{IR} = {8 i \over (2 \pi)^{d-1}} \, \left( - \underbrace{{2p_0^2 \over 4} \int_0^\mu d|\kbf | |\kbf |^{d-5} \int_{-1}^{1} dt \, \frac{(1-t^2)^{\frac{d}{2}-1}}{1-t^2}}_{2 \, (p_1 \cdot p_2) \, H^{(f,\delta,f)}[p_1 \cdot k]} + \underbrace{{p_0^2 \over 2} \int_0^\mu d|\kbf | |\kbf |^{d-5} \int_{-1}^{1} dt \, \frac{(1-t^2)^{\frac{d}{2}-1}}{1-t^2}}_{2 \, H^{(f,\delta,f)}[(p_1 \cdot k )^2]} \right) 
\end{equation}
}
and therefore vanishes.\\ 
Using the integrals in the Appendix \ref{appIn}, one is able to show that the whole \eqref{blexpre} combination vanishes in the limit $d\to 4$, leaving us with the conclusion that
\begin{equation}
\omega_L=\omega_{\mu=0}+\omega_{\mu\neq0}=\omega_{\mu=0}.
\end{equation}
We conclude that the anomaly contribution, at least in the on-shell photon case, is identified by an anomaly pole which does not acquire any correction even at finite fermion density. By covariance, obviously, the result remains valid in any frame. Therefore, we have shown that the entire diagram is uniquely identified by the anomaly pole, since the transverse part is zero. Obviously the anomaly constraint
 \beq
 \langle \partial_\mu J_5^\mu\rangle= a_n F\tilde{F}
 \eeq
 for on-shell photons, is satisfied also for $\mu\neq 0$ and $m=0$, and no extra corrections are present in the equation. In addition, we have shown that the expression reduces only to the pole contribution. The result, as discussed in the previous section, can be viewed as a consequence of conformal symmetry, since the amplitude, in this special kinematical limit, satisfies conformal Ward identities, and the anomaly is simply given by the residue of the massless $1/q^2$ pole. 
 
\section{The transverse sector }
As we have shown in the previous section by an explicit computation in the on-shell case, the independence of the anomaly - and hence of  $\Gamma_{\text{long}}$ - from the chemical potential, claimed by previous analyses \cite{Qian:1994pp,Itoyama:1982up} 
is correct. For this reason we are now going to rely on such previous results, which are valid also off-shell, in order to reduce the number of form factors in our parametrization. \\
 Once we have verified that the chemical potential does not affect the chiral anomaly part of the correlator, we can reduce \eqref{FFDecompB2} even further, by imposing the axial WI. Our explicit evaluation of the anomaly was worked out in the on-shell case $p^2=0$, nonetheless we will now proceed in the general $p^2 \neq  0$ case. \\
 Formally, we impose the constraint
\beq
\label{cons}
q_{\lambda}\Gamma^{\lambda\mu\nu}=q_{\lambda}\Gamma^{(0)\lambda\mu\nu}=a_n \epsilon^{\mu\nu p_1 p_2}.
\eeq 
which is valid in general. 
This constraint is the result of the analysis presented in \cite{Qian:1994pp} that we have independently verified. The proof does not require the explicit evaluation of the integrals, and can be used to impose additional constraints on our explicit tensorial parametrizations.  \\
This implies that the pure finite density corrections contained in $\Gamma^{(1)}$  are transverse with respect to the axial-vector channel
\begin{equation}
	\Gamma^{\lambda\mu\nu}=a_n \frac{q^\lambda}{q^2} \epsilon^{\mu\nu p_1 p_2}+\Gamma_T^{\lambda\mu\nu}=a_n \frac{q^\lambda}{q^2} \epsilon^{\mu\nu p_1 p_2}+\Gamma_T^{(0)\,\lambda\mu\nu}+\Gamma_T^{(1)\,\lambda\mu\nu}.
\end{equation}
Therefore, we now focus on the transverse part $\Gamma_T^{\lambda\mu\nu}$ which can also be decomposed as in eq$.$ \eqref{FFDecompB2} but with the following additional constraint
\begin{align}
	0=q_{\lambda}\Gamma_T^{\lambda\mu\nu} &=
	\left(2 \bar B_6 p^2+2 \bar B_{3} (p\cdot \eta )+2 \bar B_7	\left(p_1\cdot p_2\right)\right) \epsilon^{\mu \nu {p_1}{p_2}} \notag \\ &
	+\epsilon^{\mu {p_1}{p_2} \eta } \Bigl[\left(\bar B_{4} p^2+\bar B_2\right) {p_1}^{\nu}+{p_2}^{\nu } \left(-\bar B_{5}(p\cdot \eta )-\bar B_{4} \, p_1\cdot	p_2 + \bar B_1\right)  \notag \\ & \qquad \qquad 
			+\eta^{\nu } \Big(- \bar B_1\frac{p^2}{p\cdot \eta } +\bar B_{5} \,	p^2- \bar B_2\frac{p_1\cdot p_2 }{p\cdot \eta	}\Big)\Bigr] \notag \\ &
 	+\epsilon^{\nu{p_1}{p_2}\eta} \Bigl[ {p_1}^{\mu } \left(\bar B_{5} \, p\cdot \eta  +\bar B_{4}	p_1\cdot	p_2 -\bar B_1\right)+\left(-\bar B_{4}\, p^2-\bar B_2\right) {p_2}^{\mu	} \notag \\ & \qquad\qquad
 			+\eta^{\mu } \Big( \bar B_1 \frac{p^2}{p\cdot \eta }-\bar B_{5} \, p^2+\bar B_2\frac{p_1\cdot p_2 }{p\cdot \eta	}\Big)\Bigr].
\end{align}
The standard procedure that we have used for the vector WIs, in this case leads us to the relations
\begin{eqnarray}
	\bar B_1 &=& \bar B_{5} \, p \cdot \eta + \bar B_{4} \, p_1 \cdot p_2, \notag \\
	\bar B_2 &=& - \bar B_{4} \, p^2, \notag \\
	\bar B_{3} &=& - {1 \over p \cdot \eta} \left(\bar B_6 \, p^2 +\bar  B_7 \, p_1 \cdot p_2  \right).
\end{eqnarray}
The final form of the decomposition (for $p_1^2=p_2^2\equiv p^2$, $p_1\cdot\eta=p_2\cdot\eta$) can then be written as 
\begin{eqnarray}\label{offshelff}
 \Gamma^{\lambda\mu\nu}_{\text{sym}}&=& \Gamma_L^{ \lambda\mu\nu} +
\Gamma_{T}^{\lambda\mu\nu}=a_n \frac{q^\lambda}{q^2} \epsilon^{\mu\nu p_1 p_2}+\hat B_1 \chi_1^{\lambda\mu\nu}+\hat B_{2} \chi_2^{\lambda\mu\nu}   +\hat B_{3}\chi_3^{\lambda\mu\nu}+\hat B_4 \chi_4^{\lambda\mu\nu},\end{eqnarray}
(on-shell symmetric) 
where we have defined new form factors ($\hat B$) corresponding to the recombined structures 
\begin{eqnarray}
 \chi_1^{\lambda\mu\nu} &=&\biggl(\frac{{{p_2}}^{\mu } {{p_2}}^{\nu } {\epsilon }^{\lambda {{p_1}}{{p_2}}{\eta }}}{{p}^{2}}-\frac{{{p_1}}^{\mu } {{p_1}}^{\nu } {\epsilon }^{\lambda {{p_1}}{{p_2}}{\eta }}}{{p}^{2}}-\frac{{{p_1}}^{\mu } \left({{p_1}}\cdot {{p_2}}\right) {\epsilon }^{\lambda \nu {{p_1}}{\eta }}}{{p}^{2}}-\frac{{{p_1}}^{\mu } \left({{p_1}}\cdot {{p_2}}\right) {\epsilon }^{\lambda \nu {{p_2}}{\eta }}}{{p}^{2}}\notag \\
 &&-\frac{{{p_2}}^{\nu} \left({{p_1}}\cdot {{p_2}}\right) {\epsilon }^{\lambda \mu {{p_1}}{\eta }}}{{p}^{2}}-\frac{{{p_2}}^{\nu } \left({{p_1}}\cdot {{p_2}}\right) {\epsilon }^{\lambda \mu {{p_2}}{\eta }}}{{p}^{2}}+{{p_2}}^{\mu } {\epsilon }^{\lambda \nu {{p_1}}{\eta }}+{{p_1}}^{\nu } {\epsilon }^{\lambda \mu {{p_2}}{\eta }}+{{p_1}}^{\nu } {\epsilon }^{\lambda \mu {{p_1}}{\eta }}+{{p_2}}^{\mu } {\epsilon }^{\lambda \nu {{p_2}}{\eta }}\biggl),\notag \\
  \chi_2^{\lambda\mu\nu} &=& \left(-\frac{{{p_1}}^{\mu } \left({k}\cdot {\eta }\right) {\epsilon }^{\lambda \nu {{p_1}}{{p_2}}}}{{p}^{2}}+\frac{{{p_2}}^{\nu } \left({k}\cdot {\eta }\right) {\epsilon }^{\lambda \mu {{p_1}}{{p_2}}}}{{p}^{2}}+{\eta }^{\mu } {\epsilon }^{\lambda \nu {{p_1}}{{p_2}}}-{\eta }^{\nu } {\epsilon }^{\lambda \mu {{p_1}}{{p_2}}}\right),\notag
\\
 \chi_3^{\lambda\mu\nu}&=& \biggl({{p_2}}^{\nu } \left({k}\cdot {\eta }\right) {\epsilon }^{\lambda \mu {{p_1}}{\eta
   }}+{{p_1}}^{\mu } \left({k}\cdot {\eta }\right) {\epsilon }^{\lambda \nu {{p_2}}{\eta }}-{p}^{2} {\eta }^{\mu } {\epsilon }^{\lambda \nu {{p_1}}{\eta }}-{p}^{2} {\eta }^{\nu } {\epsilon }^{\lambda \mu {{p_1}}{\eta }}+{{p_1}}^{\mu } \left({k}\cdot {\eta }\right) {\epsilon }^{\lambda \nu {{p_1}}{\eta }}-{p}^{2} {\eta }^{\mu } {\epsilon }^{\lambda \nu {{p_2}}{\eta }}\notag\\
   && -{p}^{2} {\eta }^{\nu } {\epsilon }^{\lambda \mu {{p_2}}{\eta
   }}+{{p_2}}^{\nu } \left({k}\cdot {\eta }\right) {\epsilon }^{\lambda \mu {{p_2}}{\eta }}-{\eta }^{\mu } {{p_2}}^{\nu } {\epsilon }^{\lambda {{p_1}}{{p_2}}{\eta }}+{\eta }^{\nu } {{p_1}}^{\mu } {\epsilon }^{\lambda {{p_1}}{{p_2}}{\eta }}\biggl),\notag\\
   \chi_4^{\lambda\mu\nu} &=& \left(-\frac{{{p_1}}^{\mu } \left({{p_1}}\cdot {{p_2}}\right) {\epsilon }^{\lambda \nu {{p_1}}{{p_2}}}}{{p}^{2}}+\frac{{{p_2}}^{\nu } \left({{p_1}}\cdot
   {{p_2}}\right) {\epsilon }^{\lambda \mu {{p_1}}{{p_2}}}}{{p}^{2}}+{{p_2}}^{\mu } {\epsilon }^{\lambda \nu {{p_1}}{{p_2}}}-{{p_1}}^{\nu } {\epsilon }^{\lambda \mu {{p_1}}{{p_2}}}\right).
\end{eqnarray}
In summary, eq. \eqref{offshelff} describes the entire vertex at nonzero $\mu$ in the symmetric case in the off-shell case. 
\subsection{General expressions of the form factors in the transverse sector}

While the explicit computation of the scalar integrals in the off-shell case cannot be evaluated explicitly, it is still interesting to determine the structure of the form factors in the parametrization worked out in \eqref{offshelff}. Therefore, here we present the expressions of all the form factors in terms of integrals that can be computed numerically, since all the integrals are finite. This is guaranteed since the nonzero virtualities of the external vector lines are sufficient to remove both the collinear and infrared divergences. As we have shown  in the previous sections, these are only present in some of the integrals  appearing in the on-shell case, canceling in the complete expression. Here instead they are all finite. \\
Starting from \eqref{offshelff}, we derive a system of equations that allow to compute the form factors in terms of the scalar integrals. We obtain
{ \allowdisplaybreaks
\begin{eqnarray}\label{OFFFF}
	\hat B_{1}
	&=&-\frac{8 i}{M^2 (M^4-4 p_0^4) }  \Big[ 
	(M^6+4 M^2 p_0^4+16 p_0^6) J[k \cdot \eta]
	+(4 M^4 p_0-16 p_0^5) J[(k \cdot \eta)^2]  \notag\\&& 
	+(-M^4-6 M^2 p_0^2+8 p_0^4) J[k \cdot \eta \, k \cdot p_1]  
	+(-M^4-6 M^2 p_0^2+8 p_0^4) J[k \cdot \eta \, k \cdot p_2]  \notag\\&& 
	+(-3 M^4 p_0-2 M^2 p_0^3-8 p_0^5) J[k \cdot p_1]
	-3 M^2 p_0 J[k^2 \, k \cdot p_1]  
	+7 M^2 p_0 J[(k \cdot p_1)^2]  \notag\\&& 
	-4 M^2 p_0 J[k \cdot p_1 \, k \cdot p_2]  
	+(3 M^4 p_0+2 M^2 p_0^3-8 p_0^5) J[k \cdot p_2]
	+M^2 p_0 J[(k \cdot p_2)^2] \Big], \notag\\
	\hat B_{2}
	&=&\frac{8 i}{M^2 (2 p_0^2-M^2)}  \Big[ 
	(4 p_0^4-M^4) J[k \cdot \eta](
	+(M^2-2 p_0^2) J[k \cdot \eta \, k \cdot p_1]
	+(M^2-2 p_0^2) J[k \cdot \eta \, k \cdot p_2]  \notag\\&&
	+(M^2 p_0+2 p_0^3) 	J[k \cdot p_1]  
	+p_0 J[k^2 \, k \cdot p_1]  
	-p_0 J[(k \cdot p_1)^2] \notag\\&&
	+(-M^2 p_0-2 p_0^3) J[k \cdot p_2]
	+p_0 J[(k \cdot p_2)^2]) \Big],\notag\\
	\hat B_{3}
	&=&\frac{8i}{M^4 \left( M^2-2 p_0^2 \right)}  
	\Big[ \left(4 M^4 p_0-16 p_0^5\right) J[k \cdot \eta]
	+\left(16 p_0^4-4 M^4\right) J[(k \cdot \eta)^2]
	+\left(4 M^2 p_0-8 p_0^3\right) J[k \cdot \eta \, k \cdot p_1] \notag\\&& 
	+\left(4 M^2 p_0-8 p_0^3\right) J[k \cdot \eta \, k \cdot p_2]
	+8 p_0^4 J[k \cdot p_1]
	+M^2 J[k^2 \, k \cdot p_2]  \notag\\&& 
	-2 M^2	J[( k \cdot p_1)^2]
	+2 M^2J[k \cdot p_1 \, k \cdot p_2]
	+\left(-2 M^4-4 M^2 p_0^2+8 p_0^4\right)J[ k \cdot p_2]\Big) \Big],\notag\\
	\hat B_{4}
	&=&-\frac{8 i}{M^2 (M^2-2 p_0^2)^2}  \Big[ 
	(3 M^4 p_0-8 M^2 p_0^3+4 p_0^5) J[k \cdot \eta]
	+(4 M^2 p_0^2-2 M^4) J[(k \cdot \eta)^2]  \notag\\&& 
	+(M^2 p_0-2 p_0^3) J[k \cdot \eta \, k \cdot p_1] 
	+(M^2 p_0-2 p_0^3) J[k \cdot \eta \, k \cdot p_2]  \notag\\&& 
	+(-M^4+M^2 p_0^2+2 p_0^4) J[k \cdot p_1]  
	+(p_0^2-M^2) J[k^2 \, k \cdot p_1]    \notag\\&& 
	+(M^2-p_0^2) J[(k \cdot p_1)^2]
	+(-M^4+3 M^2 p_0^2-2 p_0^4) J[k \cdot p_2]
	+(p_0^2-M^2) J[(k \cdot p_2)^2] \Big]. 
	\end{eqnarray}}
Notice that in the expressions above, the integrals $H[f(p_1,p_2,\eta,k)]$ are not present due to relations such as $H[p_1\cdot k]=J[p_2\cdot k]$. Some additional details on the manipulations of these integrals are given in Appendix \ref{OFF SHELL INT}.

\subsection{$\mu$ independence of the amplitude $\Gamma_{T}$ in the on-shell symmetric case}
At this point, using the simplifications presented above, which allow to reduce the number of form factors in the general vertex, we move to compute explicitly the finite density corrections in the on-shell case.
Performing the  $p^2\to 0$ limit we obtain
\begin{eqnarray}\label{B_10}
\hat B_{2} &=& -\frac{4i}{\left({{p_1}}\cdot {{p_2}}\right)
   \left({{p_1}}\cdot {{p_2}}-2 ({k}\cdot {\eta })^{2}\right)} \, 
\biggl(-(H[\eta\cdot k]+J[\eta\cdot k]) ({{p_1}}\cdot {{p_2}})^{2}\notag\\
&&+\left({{p_1}}\cdot {{p_2}}\right) \biggl(-\left({k}\cdot {\eta }\right)(-H[p_1\cdot k]+H[p_2\cdot k]+H[k\cdot k]+J[p_1\cdot k]-J[p_2\cdot k]+J[k\cdot k])\notag\\&&+H[(p_1\cdot k)(\eta\cdot k)]+H[(p_2\cdot k)(\eta\cdot k)]+J[(p_1\cdot k)(\eta\cdot k)]+J[(p_2\cdot k)(\eta\cdot k)]\biggl)\notag\\
&&+\left({k}\cdot {\eta }\right) (H[(k\cdot k)(p_1\cdot k)]-H[(p_1\cdot k)^2]-H[(k\cdot k)(p_2\cdot k)]+H[(p_2\cdot k)^2]\notag\\&&
-J[(k\cdot k)(p_1\cdot k)]+J[(p_1\cdot k)^2]+J[(k\cdot k)(p_2\cdot k)]-J[(p_2\cdot k)^2] \biggl),
\end{eqnarray}
\begin{align}
\hat B_4= \hat B_{2},\end{align}
\begin{equation}
\hat B_3=0.
\end{equation}
$\hat B_1$ does not play any role in the on-shell case, given that $t_1^{\lambda\mu\nu}$ is always perpendicular to the photon polarization $\epsilon^\mu$ and $\epsilon^\nu$. In this case, the denominator of the equation \eqref{B_10} happens to be singular, since
\begin{equation}
p_1\cdot p_2=2 \, (p\cdot\eta)^2.
\end{equation} 
To overcome this singularity we slightly modify the heat-bath velocity by setting $\eta=(1+\epsilon,0,0,0)$, with $\epsilon=d-4$. The denominator in \eqref{B_10} is now proportional to ${1}/{(2p_0^2\epsilon)}$. This allows us to proceed with the analytic computation of the integrals.\\
For example, in this new reference frame a typical integral takes the form
\begin{eqnarray}
H^{(\delta,f,f)}[p_1\cdot k] &=& i\int \frac{d^dk}{(2\pi)^{d-1}} \, k\cdot p_1 \, \biggl (\theta(\mu- \eta \cdot k ) \, \theta (\eta \cdot k ) \, \delta(k^2) \biggl )\frac{1}{(k-q)^2}\frac{1}{(k-p_2)^2}\notag\\
&=&i\int \frac{d^dk}{(2\pi)^{d-1}} \, (k_0 \, p_0- |\kbf| p_0 \cos\theta) \, \biggl (\theta(\mu-k_0[1+\epsilon]) \, \frac{\delta(k_0-|\mathbf k|)}{2|\mathbf k|} \biggl )\notag \\ && \qquad \times \frac{1}{4p_0^2+k^2-4k_0 p_0}\frac{1}{k^2-2k_0p_0 + 2 |\kbf| p_0 \cos \theta}.\notag\\
\end{eqnarray}
The next step, in this example, is to perform the $k^0$ integration 
\begin{equation}
H^{(\delta,f,f)}[p_1\cdot k]=\frac{i}{16 \, p_0}\int {d^{d-1}k \over (2\pi)^{d-3}} \, \frac{1}{{|\mathbf k|} \, (|\mathbf k|-p^0)}\frac{1-\cos\theta}{1+\cos\theta} \, \theta(\mu - |\kbf|[1+\epsilon]) 
\end{equation}
and use spherical coordinates in $d-1$ dimensions. The integral can be re-expressed in the form 
\begin{equation}
H^{(\delta,f,f)}[p_1\cdot k]= \frac{i\Omega(d-3)}{16\, p_0 \, (2\pi)^{d-1}}\int_0^{\mu/[1+\epsilon]} d|\kbf | \frac{ |\kbf |^{d-3}}{|\mathbf k|-p_0}\int_{-1}^{1} dt \, \frac{1-t}{1+t} \,(1-t^2)^{\frac{d}{2}-1}.
\end{equation}
If we redefine $\tilde \mu=\mu/[1+\epsilon]$, the computation is identical to the one encountered in eq.  \eqref{vv}. Therefore we get 
\begin{equation}
H^{(\delta,f,f)} [p_1\cdot k] = 
-\frac{2^{-d-4}\,  \pi ^{-d/2} \, \tilde\mu ^{d-2} \, \Gamma \left(\frac{d}{2}-2\right) \, _2F_1\left(1,d-2;d-1;\frac{\tilde\mu }{p_0}\right)}{p_0^2 \,  \Gamma \left(\frac{d-1}{2}\right)^2}.
\end{equation} 
It is easy to show that the linear combinations of integrals in \eqref{B_10} go as $\sim O((d-4)^2)$, giving \begin{equation}
\tilde B_2=\frac{1}{2p_0^2\epsilon}O((d-4)^2).
\end{equation}
This shows that the transverse part of the $AVV$ at finite density, $\Gamma$, is not modified by the 
chemical potential in the on-shell limit. By separating $\Gamma$ into its longitudinal and transverse contributions 
\beqa
\Gamma&=&\Gamma^{(0)}_L +  \Gamma^{(0)}_T + \Gamma^{(1)}_L +  \Gamma^{(1)}_T 
\eeqa
we have 
\beq
\Gamma^{(1)}=\Gamma^{(1)}_L +  \Gamma^{(1)}_T=0
\eeq
and using the result of section \ref{vacuum} $\Gamma^{(0)}_T=0$ (the on-shell vacuum amplitude in the symmetric case, is just given by the pole), we obtain 
\begin{equation}
\Gamma^{\lambda\mu\nu}=\Gamma^{\lambda\mu\nu}_{\text{L}}(\mu=0)=-\frac{i}{2\pi^2}\epsilon^{\mu\nu p_1 p_2}\frac{q^\lambda}{q^2}.
\end{equation}
Therefore we have shown that the on-shell vertex does not acquire any correction from a chemical potential. \\
The implication at the level of the on-shell effective action is that this takes the exact form 
\beq
 \mathcal{S}_{JJJ_5} =\int d^4 x\ d^4 y \ \partial \cdot B \,\Box^{-1}(x,y)\ F \tilde F (y), 
 \label{chir}
 \eeq 
 with $B_\lambda$ denoting the external axial-vector source field.

\subsection{Scaling violations in $\mu$ in the off-shell case}
The extension of previous studies to the off-shell case, along with the challenge of explicitly evaluating perturbative form factors, leaves us with an unanswered question on  whether this dependence also cancels out in the off-shell case. We are going to show that this is not the case and we will do it applying the operator $\partial/\partial{\mu}$ to the transverse part of the vertex. \\
The derivative operator acts only on the hot propagator term, in particular on one of the step functions
\begin{equation}
	\theta(\mu - \eta \cdot k), \qquad \theta(\mu - \eta \cdot (k - p_1)),  \qquad \theta(\mu - \eta \cdot (k - q)) .
\end{equation}
Deriving these functions, we get different new Dirac Deltas. In the $(\delta, f, f)$ cases this leads to a further saturation of the integration variables, as it can be seen in
\begin{equation}
{\partial \over \partial \mu} \Big( \delta(k^2)  \,  \theta(\eta \cdot k ) \, \theta(\mu - \eta \cdot k) )\Big)= \delta(k^2) \, \theta(\eta \cdot k) \, \delta(\mu - \eta \cdot k) = {1 \over 2 | \kbf| } \, \delta(k_0 - |\kbf|) \, \delta(\mu - |\kbf|)
\end{equation}
The first delta saturates the temporal integral while the second delta eliminates the $|\kbf|$ one, leaving us with only angular integrals. This process happens in all integrals with only one hot propagator, while for $(\delta, \delta, f) $ cases and permutations, the additional delta generates conflicting conditions that cancel the integral alltogether. We take the following case as an example
\begin{align}
	&{\partial \over \partial \mu} \Big( \delta(k^2) \, \theta(\mu - \eta \cdot k) \, \delta \big((k-q)^2 \big) \, \theta \big(\mu - \eta \cdot (k-q)\big) \, \theta(\eta \cdot k) \, \theta(\eta \cdot (k - q)) \Big) = \notag \\
	& \delta(k^2) \,  \delta \big((k-q)^2\big) \, \theta (\eta \cdot k) \, \theta \big(\eta \cdot (k-q) \big) 
	 \Big[ \delta(\mu - \eta \cdot k) \, \theta \big( \mu - \eta \cdot (k-q) \big) +  \theta(\mu - \eta \cdot k) \, \delta \big( \mu - \eta \cdot (k-q) \big)  \Big]
\end{align}
We note that, for the first term of the product
\begin{equation}\delta(k^2) \, \delta(\mu - \eta \cdot k) \, \delta((k-q)^2 ) \, \theta(\eta \cdot k) \, \theta(\eta \cdot (k - q))
\propto \delta(k^0 - |\kbf|) \, \delta(\mu - |\kbf|) \, \delta\big(-4 p_0 (|\kbf| - p_0 )\big)
\end{equation}
it becomes apparent that the two deltas of $|\kbf|$ cannot be satisfied at the same time, thus the whole contribution vanishes. Similar reasoning works for the other term in the sum and for $(f,\delta, \delta)$ and $(\delta , \delta, f) $, as well as for the $(\delta, \delta, \delta)$. \\
Using these principles, one can flesh out the  whole evaluation of this derivative for all the form factors written as linear combinations of scalar integrals as in \eqref{OFFFF}.
We present here some examples of integrals that appear in this evaluation
\begin{align}
&J^{(\delta,f,f)}[p_1\cdot k]=\frac{\mu}{16p_0(p_0+\mu)}\int_{-1}^{1} dt \frac{A-t}{B_p+t}+\frac{p_0}{8(p_0+\mu)\sqrt{p_0^2-M^2}}\int_{-1}^{1}dt \frac{1}{B_p+t},
\end{align}
where 
\begin{align}
&A=\frac{p_0}{\sqrt{p_0^2-M^2}}\notag\\&
\frac{M^2}{2\mu\sqrt{p_0^2-M^2}}-A=B_m\notag\\&
\frac{M^2}{2\mu\sqrt{p_0^2-M^2}}+A=B_p
\end{align}
\begin{align}
&J^{(f,\delta,f)}[p_1\cdot k]=\frac{1}{8\sqrt{p_0^2-M^2}}\int_{-1}^{1} dt \frac{1}{B_m-t}
\end{align}

\begin{align}
&J^{(f,f,\delta)}[p_1\cdot k]=\frac{\mu}{p_0(p_0-\mu)}\int_{-1}^{1} dt \frac{A-t}{B_m+t}
\end{align}

\begin{align}
J[(p_1\cdot k)^2]^{(\delta,f,f)}=\frac{\mu^2\sqrt{p_0^2-M^2}}{16p_0(p_0-\mu)}\int_{-1}^{1}dt \frac{(A-t)^2}{B_m+t}
\end{align}

\begin{align}
J[(p_1\cdot k)^2]^{(f,\delta,f)}=\frac{\mu}{8}\int_{-1}^{1}dt \frac{(A-t+\frac{M^2}{\mu\sqrt{p_0^2-M^2}})^2}{(B_p-t)(B_m-t)}
\end{align}

\begin{align}
J[(p_1\cdot k)^2]^{(f,f\delta)}=\frac{\mu^2\sqrt{p_0^2-M^2}}{16p_0(p_0-\mu)}\int_{-1}^{1}dt \frac{(A-t+\frac{2p_0^2}{\mu\sqrt{p_0^2-M^2}})^2}{B_p+t}
\end{align}
Performing the full computation of the derivative with respect to the chemical potential for the $\bar B_2$ form factor, we get 
\begin{equation}
\frac{\partial}{\partial\mu} \bar B_2\neq 0.
\end{equation}
In fact the log independent part reads
\begin{equation}
\frac{\partial}{\partial \mu}\bar B_2\propto \frac{\mu(-\mu M^2+4 p_0^3+2\mu p_0^2)}{4\pi^3 p_0(M^2-2p_0^2)(\mu+p_0)}+\text{Log terms}
\end{equation}
This result clearly means that for the off-shell case the chemical potential plays an important role, differently from the on shell case, although its contribution is limited to the transverse sector. 

\section{ Furry's theorem and more general correlators}

 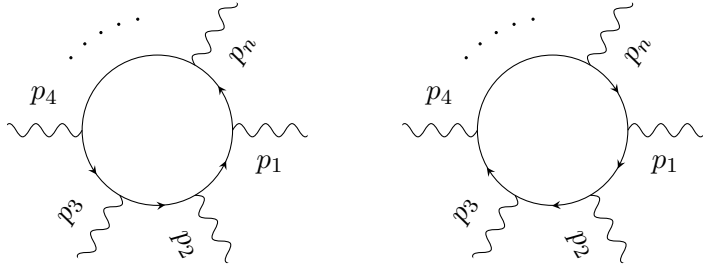
\begin{figure}[t]
	\centering
	\begin{tikzpicture}[scale=1]
		\clip (0,0) circle (2);
		\draw[rotate=180, postaction={decorate, decoration={
				markings,
				mark=between positions 0.1 and .6 step 1/6 with {\arrow{stealth} } } }
		] (0,0) circle (1);
		\draw[snake=coil, segment aspect=0] (60:1) -- (60:2.5) node[sloped, below=.2, pos=1/3] {$p_n$};
		\draw[snake=coil, segment aspect=0]  (0:1) -- (0:2.5) node[sloped, below=.2 , pos=1/3] {$p_1$};;
		\draw[snake=coil, segment aspect=0]  (-60:1) -- (-60:2.5) node[sloped, below=.2 , pos=1/3] {$p_2$};;
		\draw[snake=coil, segment aspect=0]  (-120:1) -- (-120:2.5) node[sloped, above=.2 , pos=1/3] {$p_3$};;
		\draw[snake=coil, segment aspect=0]  (180:1) -- (180:2.5) node[sloped, above=.2, pos=1/3] {$p_4$};;	
		\filldraw[black]  (100:1.5) circle (.4pt) ;
		\filldraw[black]  (110:1.5) circle (.4pt) ;
		\filldraw[black]  (120:1.5) circle (.4pt) ;
		\filldraw[black]  (130:1.5) circle (.4pt) ;
		\filldraw[black]  (140:1.5) circle (.4pt) ;
	\end{tikzpicture} 
	\hspace{1cm}
	\begin{tikzpicture}[scale=1]
		\clip (0,0) circle (2);
		\draw[rotate=180, postaction={decorate, decoration={
				markings,
				mark=between positions 0.1 and .6 step 1/6 with {\arrow{stealth[reversed]} } } }
		] (0,0) circle (1);
		\draw[snake=coil, segment aspect=0] (60:1) -- (60:2.5) node[sloped, below=.2, pos=1/3] {$p_n$};
		\draw[snake=coil, segment aspect=0]  (0:1) -- (0:2.5) node[sloped, below=.2 , pos=1/3] {$p_1$};
		\draw[snake=coil, segment aspect=0]  (-60:1) -- (-60:2.5) node[sloped, below=.2 , pos=1/3] {$p_2$};
		\draw[snake=coil, segment aspect=0]  (-120:1) -- (-120:2.5) node[sloped, above=.2 , pos=1/3] {$p_3$};
		\draw[snake=coil, segment aspect=0]  (180:1) -- (180:2.5) node[sloped, above=.2, pos=1/3] {$p_4$};
		\filldraw[black]  (100:1.5) circle (.4pt) ;
		\filldraw[black]  (110:1.5) circle (.4pt) ;
		\filldraw[black]  (120:1.5) circle (.4pt) ;
		\filldraw[black]  (130:1.5) circle (.4pt) ;
		\filldraw[black]  (140:1.5) circle (.4pt) ;
	\end{tikzpicture}
	\caption{Diagrams with opposite fermion flow in the symmetric vertex}
	\label{fig:DiagrammiFurry}
\end{figure}
Before coming to a discussion of other chiral correlators, we turn to investigate the role of finite density effects in $C$-violating backgrounds that may be present in out-of-equilibrium configurations of the background sources. The relation 
of the $AVV$  with other diagrams built with the $J_L$ and $J_R$ currents, require an investigation of the role played by charge conjugation in the presence of both chiral and vector currents, including the $VVV$ correlator of three vector currents. This vanishes by $C$ invariance in the ordinary vacuum sector, but is nonzero in the presence of finite density backgrounds. \\
According to Furry's theorem, Feynman diagrams containing a closed electron loop with an odd number of photon vertices vanishes. 
In a closed loop there can be an electron as well as a positron “circling around”. These particles interact with the electromagnetic field with an opposite sign of the charge. 
Thus, their contributions cancel each other for an odd number of vertices.
From a Feynman diagram perspective, denoting with $M_a$ and $M_b$ the diagrams in Fig$.$ \ref{fig:DiagrammiFurry}, one is able to show that 
\begin{equation}
	M_{\text{tot}}=M_a+M_b=0
\end{equation}
However, as we will see, Furry's theorem does not hold anymore in the presence of a chemical potential due to the breaking of charge conjugation invariance by the background.\\
Let us first write the expression of the diagrams displayed in Fig$.$ \ref{fig:DiagrammiFurry}
\begin{equation}
	\begin{aligned}
		&M_a(\beta ,\mu )=\int \frac{d^4k}{(2\pi)^4}\text{Tr}\Big\{ S_F(k,\beta,\mu)\,\gamma_{\mu_n}\, \dots\, S_F(k+p_1+p_2,\beta,\mu)\,\gamma_{\mu_2}\,
		S_F(k+p_1,\beta,\mu)\,\gamma_{\mu_1}\Big\}\\&
		M_b(\beta ,\mu )=\int \frac{d^4k}{(2\pi)^4}\text{Tr}\Big\{ \gamma_{\mu_1} \,S_F(k-p_1,\beta,\mu)\,\gamma_{\mu_2}
		\,S_F(k-p_1-p_2,\beta,\mu)\, \dots\,\gamma_{\mu_n}\,S_F(k,\beta,\mu)\Big\}
	\end{aligned}
\end{equation}
Such diagrams are closely related to each other. To see this we make use of the charge conjugation matrix $\hat{C}=i \gamma^2 \gamma^0$ with the property 
\begin{equation}\label{eq:cgamma}
	\hat{C} \gamma_\mu \hat{C}^{-1}=-\gamma_{\mu}^T
\end{equation}
Applied to the Feynman propagator at finite temperature and density, this transformation yields
\begin{equation}\label{eq:sfgamma}
	\begin{aligned}
		\hat{C}\, S_F(k,\beta,\mu)\, \hat{C}^{-1}&=\left(k^\mu	\hat{C}\, \gamma_\mu \, \hat{C}^{-1} +m\right)
		G_F(k,\beta,\mu)=\left(-k^\mu	\gamma_\mu^T +m\right)
		G_F(k,\beta,\mu)=\\ &\left(-k^\mu	\gamma_\mu^T +m\right)
		G_F(-k,\beta,-\mu)={S_F}(-k,\beta,-\mu)^T
	\end{aligned}
\end{equation}
Now, inserting multiple factors of $\hat{C}^{-1}\hat{C}$ into the exchanged diagram $M_b$, we get
\begin{equation}
	\begin{aligned}
		M_b(\beta ,\mu )=\int \frac{d^4k}{(2\pi)^4}\text{Tr}&\Big\{\hat{C}^{-1}\hat{C} \, \gamma_{\mu_1}\, \hat{C}^{-1}\hat{C} \, S_F(k-p_1,\beta,\mu)\, \hat{C}^{-1}\hat{C}\,\gamma_{\mu_2}\, \hat{C}^{-1}\hat{C}
		\,S_F(k-p_1-p_2,\beta,\mu) \, \dots\\&\qquad
		\times \hat{C}^{-1}\hat{C} \, \gamma_{\mu_n}\, \hat{C}^{-1}\hat{C}\, S_F(k,\beta,\mu)\Big\}
	\end{aligned}
\end{equation}
Using the properties \eqref{eq:cgamma} and  \eqref{eq:sfgamma}, we can then write
\begin{equation}
	\begin{aligned}
		&M_b(\beta ,\mu )=\\&(-1)^n \int \frac{d^4k}{(2\pi)^4}\text{Tr}\Big\{ \gamma^T_{\mu_1}\, S_F(-k+p_1,\beta,-\mu)^T\, \gamma_{\mu_2}^T
		\,S_F(-k+p_1+p_2,\beta,-\mu)^T \, \dots\, \gamma_{\mu_n}^T \, S_F(-k,\beta,-\mu)^T\Big\}=\\&
		(-1)^n \int \frac{d^4k}{(2\pi)^4}\text{Tr}\Big\{ 
		S_F(-k,\beta,-\mu)\, \gamma_{\mu_n}
		\,\dots \,S_F(-k+p_1+p_2,\beta,-\mu)\,
		\gamma_{\mu_2}
		\, S_F(-k+p_1,\beta,-\mu)\, \gamma_{\mu_1}  \Big\}^T
	\end{aligned}
\end{equation}
Applying the change of variable $k\rightarrow -k$ in the integral, we arrive to
\begin{equation}
	M_b(\beta ,\mu )=(-1)^n M_a(\beta ,-\mu )
\end{equation}
For an odd number $n$  of vertices, this condition reduces to the Furry's theorem for any value of the temperature when $\mu=0$
\begin{equation}
	M_{\text{tot}}(\beta ,\mu=0 )=M_a(\beta ,0 )-M_a(\beta ,0 )=0
\end{equation}
However in general the sum of the two diagrams $M_{\text{tot}}$ will not vanish
\begin{equation}
	M_{\text{tot}}(\beta ,\mu )=M_a(\beta ,\mu )-M_a(\beta ,-\mu )
\end{equation}
The presence of a chemical potential manifestly breaks charge conjugation invariance and leads to the appearance of new processes in a perturbative expansion.
Note that there is still a case where $\mu\neq 0$ and Furry's theorem is satisfied. Indeed, in the limit $\beta \rightarrow 0$ (infinite temperature) the fermion propagator \eqref{eq:propmassivtemp} does not depend on the value of the chemical potential and therefore $M_{\text{tot}}=0$.\\
Using this result, we can address the extension of the analysis of the previous sections to more general correlators built with left and right handed currents 
\begin{equation} J_L=\bar \psi_L \gamma^\mu \psi_L ,
\qquad \qquad J_R=\bar \psi_R \gamma^\mu \psi_R.
 \end{equation}
The propagator of a chiral fermion is modified by the inclusion of the chiral projector as
\begin{equation}\label{3} S_{L,R}=\int d^4 k e^{-ik(x'-x)}P_{L,R} \biggl  [  \frac{\slashed k}{k^2}+2i\pi \delta (k^2) \theta(k\cdot \eta)\theta(\mu_{L/R} -k\cdot \eta)  \biggl ]  \end{equation}
where we introduce two different chemical potentials $\mu_L,\mu_R$ for the chiral modes.
We are going to analyze this point, that requires a careful look at the role of Furry's theorem at finite temperature and density.\\
Consider the correlator $\Delta_L\equiv \langle J_L J_L J_L \rangle_{\mu_L}  $ that we expand using $J_L=\frac{1}{2}(V-A)$, where we denote with $V$ the vector part of the chiral $J_L$ interaction, and $A$ the corresponding axial-vector part. 
In the perturbative realization of the diagram we insert three chiral projectors at each vertex, reducing the computation to the former AVV one as
\begin{equation}\label{4} \Delta_L=\frac{1}{8} \langle (V-A)(V-A)(V-A)   \rangle_{\mu_L}.  \end{equation}
In this expression we will be using the propagator in the form given by \eqref{3} with $\mu\equiv \mu_L$. Expanding on the components of \eqref{4} we obtain 
\begin{align}&\Delta_L=\frac{1}{8}\bigg(\langle VVV\rangle- \langle VVA\rangle- \langle VAV\rangle- \langle AVV\rangle+ \langle VAA\rangle+\langle AVA\rangle+ \langle AAV\rangle- \langle AAA\rangle\bigg)\end{align}
The relation between the $\langle AAA \rangle$ interactions, as well as the $\langle AAV \rangle$ ones, with the $\langle AVV\rangle$ and the $\langle VVV \rangle$ are easily worked out in the form 
 \begin{equation}\label{6}
	\begin{aligned}
		&\langle VVV\rangle_{\mu_L}=\frac{1}{3}\bigg( \langle AAV\rangle+\langle AVA\rangle+ \langle VAA\rangle \bigg)_{\mu_L}\\&
		\langle AAA\rangle_{\mu_L}=\frac{1}{3}\bigg(\langle VVA\rangle+ \langle VAV\rangle+ \langle AVV\rangle\bigg)_{\mu_L}
	\end{aligned}
\end{equation}
One may re-express the result in terms of only contribution for $\Delta_{L,R}$ in the forms of $\langle AAA\rangle$ and $\langle VVV\rangle$ in the form
\begin{equation}\label{eq:deltalw} \Delta_L\equiv \frac{1}{2} \biggl (\langle VVV\rangle- \langle AAA\rangle \biggl)_{\mu_L} \end{equation}
Proceeding in a similar manner for right-handed fermions, we obtain
\begin{equation}\label{eq:deltarw}{\Delta_R}=\frac{1}{2} \biggl(\langle VVV\rangle + \langle AAA\rangle  \biggl )_{\mu_R}\end{equation}
We now introduce a symmetric notation for the external momenta $(p_1,p_2,p_3)$ all of them off-shell.
By symmetries, the AAA diagram can be organised in the from
\begin{equation}\label{11}
\langle AAA\rangle_{\mu}^{\lambda_1\lambda_2\lambda_3}=\frac{a_n}{3}\biggl( \frac{p_1^{\lambda_1}}{p_1^2}\epsilon^{\lambda_2 \lambda_3 p_2 p_3}+\frac{p_2^{\lambda_2}}{p_2^2}\epsilon^{\lambda_1\lambda_3 p_1 p_3}+\frac{p_3^{\lambda_3}}{p_3^2}\epsilon^{\lambda_1\lambda_2 p_1 p_2}\biggl)+\Delta_T^{\lambda_1\lambda_2\lambda_3}(\mu,p_1^2,p_2^2,p_3^2),
\end{equation}
with $\Delta_T^{\lambda_1\lambda_2\lambda_3}$ transverse in $p_1,p_2,p_3$. The three anomaly poles distribute the anomaly equally across the three axial-vector vertices.
Equation \eqref{11} is a natural consequences of \eqref{6}, since each permutation of the $\langle AVV\rangle$ in \eqref{6} is characterized by a $\mu$-independent longitudinal sector in the axial vector channel. By symmetry, the result can be extended to the $\langle AAA\rangle$ case. \\
 Similarly, recalling eq$.$ \eqref{eq:deltalw} and \eqref{eq:deltarw}, one can verify that the longitudinal sector of $\Delta_L$ and $\Delta_R$ is $\mu$-independent
\begin{equation}
	\begin{aligned}
			\Delta_L=&-\frac{a_n}{6}\biggl( \frac{p_1^{\lambda_1}}{p_1^2}\epsilon^{\lambda_2 \lambda_3 p_2 p_3}+\frac{p_2^{\lambda_2}}{p_2^2}\epsilon^{\lambda_1\lambda_3 p_1 p_3}+\frac{p_3^{\lambda_3}}{p_3^2}\epsilon^{\lambda_1\lambda_2 p_1 p_2}\biggl)+\Delta_{L,\,T}(\mu_L,p_1^2,p_2^2,p_3^2),\\
		\Delta_R=&\, \frac{a_n}{6}\biggl( \frac{p_1^{\lambda_1}}{p_1^2}\epsilon^{\lambda_2 \lambda_3 p_2 p_3}+\frac{p_2^{\lambda_2}}{p_2^2}\epsilon^{\lambda_1\lambda_3 p_1 p_3}+\frac{p_3^{\lambda_3}}{p_3^2}\epsilon^{\lambda_1\lambda_2 p_1 p_2}\biggl)+\Delta_{R,\,T}(\mu_R,p_1^2,p_2^2,p_3^2)
	\end{aligned}
\end{equation}
with $\Delta_{L,\,T}$ and $\Delta_{R,\,T}$ transverse in $p_1,p_2,p_3$. 
\section{Comments}
A distinct disparity exists between the asymptotic state associated with a chiral anomaly interaction, the axion, and its perturbative description. The former is characterized by a direct (local) coupling denoted as $\varphi F \tilde {F}$, representing the interaction between the pseudoscalar (asymptotic) degree of freedom $(\varphi)$ and photons. The coupling of the axion field occurs through the divergence of the axial-vector current $(\varphi\, \partial J_5)$ (see \cite{Bai:2023bbg} for recent developments). Generally, an asymptotic axion is conceptualized as an elementary state, although composite models have also been formulated. In contrast, the perturbative description of this interaction assumes a nonlocal form. It arises in the perturbative analysis of a chiral anomaly diagram in momentum space, wherein the interaction is defined by the exchange of a massless pole. In this context, interpreting this state as the topological response of a physical system to an external chiral perturbation, essentially a quasiparticle, is quite natural. \\
While the anomaly vertex remains identical in both descriptions from a perturbative standpoint, stemming from an explicit symmetry breaking of a global symmetry, the two descriptions exhibit partial overlap. An ordinary axion is an asymptotic state whose decay into photons is suppressed by a large coupling constant 
($f_a$) and whose dynamics is linked with the nonperturbative vacuum of a nonabelian gauge theory, such as QCD. Its mass is generated by a periodic instanton potential at the hadronic scale, thereby linking two separate scales. \\
On the other end, as mentioned in the Introduction, the chiral anomaly interaction in a topological material, from a relativistic perspective, can be classified only as an analog one.
The effective action, in both cases, can be articulated by incorporating both a local and a nonlocal operator, yet the underlying physical manifestations remain distinctive for each scenario. For instance, in the context of a topological material, the interpolating state is naturally perceived as non-elementary, with a pseudoparticle interpretation. This is the description that we have investigated in this work, by resorting to a complete perturbative analysis of the chiral anomaly interaction.

\section{Conclusions}
In this work we have classified all the tensor structures that are part of the interaction, and provided a direct perturbative identification of its corresponding form factors. The parametrization that we have presented is minimal, since we have kept into account all the symmetries and the Ward identities that are part of its definition. We have shown that the on-shell vertex, where the photons are physical, reduces to a massless anomaly pole even at finite density, similarly to the ordinary  $AVV$ at zero density.  \\
In the off-shell case, the longitudinal part of the vertex is not modified by the chemical potential, while the transverse part is. 
Therefore, from the perspective of the 1PI effective action, our main conclusion is that such action acquires a form identical to \eqref{nonl} in the on-shell case, even at finite density.  Nonlocal actions of such type cover, at least in the vacuum case, all the chiral and conformal anomaly interactions, thereby characterizing a unique trend. \\
The current result is therefore in line with a previous analysis of other important correlators in the vacuum, such as the $J_5 TT$, responsible for the generation of the gravitational chiral anomaly. 
A complete nonperturbative analysis, based on the solution of the trace Ward identity by the inclusion of a $1/\Box$  "anomaly pole" in the trace part of this correlator, allows  the reconstruction of its entire effective action, by solving the related conformal Ward identities \cite{Coriano:2023gxa,Coriano:2023hts}.\\
The interpolating states emerging in the conformal case are similar. The related 1PI vertex, in this other case, is the $TJJ$, with one graviton and two photons. For instance, in this vertex, a scalar dilaton field couples to $FF$, generating a dilaton-like intermediate pseudoparticle. 
 This interaction can be reproduced, in an analog setting, by subjecting a topological material to a thermal gradient, using Luttinger's relation \cite{Luttinger:1964zz}, where gravitational interactions are related to thermal fluctuations.  \\
From the theoretical perspective, in recent years, it has become increasingly evident that chiral anomalies, associated with the breaking of classical global symmetries by quantum corrections, play a crucial role in the dynamics of fundamental interactions, not only in the vacuum $(\mu=0)$ case, but also 
in the presence of chiral chemical potentials, as shown in the case of the chiral magnetic effect. The possibility of performing experimental tests of these interactions, that could identify the nature of the pseudoparticle emerging from the virtual corrections, as predicted by the current and previous analyses \cite{Armillis:2009im,Giannotti:2008cv, Armillis:2010qk,Armillis:2009pq}  should be taken very seriously  at experimental level.  We have pointed out that this behavior is directly related to conformal symmetry. Along this line, one could envision possible tests of chirally-odd trace anomalies in table-top experiments, in analog gravitational setups involving topological materials. For instance, one unanswered question is if gravity may be the source of $CP$ violation at nonperturbative level, a point which is undergoing a new debate. Analog gravitational systems could play an extraordinary role in settling the recent debate concerning such anomalies from the experimental viewpoint \cite{Armillis:2010pa,Bastianelli:2018osv,Bastianelli:2019fot,Abdallah:2023cdw,Bonora:2014qla,Coriano:2023cvf,Ferrero:2023unz} and offer also an avenue for even more exotic realizations, by investigating the conformal behavior of analog correlators \cite{Bermond:2022mjo} in gravity in  general, and in particular in de Sitter space, in an engaging correspondence with recent analysis \cite{Sleight:2021plv,Sleight:2020obc,Sleight:2019hfp}. 
  
\centerline{\bf Acknowledgements}
We thank Emil Mottola for discussions and for suggesting this investigation. We thank Alessandro Bramanti and Stefania D'Agostino for discussions.
This work is partially funded by the European Union, Next Generation EU, PNRR project "National Centre for HPC, Big Data and Quantum Computing", project code CN00000013; by INFN, inziativa specifica {\em QG-sky} and by the grant PRIN 2022BP52A MUR "The Holographic Universe for all Lambdas" Lecce-Naples. We thank Anastasios Petkou and Kostantinos Siampos and all the organizers of the workshop "Recent Developments in CFT" at Thessalonikki for hospitality, where results concerning this work have been presented.

\appendix

\section{Summary on the tensorial decompositions}

The original tensor structures are 60, and have been introduced in Eq$.$ \eqref{fdecom}. Then we impose the Schouten identities as in equation \eqref{schouten1} and also on the tensorial structures reported in \eqref{schouten2}. In total we impose $\textbf{32}$ Schouten identities that bring the number of structures down to $\textbf{28}$ as in \eqref{tensorfinali}
\begin{center}
\textbf{Schouten Identities}:\\
$\textbf{60}\rightarrow\textbf{28}.$
\end{center}
The second step consists in imposing the Bose simmetry on the external photons. The starting point are the \textbf{28} form factor in \eqref{le28}. We request that under $\{p_1,\mu\}\longleftrightarrow \{p_2,\nu \}$, $\Gamma^{\lambda\mu\nu}$ is invariant, obtaining
\begin{center}
\textbf{Bose symmetry}:\\
$\textbf{28}\rightarrow\textbf{16}.$
\end{center}
The third step is given by imposing the {vectorial Ward identities}. In fact, imposing \eqref{wicasononsimm1} and \eqref{wicasononsimm2} we obtain \eqref{gen}, so the scheme for this third step is
\begin{center}
\textbf{Vectorial WIs}:\\
$\textbf{16}\rightarrow\textbf{10}.$
\end{center}
When we impose the symmetry constraints, namely
\begin{equation}
p_1\cdot\eta=p_2\cdot\eta=p\cdot\eta,\end{equation}\begin{equation}
p_1^2=p_2^2=p^2,
\end{equation}
we get further reductions on the number of form factors. The reduction by Schouten identities remains the same, so the difference in the number of final form factors is generated by the other two steps, i.e. the Bose symmetry constraint and the vectorial WIs constraint. In fact, as we can see in \eqref{BosSybSubON}, the reduction in form factors is obtained  by imposing
\begin{center}
\textbf{Bose symmetry}:\\
$\textbf{28}\rightarrow\textbf{12}.$
\end{center}
and using the set of equations \eqref{eqreffff}
\begin{center}
\textbf{Vectorial WIs}:\\
$\textbf{12}\rightarrow\textbf{7}.$
\end{center}
This reduction process  generates \eqref{gen1}.

\section{Structures}
\label{structures}
We list the tensor structures that appear in the expansion of the $AVV$ vertex. They are: 

{{

\begin{align} \label{60tensors}
	&\epsilon^{\lambda \mu \nu{p_1}} & 
	&\epsilon^{\lambda \mu \nu {p_2}} & 
	&\epsilon^{\lambda \mu \nu \eta} &
	&{p_1}^{\lambda } \epsilon^{\mu \nu {p_1}{p_2}} &
	&{p_1}^{\lambda }\epsilon^{\mu \nu{p_1}\eta} \notag \\
	&{p_1}^{\lambda } \epsilon^{\mu \nu {p_2}\eta} &
	&{p_2}^{\lambda } {\epsilon}^{\mu \nu{p_1}{p_2}} &
	&{p_2}^{\lambda } \epsilon^{\mu \nu{p_1}\eta} &
	&{p_2}^{\lambda } \epsilon^{\mu \nu {p_2}\eta} &
	 &\eta^{\lambda } \epsilon^{\mu	\nu{p_1}{p_2}} \notag \\
	&\eta^{\lambda } \epsilon^{\mu \nu{p_1}\eta} &
	& \eta^{\lambda } \epsilon^{\mu \nu {p_2}\eta} &
	&{p_1}^{\mu } {\epsilon}^{\lambda \nu{p_1}{p_2}} &
	&{p_1}^{\mu }	\epsilon^{\lambda \nu{p_1}\eta} &
	&{p_1}^{\mu } \epsilon^{\lambda \nu {p_2}\eta}  \notag \\
	&{p_2}^{\mu } \epsilon^{\lambda \nu {p_1}{p_2}} &
	&{p_2}^{\mu } {\epsilon}^{\lambda \nu{p_1}\eta} &
	&{p_2}^{\mu } \epsilon^{\lambda \nu{p_2}\eta} &
	&{\eta}^{\mu } \epsilon^{\lambda \nu{p_1}{p_2}} &
	&\eta^{\mu } \epsilon^{\lambda \nu{p_1}{\eta}}  \notag \\
	&\eta^{\mu } \epsilon^{\lambda \nu{p_2}\eta}&
	&{p_1}^{\nu } \epsilon^{\lambda \mu {p_1}{p_2}} &
	&{p_1}^{\nu } {\epsilon }^{\lambda \mu{p_1}\eta} &
	&{p_1}^{\nu }\epsilon^{\lambda \mu{p_2}\eta} &
	&{p_2}^{\nu } \epsilon^{\lambda \mu{p_1}{p_2}}  \notag \\
	&{p_2}^{\nu } \epsilon^{\lambda \mu {p_1}\eta} &
	&{p_2}^{\nu } {\epsilon}^{\lambda \mu{p_2}\eta} &
	 &\eta^{\nu }\epsilon^{\lambda \mu{p_1}{p_2}} &
	 &{\eta}^{\nu } \epsilon^{\lambda \mu{p_1}\eta} &
	&\eta^{\nu } \epsilon^{\lambda \mu{p_2}{\eta}}  \notag \\
	&{p_1}^{\lambda }{p_1}^{\nu } \epsilon^{\mu{p_1}{p_2}\eta} &
	&{p_1}^{\nu }{p_2}^{\lambda } \epsilon^{\mu{p_1}{p_2}\eta} & 
	&\eta^{\lambda }{p_1}^{\nu } \epsilon^{\mu{p_1}{p_2}\eta} &
	&{p_1}^{\lambda }{p_2}^{\nu } \epsilon^{\mu{p_1}{p_2}\eta} &
	&{p_2}^{\lambda }{p_2}^{\nu } \epsilon^{\mu{p_1}{p_2}\eta} \notag \\
	&\eta^{\lambda }{p_2}^{\nu } \epsilon^{\mu{p_1}{p_2}\eta} &
	 &\eta^{\nu } {p_1}^{\lambda } \epsilon^{\mu{p_1}{p_2}\eta} & 
	&\eta^{\nu }	{p_2}^{\lambda } \epsilon^{\mu{p_1}{p_2}\eta} &
	 &\eta^{\lambda }\eta^{\nu } \epsilon^{\mu{p_1}{p_2}\eta} &
	& \delta^{\lambda \nu } \epsilon^{\mu{p_1}{p_2}\eta}  \notag \\
	&{p_1}^{\lambda }{p_1}^{\mu } \epsilon^{\nu{p_1}{p_2}\eta} &
	&{p_1}^{\mu } {p_2}^{\lambda } \epsilon^{\nu{p_1}{p_2}\eta} &
	 &\eta^{\lambda }	{p_1}^{\mu } \epsilon^{\nu{p_1}{p_2}\eta} &
	&{p_1}^{\lambda }{p_2}^{\mu } \epsilon^{\nu{p_1}{p_2}\eta} &
	&{p_2}^{\lambda }{p_2}^{\mu } \epsilon^{\nu{p_1}{p_2}\eta}  \notag \\
	&\eta^{\lambda }{p_2}^{\mu } \epsilon^{\nu {p_1}{p_2}\eta} &
	 &\eta^{\mu } {p_1}^{\lambda } \epsilon^{\nu	{p_1}{p_2}\eta} & 
	&\eta^{\mu }	{p_2}^{\lambda } \epsilon^{\nu {p_1}{p_2}\eta} & 
	&\eta^{\lambda }	\eta^{\mu } \epsilon^{\nu {p_1}{p_2}\eta} &
	 &\delta^{\lambda \mu }	\epsilon^{\nu{p_1}{p_2}\eta} \notag \\
	&{p_1}^{\mu }{p_1}^{\nu } \epsilon^{\lambda	{p_1}{p_2}\eta} &
	&{p_1}^{\mu } {p_2}^{\nu } \epsilon^{\lambda {p_1 }{p_2}\eta} &
	 &\eta^{\nu } {p_1}^{\mu } \epsilon^{\lambda {p_1}{p_2}\eta} &
	 &{p_1}^{\nu } {p_2}^{\mu } \epsilon^{\lambda {p_1}{p_2}\eta} &
	 &{p_2}^{\mu }{p_2}^{\nu } \epsilon^{\lambda{p_1}{p_2}\eta}  \notag \\
	&\eta^{\nu }{p_2}^{\mu } \epsilon^{\lambda{p_1}{p_2}\eta} & 
	&\eta^{\mu }{p_1}^{\nu } \epsilon^{\lambda{p_1}{p_2}\eta} &
	& \eta^{\mu }{p_2}^{\nu } \epsilon^{\lambda	{p_1}{p_2}\eta} & 
 	&\eta^{\mu }\eta^{\nu } \epsilon^{\lambda {p_1}{p_2}\eta} &
  	&\delta^{\mu \nu } \epsilon^{\lambda{p_1}{p_2}\eta}
\end{align}

\section{Bose Symmetry and Ward Identity}
\label{bose}
Imposing the Bose symmetry, we find the following set of relations between the form factors in the expansion
\begin{align} \label{BosSybSub}
		B_2(p_1,p_2,\eta )&= -B_1(p_2,p_1,\eta ) &
		B_3(p_2,p_1,\eta )&= -B_3(p_1,p_2,\eta )  \notag \\
		B_{13}(p_1,p_2,\eta )&= -B_7(p_2,p_1,\eta ) &
		B_{14}(p_1,p_2,\eta )&= B_9(p_2,p_1,\eta )  \notag \\
		B_{15}(p_1,p_2,\eta )&= B_8(p_2,p_1,\eta ) &
		B_{16}(p_1,p_2,\eta )&= -B_4(p_2,p_1,\eta )  \notag \\
		B_{17}(p_1,p_2,\eta )&= B_6(p_2,p_1,\eta ) &
		B_{18}(p_1,p_2,\eta )&= B_5(p_2,p_1,\eta ) \notag \\
		B_{19}(p_1,p_2,\eta )&= -B_{10}(p_2,p_1,\eta ) &
		B_{20}(p_1,p_2,\eta )&= B_{12}(p_2,p_1,\eta ) \notag \\
		B_{21}(p_1,p_2,\eta )&= B_{11}(p_2,p_1,\eta ) &
		B_{23}(p_2,p_1,\eta )&= -B_{23}(p_1,p_2,\eta ) \notag \\
		B_{25}(p_1,p_2,\eta )&= -B_{22}(p_2,p_1,\eta ) &
		B_{26}(p_1,p_2,\eta )&= -B_{24}(p_2,p_1,\eta )  \notag \\
		B_{27}(p_2,p_1,\eta )&= -B_{27}(p_1,p_2,\eta ) &
		B_{28}(p_2,p_1,\eta )&= -B_{28}(p_1,p_2,\eta ) 
\end{align}
that reduce their number to 16. 
\\
Further simplifications are introduced by requiring that the interaction satisfies the vector Ward identities. These give the relations
\begin{eqnarray}\label{VecWIoff}
	B_1\left(p_2,p_1,\eta \right) & = &  -{p_1}^2 B_4\left(p_1,p_2,\eta
	\right)-\left({p}_1\cdot {p}_2\right) B_7\left(p_1,p_2,\eta
	\right)-\left({\eta }\cdot {p}_1\right) B_{10}\left(p_1,p_2,\eta
	\right) \notag \\
	B_3\left(p_1,p_2,\eta \right) & = &  -{p_1}^2 {p_2}^2
	B_{23}\left(p_2,p_1,\eta \right)+\left({p}_1\cdot {p}_2\right)
	B_8\left(p_1,p_2,\eta \right)+{p_1}^2 \left({p}_1\cdot
	{p}_2\right) B_{22}\left(p_1,p_2,\eta \right)+ \notag \\ && 
	{p_1}^2 \left({\eta }\cdot {p}_2\right) B_{24}\left(p_1,p_2,\eta
	\right)-{p_2}^2 \left({\eta }\cdot {p}_1\right)
	B_{24}\left(p_2,p_1,\eta \right)-\left({\eta }\cdot {p}_1\right)
	\left({\eta }\cdot {p}_2\right) B_{27}\left(p_2,p_1,\eta \right) \notag \\
	B_5\left(p_2,p_1,\eta \right) & = &  -{p_1}^2 B_{23}\left(p_1,p_2,\eta
	\right)+\left({p}_1\cdot {p}_2\right) B_{22}\left(p_2,p_1,\eta
	\right)+\left({\eta }\cdot {p}_1\right) B_{24}\left(p_2,p_1,\eta
	\right)\notag \\
	B_{11}\left(p_2,p_1,\eta \right) & = &  -{p_1}^2 B_{24}\left(p_1,p_2,\eta
	\right)-\left({\eta }\cdot {p}_1\right) B_{27}\left(p_1,p_2,\eta
	\right) \notag \\
	B_{12}\left(p_1,p_2,\eta \right) & = &  -\frac{{p_1}^2 B_6\left(p_1,p_2,\eta
		\right)}{{\eta }\cdot {p}_1}-\frac{\left({p}_1\cdot
		{p}_2\right) B_9\left(p_1,p_2,\eta \right)}{{\eta }\cdot
		{p}_1} \notag \\
	B_{28}\left(p_1,p_2,\eta \right) & = &  -{p_1}^2 B_{22}\left(p_1,p_2,\eta
	\right)-B_8\left(p_2,p_1,\eta \right). 
\end{eqnarray}

\section{Scalar Integrals: On-shell Case}
\label{appIn}
In this Appendix we provide some details concerning the analysis of all the integrals that appear in \eqref{blexpre}. 
The procedure is based on the factorization of the radial and angular integrals in dimensional regularization.  
As explained in the main sections, we indicate by "$\delta$" the finite density contribution coming from the insertion of the finite density part of each propagator in a given internal leg of the loop. "$f$" stands for the ordinary propagator. 
The resulting final integrals will be written using the following basic integrals 

\begin{equation}
	\Theta[ f(t) ] = \frac{\pi^{d/2-3/2}}{\Gamma[d/2-1/2]} \int_{-1}^1 dt \, f(t) \, (1-t^2)^{d/2-2} \qquad R [ f(|\kbf|) ] = \int_0^\mu \, d |\kbf| f(|\kbf|)
\end{equation}
$J \, [p_1 \cdot k] =H \, [p_2\cdot k]$
{\allowdisplaybreaks
\begin{eqnarray}
	J^{(\delta,f,f)} [p_1 \cdot k]&=& -{1 \over (2 \pi)^{d-1}} {1 \over 16 p_0} R\left[\frac{|\kbf|^{d-3}}{p_0+k}\right] \,  \Theta[1] \notag \\
	J^{(f,\delta,f)} [p_1 \cdot k]&=& - {1 \over (2 \pi)^{d-1}} {1 \over 8 p_0} \ \,  \Theta \left[{1 \over 1 + t}\right]  \notag \\
	J^{(f,f,\delta)}  [p_1 \cdot k]&=& {1 \over (2 \pi)^{d-1}} {1 \over 16 p_0} R\left[\frac{|\kbf|^{d-3}}{p_0+k}\right] \,  \Theta \left[{1-t \over 1 + t}\right] +{1 \over (2 \pi)^{d-1}} {1 \over 8} R\left[\frac{|\kbf|^{d-4}}{p_0+k}\right] \,  \Theta \left[{1 \over 1 + t}\right]  \notag \\
	J^{(\delta,f,\delta)}  [p_1 \cdot k]&=&{1 \over (2 \pi)^{d-2}} {p_0^{d-4} \over 16} \Theta [1]  \, \theta(\mu - p_0) \notag \\
\end{eqnarray}

$H \, [p_1 \cdot k]=J \, [p_2 \cdot k]$
\begin{eqnarray}
	H^{(\delta,f,f)} [p_1 \cdot k]&=& -{1 \over (2 \pi)^{d-1}} {1 \over 16 p_0} R\left[\frac{|\kbf|^{d-3}}{p_0-k}\right] \,  \Theta\left[{1-t \over 1 +t}\right] \notag \\
	H^{(f,\delta,f)} [p_1 \cdot k])&=& - {1 \over (2 \pi)^{d-1}} {1 \over 8 p_0} R [|\kbf|^{d-4}] \,  \Theta \left[{1 \over 1 + t}\right] - {1 \over (2 \pi)^{d-1}} {1 \over 4} R[|\kbf|^{d-5}] \, \Theta \left[1 \over 1-t^2 \right]   \notag \\
	H^{(f,f,\delta)}  [p_1 \cdot k])&=& {1 \over (2 \pi)^{d-1}} {1 \over 16 p_0} R\left[\frac{|\kbf|^{d-3}}{p_0+k}\right] \,  \Theta [1] +{1 \over (2 \pi)^{d-1}} {1 \over 8}  R\left[\frac{|\kbf|^{d-4}}{p_0+k}\right] \,  \Theta \left[1 \over 1-t\right] \notag \\
	H^{(\delta,f,\delta)}  [p_1 \cdot k]&=&{1 \over (2 \pi)^{d-2}} {p_0^{d-4} \over 16} \Theta \left[1-t \over 1+t\right]  \, \theta(\mu - p_0) \notag \\
\end{eqnarray}
}
$J \, [p_1 \cdot k \, k^2]=H \,[p_2 \cdot k \, k^2]$
\begin{eqnarray}
	J^{(\delta,f,f)}  [p_1 \cdot k \, k^2]&=& 0 \notag \\
	J^{(f,\delta,f)}  [p_1 \cdot k \, k^2]&=& - {1 \over (2 \pi)^{d-1}} {1 \over 4} R [|\kbf|^{d-3}] \,  \Theta \left[{1-t \over 1 + t}\right]  \notag \\
	J^{(f,f,\delta)}  [p_1 \cdot k \, k^2]&=& {1 \over (2 \pi)^{d-1}} {1 \over 4} R [|\kbf|^{d-3}] \,  \Theta \left[{1-t \over 1 + t}\right] +{1 \over (2 \pi)^{d-1}} {p_0 \over 2} R [|\kbf|^{d-4}] \,  \Theta \left[{1 \over 1 + t}\right]  \notag \\
\end{eqnarray}

$H \, [p_1 \cdot k \, k^2]= J \, [p_2 \cdot k \, k^2] = 0$

$J\, [(p_1 \cdot k)^2]=H \, [(p_2 \cdot k)^2]$
\begin{eqnarray}
	J^{(\delta,f,f)} [(p_1 \cdot k)^2]&=& -{1 \over (2 \pi)^{d-1}} {1 \over 16} R\left[\frac{|\kbf|^{d-2}}{p_0-k}\right] \,  \Theta[1-t] \notag \\
	J^{(f,\delta,f)} [(p_1 \cdot k)^2]&=& - {1 \over (2 \pi)^{d-1}} {1 \over 8} R [|\kbf|^{d-3}] \,  \Theta \left[{1 - t \over 1 + t}\right]  \notag \\
	J^{(f,f,\delta)}[(p_1 \cdot k)^2]&=& {1 \over (2 \pi)^{d-1}} {1 \over 16} R\left[\frac{|\kbf|^{d-2}}{p_0+k}\right] \,  \Theta \left[{(1-t)^2 \over 1 + t}\right] 
	+{1 \over (2 \pi)^{d-1}} {p_0 \over 4} R\left[\frac{|\kbf|^{d-3}}{p_0+k}\right] \,  \Theta \left[{1-t \over 1 + t}\right]    \notag \\ &&
	+{1 \over (2 \pi)^{d-1}} {p_0^2 \over 4} R\left[\frac{|\kbf|^{d-4}}{p_0+k}\right] \,  \Theta \left[{1 \over 1 + t}\right]  \notag \\
	J^{(\delta,f,\delta)} [(p_1 \cdot k)^2]&=&{1 \over (2 \pi)^{d-2}} {p_0^{d-2} \over 16} \Theta [1-t]  \, \theta(\mu - p_0) \notag \\
\end{eqnarray}

{\allowdisplaybreaks
$J \, [p_1 \cdot k \, p_2 \cdot k]=H \,[p_1 \cdot k \, p_2 \cdot k]$
\begin{eqnarray}
	J^{(\delta,f,f)}[p_1 \cdot k \, p_2 \cdot k]&=& -{1 \over (2 \pi)^{d-1}} {1 \over 16} R\left[\frac{|\kbf|^{d-2}}{p_0-k}\right] \,  \Theta[1+t] \notag \\
	J^{(f,\delta,f)} [p_1 \cdot k \, p_2 \cdot k]&=& - {1 \over (2 \pi)^{d-1}} {1 \over 8} R [|\kbf|^{d-3}] \,  \Theta [1] 
	- {1 \over (2 \pi)^{d-1}} {p_0 \over 4} R [|\kbf|^{d-4}] \,  \Theta \left[1 \over 1+t \right] \notag \\
	J^{(f,f,\delta)}[p_1 \cdot k \, p_2 \cdot k]&=& {1 \over (2 \pi)^{d-1}} {1 \over 16} R\left[\frac{|\kbf|^{d-2}}{p_0+k}\right] \,  \Theta [1-t] 
	+{1 \over (2 \pi)^{d-1}} {p_0 \over 4} R\left[\frac{|\kbf|^{d-3}}{p_0+k}\right] \,  \Theta \left[{1 \over 1 + t}\right]  \notag \\ &&
	+{1 \over (2 \pi)^{d-1}} {p_0^2 \over 4} R\left[\frac{|\kbf|^{d-4}}{p_0+k}\right] \,  \Theta \left[{1 \over 1 + t}\right]  \notag \\
	J^{(\delta,f,\delta)}[p_1 \cdot k \, p_2 \cdot k]&=&{1 \over (2 \pi)^{d-2}} {p_0^{d-2} \over 16} \Theta [1]  \, \theta(\mu - p_0) \notag \\
\end{eqnarray}

$H \, [(p_1 \cdot k)^2]=J \,[(p_2 \cdot k)^2]$
\begin{eqnarray}
	H^{(\delta,f,f)}[(p_1 \cdot k)^2]&=& -{1 \over (2 \pi)^{d-1}} {1 \over 16} R\left[\frac{|\kbf|^{d-2}}{p_0-k}\right] \,  \Theta\left[(1-t)^2 \over 1+t\right] \notag \\
	H^{(f,\delta,f)}[(p_1 \cdot k)^2]&=& - {1 \over (2 \pi)^{d-1}} {1 \over 8} R [|\kbf|^{d-3}] \,  \Theta \left[{1 - t \over 1 + t}\right]
	- {1 \over (2 \pi)^{d-1}} {p_0 \over 2} R [|\kbf|^{d-4}] \,  \Theta \left[{1  \over 1 + t}\right]   \notag \\ &&
	- {1 \over (2 \pi)^{d-1}} {p_0^2 \over 2} R [|\kbf|^{d-5}] \,  \Theta \left[{1 \over 1 - t^2}\right] 	\notag \\
	H^{(f,f,\delta)}[(p_1 \cdot k)^2]&=& {1 \over (2 \pi)^{d-1}} {1 \over 16} R\left[\frac{|\kbf|^{d-2}}{p_0+k}\right] \,  \Theta [1-t] 
	+{1 \over (2 \pi)^{d-1}} {p_0 \over 4} R\left[\frac{|\kbf|^{d-3}}{p_0+k}\right] \,  \Theta [1]   \notag \\ &&
	+{1 \over (2 \pi)^{d-1}} {p_0^2 \over 4} R\left[\frac{|\kbf|^{d-4}}{p_0+k}\right] \,  \Theta \left[{1 \over 1 - t}\right]  \notag \\
	H^{(\delta,f,\delta)} [(p_1 \cdot k)^2]&=&{1 \over (2 \pi)^{d-2}} {p_0^{d-2} \over 16} \Theta \left[(1-t)^2 \over 1+t\right]  \, \theta(\mu - p_0) \notag \\
\end{eqnarray}
$(\delta, \delta, f)$ integrals are identically zero, since the two deltas can be written as
\begin{equation}
	\delta(k^2) \, \delta((k-p_1)^2) = \delta(k^2) \, \delta(- 2 k \cdot p_1 ) 
	=   \delta(k^2) \, \delta \left(- 2 | \kbf | p_0 (1 - \cos \theta) \right) =\delta(k^2) \, {\delta(1-\cos \theta) \over 2 |\kbf| p_0}
\end{equation}
The second delta applied on the dimensional regularization factor $(1-t^2)^{d/2-2}$ yields zero, making the whole integral vanish. A similar argument can be made for $	\delta(k^2) \, \delta((k-p_2)^2)$, that will be transformed into
\begin{equation}
	\delta(k^2) \, \delta((k-p_2)^2) = \delta(k^2) \, {\delta(1+\cos \theta) \over - 2 |\kbf| p_0}
\end{equation}
and for $(f, \delta, \delta)$ integrals as well, since after a translation $k \to k+q$ can be transformed into
\begin{equation}
	\delta((k-q)^2) \, \delta((k-p_1)^2) = 	\delta((k-q)^2) \, \delta((k+p_1)^2)= \delta(k^2) \, \delta(2 k \cdot p_1 ) 
 =\delta(k^2) \, {\delta(1-\cos \theta) \over -2 |\kbf| p_0}.
 \end{equation}
$(\delta, \delta, \delta)$ integrals obviously vanish.
\subsection{$d \to 4$ expansions}
\subsubsection{Angular Integrals}
\begin{eqnarray}
	\Theta[1] &=& \pi^{d/2-1} \frac{\Gamma({d \over 2}-1)}{\Gamma({d-1 \over 2})^2}\to 4 \notag \\
	\Theta[t] &=& 0 \notag \\
	\Theta[1+t] &=& \Theta[1-t] = \Theta[1] \notag \\
	\Theta \left[1 \over 1 + t \right] &=& \Theta\left[1 \over 1 - t \right] = \pi^{d/2-1} \frac{\Gamma({d \over 2}-2)}{\Gamma({d-1 \over 2})\Gamma({d-3 \over 2})}\to {4 \over d-4}  + 2\left(\gamma - 2 + 2\log(16\pi)  \right) + O(d-4) \notag \\
	\Theta \left[{1-t \over 1 + t}\right] &=& \Theta \left[1+t \over 1 - t\right] = (d-2) \, \pi^{d/2-1} \frac{\Gamma({d \over 2}-1)}{2 \, \Gamma({d-1 \over 2})^2}\to {8 \over d-4} + 4(\gamma - 3 + \log(16 \pi))  + O(d-4) \notag \\
	\Theta \left[{(1-t)^2 \over 1 + t}\right] &=& \Theta \left[(1+t)^2 \over 1 + t\right] = d \,  \pi^{d/2-1} \frac{\Gamma({d \over 2}-1)}{2 \, \Gamma({d-1 \over 2})^2}\to {16 \over d-4} + 4(2\gamma - 7 + 2\log(16 \pi))  + O(d-4) \notag \\
	\Theta \left[1 \over 1 - t^2 \right] &=&  \pi^{d/2-1} \frac{\Gamma({d \over 2}-2)}{\Gamma({d-1 \over 2})\Gamma({d-3 \over 2})}\to {4 \over d-4}  + 2\left(\gamma - 2 + 2\log(16\pi)  \right) + O(d-4) \notag \\
\end{eqnarray}
\subsubsection{Radial Integrals}
\begin{align} \label{radialintegrals}
	 &R\left[\frac{|\kbf|^{d-2}}{p_0+k}\right]  \to {\mu \over 2}(\mu - 2p_0) + p_0^2 \log \left( 1 + {\mu \over p_0}\right) 
	 &&R\left[\frac{|\kbf|^{d-2}}{p_0-k}\right]  \to -{\mu \over 2}(\mu + 2p_0) - p_0^2 \log \left( 1 - {\mu \over p_0}\right) \notag \\
	 &R\left[\frac{|\kbf|^{d-3}}{p_0+k}\right]  \to \mu - p_0 \log \left( 1 + {\mu \over p_0}\right)
	 &&R\left[\frac{|\kbf|^{d-3}}{p_0-k}\right]  \to -\mu - p_0 \log \left( 1 - {\mu \over p_0}\right)  \notag \\
	 &R\left[\frac{|\kbf|^{d-4}}{p_0+k}\right]  \to \log \left( 1 + {\mu \over p_0}\right) 
	 &&R\left[\frac{|\kbf|^{d-4}}{p_0-k}\right]  \to -\log \left( 1 - {\mu \over p_0}\right) \notag \\
	 &R[|\kbf|^{d-2}] \to {\mu^3 \over 3} 
	 &&R[|\kbf|^{d-3}] \to {\mu^2 \over 2}   \notag \\
	 &R[|\kbf|^{d-4}] \to \mu 
	 &&R[|\kbf|^{d-5}] \to {1 \over d - 4} + \log \mu + O(d-4)  
\end{align}
}

\section{Radial integrals: Off-shell symmetric case}\label{OFF SHELL INT}
In the following we will analyze all the integrals that appears in \eqref{blexpre}. Here we introduce $\chi_\pm$ defined as
{\allowdisplaybreaks
\begin{equation}
\chi_\pm=\sqrt{p_0^2-M^2}\left(\frac{p_0}{\sqrt{p_0^2-M^2}}\pm\cos\theta \right)=|\pbf |\left (\frac{p_0}{|\pbf|}\pm\cos\theta \right).
\end{equation}
The procedure is based on factorizing radial and angular integrals and solve only the angular ones.
\\
\\
$H\,[p_1 \cdot k]$
\begin{eqnarray}
H^{(\delta,f,f)} [p_1 \cdot k]&=&-\frac{1}{8p_0}\int_0^\mu d|\kbf| \frac{|\kbf|}{p_0-|\kbf|}\int_{-1}^1 d\cos\theta\frac{\chi_-}{\chi_+}\\
H^{(f,\delta,f)} [p_1 \cdot k] &=&\frac{2p_0^2-M^2}{2}\int_0^\mu d|\kbf| |\kbf|\int_{-1}^1 d\cos\theta\frac{1}{(M^2+2|\kbf|\chi_+)(M^2-2|\kbf|\chi_-)}\\
&&+\frac{1}{2}\int_0^\mu d|\kbf| |\kbf|^2 \int_{-1}^1 d\cos\theta \frac{\chi_-}{(M^2+2|\kbf|\chi_+)(M^2-2|\kbf|\chi_-)}\\
H^{(f,f,\delta)} [p_1 \cdot k]&=&\frac{1}{8p_0}\int_0^\mu d|\kbf| \frac{|\kbf|^2}{p_0+|\kbf|}\int_{-1}^1 d\cos\theta\frac{\chi_-}{(M^2+2|\kbf|\chi_-)}\\
&&+\frac{p_0}{4}\int_0^\mu d|\kbf| \frac{|\kbf|}{p_0+|\kbf|} \int_{-1}^1 d\cos\theta \frac{1}{(M^2+2|\kbf|\chi_-)}
\end{eqnarray}
$J \, [p_1 \cdot k \, k^2]$
\begin{eqnarray}
	J^{(\delta, f, f)} [p_1 \cdot k \, k^2]&=& 0 \\
	J^{(f, \delta, f)}[p_1 \cdot k \, k^2] &=& 
	\int_0^\mu \frac{ d |\kbf|}{(2 \pi)^2} \frac{|\kbf|}{2} 
	\int_{-1}^1 d \cos \theta \, \frac{M^2 + |\kbf| \chi_- }{M^2 - 2|\kbf| \chi_-} \\
	J^{(f, f, \delta)}[p_1 \cdot k \, k^2] &=& 
	\int_0^\mu \,  \frac{ d |\kbf|}{(2 \pi)^2} \frac{ |\kbf|}{2}  
	\int_{-1}^1 d \cos \theta \, \frac{2 p_0^2 + |\kbf|\chi_- }{M^2 + 2 | \kbf| \chi_+} 
\end{eqnarray}
$J \, [p_2 \cdot k \, k^2]$
\begin{eqnarray}
	J^{(\delta, f, f)} [p_2 \cdot k \, k^2]&=& 0 \\
	J^{(f, \delta, f)} [p_2 \cdot k \, k^2] &=& 
	\int_0^\mu \frac{ d |\kbf|}{(2 \pi)^2} \frac{|\kbf|}{2} 
	\int_{-1}^1 d \cos \theta \, \frac{2p_0^2 - M^2 + |\kbf| \chi_-}{M^2- 2|\kbf| \chi_+} \\
	J^{(f, f, \delta)} [p_2 \cdot k \, k^2]&=& 
	\int_0^\mu \,  \frac{ d |\kbf|}{(2 \pi)^2} \frac{ |\kbf|}{2}  
	\int_{-1}^1 d \cos \theta \, \frac{2 p_0^2 + |\kbf|\chi_+ }{M^2 + 2 | \kbf| \chi_+} 
\end{eqnarray}
$J \, [(p_1 \cdot k)^2]$
	\begin{eqnarray}
	J^{(\delta, f, f)} [(p_1 \cdot k)^2] &=& 
	{1 \over 8 p_0} \int_0^\mu \,  \frac{ d |\kbf|}{(2 \pi)^2} \frac{ |\kbf|^3}{p_0 - |\kbf|}  
	\int_{-1}^1 d \cos \theta \, \frac{\chi_-^2}{M^2 - 2 | \kbf| \chi_-} \\
	J^{(f, \delta, f)} [(p_1 \cdot k)^2] &=& 
	\int_0^\mu \frac{ d |\kbf|}{(2 \pi)^2} \frac{|\kbf|}{2} 
	\int_{-1}^1 d \cos \theta \, \frac{(M^2 + |\kbf|\chi_- )^2}{(M^2 + 2 |\kbf| \chi_-)(M^2- 2 |\kbf| \chi_+)} \\
	J^{(f, f, \delta)} [(p_1 \cdot k)^2] &=& 
	{1 \over 8 p_0} \int_0^\mu \,  \frac{ d |\kbf|}{(2 \pi)^2} \frac{ |\kbf|}{p_0 + |\kbf|}  
	\int_{-1}^1 d \cos \theta \, \frac{(2 p_0^2 + |\kbf|\chi_- )^2}{M^2 + 2 | \kbf| \chi_+} 
	\end{eqnarray}
$J \,  [p_1 \cdot k \,  p_2 \cdot k] $
\begin{eqnarray}
	J^{(\delta, f, f)} [p_1 \cdot k \,  p_2 \cdot k] &=& 
	{1 \over 8 p_0} \int_0^\mu \,  \frac{ d |\kbf|}{(2 \pi)^2} \frac{ |\kbf|^3}{p_0 - |\kbf|}  
	\int_{-1}^1 d \cos \theta \, \frac{\chi_- \, \chi_+}{M^2 - 2 | \kbf| \chi_-} \\
	J^{(f, \delta, f)}  [p_1 \cdot k \,  p_2 \cdot k] &=& 
	\int_0^\mu \frac{ d |\kbf|}{(2 \pi)^2} \frac{|\kbf|}{2} 
	\int_{-1}^1 d \cos \theta \, \frac{(M^2 + |\kbf|\chi_- )\,(2p_0^2-M^2 + |\kbf|\chi_+ )}{(M^2 + 2 |\kbf| \chi_-)(M^2- 2 |\kbf| \chi_+)} \\
	J^{(f, f, \delta)} [p_1 \cdot k \,  p_2 \cdot k]  &=& 
	{1 \over 8 p_0} \int_0^\mu \,  \frac{ d |\kbf|}{(2 \pi)^2} \frac{ |\kbf|}{p_0 + |\kbf|}  
	\int_{-1}^1 d \cos \theta \, \frac{(2 p_0^2 + |\kbf|\chi_- )\, (2 p_0^2 + |\kbf|\chi_+ ) }{M^2 + 2 | \kbf| \chi_+} 
\end{eqnarray}
$J \, [(p_2 \cdot k)^2]$
\begin{eqnarray}
	J^{(\delta, f, f)} [(p_2 \cdot k)^2] &=& 
	{1 \over 8 p_0} \int_0^\mu \,  \frac{ d |\kbf|}{(2 \pi)^2} \frac{ |\kbf|^3}{p_0 - |\kbf|}  
	\int_{-1}^1 d \cos \theta \, \frac{\chi_+^2}{M^2 - 2 | \kbf| \chi_-} \\
	J^{(f, \delta, f)}[(p_2 \cdot k)^2] &=& 
	\int_0^\mu \frac{ d |\kbf|}{(2 \pi)^2} \frac{|\kbf|}{2} 
	\int_{-1}^1 d \cos \theta \, \frac{(2p_0^2-M^2 + |\kbf|\chi_+ )^2}{(M^2 + 2 |\kbf| \chi_-)(M^2- 2 |\kbf| \chi_+)} \\
	J^{(f, f, \delta)} [(p_2 \cdot k)^2]&=& 
	{1 \over 8 p_0} \int_0^\mu \,  \frac{ d |\kbf|}{(2 \pi)^2} \frac{ |\kbf|}{p_0 + |\kbf|}  
	\int_{-1}^1 d \cos \theta \, \frac{(2 p_0^2 + |\kbf|\chi_+ )^2}{M^2 + 2 | \kbf| \chi_+} 
\end{eqnarray}
}
\section{Angular integrals for $p^2\neq 0$}
$ H \, [p_1 \cdot k]$
\begin{eqnarray}
\int_{-1}^1 d\cos\theta\frac{\chi_-}{\chi_+}&=&-\frac{2 \left(|\pbf|+\log \left(\frac{p_0-p}{|\pbf|+p_0}\right)p_0\right)}{|\pbf|}\\
\int_{-1}^1 d\cos\theta\frac{1}{(M^2+2|\kbf|\chi_+)(M^2-2|\kbf|\chi_-)}&=&\frac{\log \left(\frac{\left(2 |\kbf| (|\pbf|-p_0)+M^2\right) \left(2 |\kbf|(p_0-|\pbf|)+M^2\right)}{M^4-4 |\kbf|^2 (|\pbf|+p_0)^2}\right)}{8 |\kbf|^2 |\pbf| p_0}\\
\int_{-1}^1 d\cos\theta \frac{\chi_-}{(M^2+2|\kbf|\chi_+)(M^2-2|\kbf|\chi_-)}&=&\frac{| \kbf | p_0 \log \left(\frac{\left(2 | \kbf | | \pbf |-2 | \kbf | p_0+M^2\right) \left(-2| \kbf | | \pbf |+2 | \kbf | p_0+M^2\right)}{M^4-4 | \kbf |^2(| \pbf |+p_0)^2}\right)}{8 | \kbf |^3 | \pbf | p_0}\notag\\
&&+\frac{\left(M^2-2 | \kbf | p_0\right) \tanh^{-1}\left(\frac{2 | \kbf | | \pbf |}{M^2-2 | \kbf | p_0}\right)}{8 | \kbf |^3 | \pbf | p_0}\notag \\
&&-\frac{\left(2 | \kbf |p_0+M^2\right) \tanh ^{-1}\left(\frac{2 | \kbf | | \pbf |}{2 | \kbf |p_0+M^2}\right)}{8 | \kbf |^3 | \pbf | p_0}\\
\int_{-1}^1 d\cos\theta\frac{\chi_-}{(M^2+2|\kbf|\chi_-)}&=&\frac{M^2 \log \frac{\left(-2 | \kbf | | \pbf |+2 | \kbf | p_0+M^2\right)}{\left(2 | \kbf | (| \pbf |+p_0)+M^2\right)}}{4 | \kbf |^2 | \pbf |}+\frac{1}{| \kbf |}\\
\int_{-1}^1 d\cos\theta \frac{1}{(M^2+2|\kbf|\chi_-)}&=&\frac{\log \frac{\left(2 | \kbf | (| \pbf |+p_0)+M^2\right)}{\left(-2 | \kbf | | \pbf |+2 | \kbf | p_0+M^2\right)}}{2 | \kbf | | \pbf |}
\end{eqnarray}
$J \, [p_1 \cdot k \, k^2]$
\begin{eqnarray}
\int_{-1}^1 d \cos \theta \, \frac{M^2 + |\kbf| \chi_- }{M^2 - 2|\kbf| \chi_-} &=&\frac{3 M^2 \log \frac{\left(2 | \kbf | | \pbf |-2 | \kbf | p_0+M^2\right)}{ \left(M^2-2 | \kbf | (| \pbf |+p_0)\right)}}{4 | \kbf | | \pbf |}-1\\
\int_{-1}^1 d \cos \theta \, \frac{2 p_0^2 + |\kbf|\chi_- }{M^2 + 2 | \kbf| \chi_+} &=&-\frac{\left(4 p_0 (| \kbf |+p_0)+M^2\right) \left(\log\frac{ \left(-2 | \kbf | | \pbf |+2 | \kbf |p_0+M^2\right)}{\left(2 | \kbf | (| \pbf |+p_0)+M^2\right)}\right)}{4 | \kbf | | \pbf |}-1
\end{eqnarray}
$J \, [p_2 \cdot k \, k^2]$
\begin{eqnarray}
\int_{-1}^1 d \cos \theta \, \frac{2p_0^2 - M^2 + |\kbf| \chi_-}{M^2- 2|\kbf| \chi_+}&=& \frac{\left(4 p_0(| \kbf |+p_0)-3 M^2\right)\left(\log \frac{\left(2 | \kbf | | \pbf |-2| \kbf | p_0+M^2\right)}{\left(M^2-2 | \kbf |(| \pbf |+p_0)\right)}\right)}{4 | \kbf | | \pbf |}+1\notag\\
\int_{-1}^1 d \cos \theta \, \frac{2 p_0^2 + |\kbf|\chi_+ }{M^2 + 2 | \kbf| \chi_+} &=& \frac{\left(M^2-4 p_0^2\right) \left(\log\frac{ \left(-2 | \kbf | | \pbf |+2 | \kbf | p_0+M^2\right)}{\left(2 | \kbf |(| \pbf |+p_0)+M^2\right)}\right)}{4 | \kbf | | \pbf |}+1
\end{eqnarray}
$J \, [(p_1 \cdot k)^2]$
\begin{eqnarray}
\int_{-1}^1 d \cos \theta \, \frac{\chi_-^2}{M^2 - 2 | \kbf| \chi_-}&=&\frac{M^4 \left(\log \frac{\left(2 | \kbf || \pbf |-2 | \kbf |p_0+M^2\right)}{\left(M^2-2 | \kbf |(| \pbf |+p_0)\right)}-\right)-4 | \kbf | | \pbf | \left(2 | \kbf |p_0+M^2\right)}{8 | \kbf |^3| \pbf |} \\
\int_{-1}^1 d \cos \theta \, \frac{(M^2 + |\kbf|\chi_- )^2}{(M^2 + 2 |\kbf| \chi_-)(M^2- 2 |\kbf| \chi_+)}&=&\frac{1}{32 |\kbf|^4 |\pbf| p_0}\biggl[16 |\kbf|^2 |\pbf|p_0\notag\\
&&+\left((1-2 |\kbf|) M^2-4 |\kbf|p_0\right)^2 \log\frac{ \left(2 |\kbf||\pbf|-2 |\kbf|p_0+M^2\right)}{\left(M^2-2 |\kbf| (|\pbf+p_0)\right)}\notag\\
&&+(1-2 |\kbf|)^2 M^4 \log\frac{ \left(-2 |\kbf|  |\pbf|+2 |\kbf| p_0+M^2\right)}{ \left(2 |\kbf| (|\pbf|+p_0)+M^2\right)}\biggl]\\
\int_{-1}^1 d \cos \theta \, \frac{(2 p_0^2 + |\kbf|\chi_- )^2}{M^2 + 2 | \kbf| \chi_+} &=&-\frac{1}{8 |\kbf|^3 |\pbf|}\biggl[4 |\kbf| |\pbf| \left(2 |\kbf|p_0 (4p_0+3)+M^2\right)\notag\\
&&+\left(4 |\kbf|p_0(p_0+1)+M^2\right)^2\left(\log \frac{\left(-2 |\kbf| |\pbf|+2 |\kbf| p_0+M^2\right)}{\left(2 |\kbf| (|\pbf|+p_0)+M^2\right)}\right)\biggl]\notag \\
\end{eqnarray}
$J \, [p_1 \cdot k \, p_2 \cdot k]$
\begin{eqnarray} 
\int_{-1}^1 d \cos \theta \, \frac{\chi_- \, \chi_+}{M^2 - 2 | \kbf| \chi_-} &=&\frac{1}{8 |\kbf|^3 |\pbf|}\biggl[4 |\kbf| |\pbf| \left(M^2-2 |\kbf| p_0\right)\notag \\
&&+M^2 \left(M^2-4 |\kbf| p_0\right)\left(\log\frac{ \left(M^2-2 |\kbf| (|\pbf|+p_0)\right)}{\left(2 |\kbf| |\pbf|-2 |\kbf|p_0+M^2\right)}\right)\biggl]\\
\int_{-1}^1 d \cos \theta \, \frac{(M^2 + |\kbf|\chi_- )\,(2p_0^2-M^2 + |\kbf|\chi_+ )}{(M^2 + 2 |\kbf| \chi_-)(M^2- 2 |\kbf| \chi_+)}&=&\frac{1}{32 |\kbf|^2 |\pbf| p_0}\biggl[-16 |\kbf|^2 |\pbf| p_0\notag\\
&&-\left(M^2-4 p_0^2\right) \left(4 |\kbf|p_0+M^2\right) \log \frac{\left(2 |\kbf| |\pbf|-2 |\kbf| p_0+M^2\right)}{  \left(M^2-2 |\kbf|(|\pbf|+p_0)\right)}\notag\\
&&+M^2 \left(4 p_0 (|\kbf|+p_0)-M^2\right) \log\frac{ \left(-2|\kbf| |\pbf|+2 |\kbf| p_0+M^2\right)}{\left(2 |\kbf| (|\pbf|+p_0)+M^2\right)}\notag \\
&&+\left(4 p_0(|\kbf|+p_0)+M^2\right)^2 \left(\log \frac{\left(-2 |\kbf| |\pbf|+2 |\kbf| p_0+M^2\right)}{\left(2 |\kbf| (|\pbf|+p_0)+M^2\right)}\right)\biggl]\notag \\
\end{eqnarray}
{\allowdisplaybreaks
$J \, [(p_2 \cdot k)^2]$
\begin{eqnarray}
\int_{-1}^1 d \cos \theta \, \frac{\chi_+^2}{M^2 - 2 | \kbf| \chi_-}&=&\frac{1}{8 |\kbf|^3 |\pbf|}\biggl[4 |\kbf| |\pbf| \left(6 |\kbf| p_0-M^2\right)\\
&&+\left(M^2-4 |\kbf| p_0\right)^2\left(\log \frac{\left(2 |\kbf| |\pbf|-2 |\kbf| p_0+M^2\right)}{\left(M^2-2 |\kbf|(|\pbf|+p_0)\right)}\right)\biggl] \notag\\
\int_{-1}^1 d \cos \theta \, \frac{(2p_0^2-M^2 + |\kbf|\chi_+ )^2}{(M^2 + 2 |\kbf| \chi_-)(M^2- 2 |\kbf| \chi_+)}&=& \frac{1}{32 | \kbf |^2 | \pbf | p_0}\biggl[16 | \kbf |^2 | \pbf | p_0+\left(M^2-4 p_0^2\right)^2 \log \frac{\left(2 | \kbf | | \pbf |-2| \kbf | p_0+M^2\right)}{ \left(M^2-2 | \kbf |(| \pbf |+p_0)\right)}\notag\\
&&+\left(M^2-4 p_0 (| \kbf |+p_0)\right)^2 \log \frac{\left(-2 | \kbf | | \pbf |+2 | \kbf | p_0+M^2\right)}{\left(2 | \kbf | (| \pbf |+p_0)+M^2\right)}\biggl]\\
\int_{-1}^1 d \cos \theta \, \frac{(2 p_0^2 + |\kbf|\chi_+ )^2}{M^2 + 2 | \kbf| \chi_+} &=&-\frac{\left(M^2-4 p_0^2\right)^2 \left(\log\frac{ \left(-2 | \kbf | | \pbf |+2 | \kbf | p_0+M^2\right)}{ \left(2 | \kbf | (| \pbf |+p_0)+M^2\right)}\right)}{8 | \kbf || \pbf |}+p_0 (| \kbf |+4 p_0)\notag\\
&&-\frac{M^2}{2} 
\end{eqnarray}
}
}

\section{Vectorial Ward identities (WIs)}
In this Appendix we will analyze the vectorial Ward identities, which must be verified so that the gauge field is preserved and the theory is consistent. \\ We can consider the contraction with the $p_{1\mu}$ photonic leg  
\begin{align}
&p_{1\mu}\Gamma^{\lambda\mu\nu}(p_1,p_2,\beta)=\notag\\
&p_{1\mu}\int d^4k \text{Tr}[\gamma^\mu\slashed k\gamma^\lambda\gamma^5(\slashed k-\slashed q)\gamma^\nu(\slashed k-\slashed p_1)]( G_0(k)+\tilde G(k)) ( G_0(k-q)+\tilde G(k-q)) ( G_0(k-p_1)+\tilde G(k-p_1)) \notag\\
&+p_{1\mu}\int d^4k \text{Tr}[\gamma^\mu\slashed k\gamma^\lambda\gamma^5(\slashed k-\slashed q)\gamma^\nu(\slashed k-\slashed p_2)]( G_0(k)+\tilde G(k)) ( G_0(k-q)+\tilde G(k-q)) ( G_0(k-p_2)+\tilde G(k-p_2)). 
\end{align}
The algebra that we will use in order to give a full computation is quite straightforward, in fact, passing $p_1$ in the trace, we can write in the first integral $\slashed p_1\to-(\slashed k-\slashed p_1)+\slashed k$ and in the second one $\slashed p_1\to -(\slashed k-\slashed q)+(\slashed k-\slashed p_2)$
\begin{align}
&p_{1\mu}\Gamma^{\lambda\mu\nu}(p_1,p_2,\beta)=\notag\\
&-\int d^4k \text{Tr}[\slashed k\gamma^\lambda\gamma^5(\slashed k-\slashed q)\gamma^\nu](k-p_1)^2( G_0(k)+\tilde G(k)) ( G_0(k-q)+\tilde G(k-q)) ( G_0(k-p_1)+\tilde G(k-p_1)) \notag\\
&+\int d^4k \text{Tr}[\gamma^\lambda\gamma^5(\slashed k-\slashed q)\gamma^\nu(\slashed k-\slashed p_1)]k^2( G_0(k)+\tilde G(k)) ( G_0(k-q)+\tilde G(k-q)) ( G_0(k-p_1)+\tilde G(k-p_1)) \notag\\
&-\int d^4k \text{Tr}[\gamma^\nu\slashed k\gamma^\lambda\gamma^5(\slashed k-\slashed p_2)](k-q)^2( G_0(k)+\tilde G(k)) ( G_0(k-q)+\tilde G(k-q)) ( G_0(k-p_2)+\tilde G(k-p_2))\notag\\ 
&+\int d^4k \text{Tr}[\gamma^\nu\slashed k\gamma^\lambda\gamma^5(\slashed k-\slashed q)](k-p_2)^2( G_0(k)+\tilde G(k)) ( G_0(k-q)+\tilde G(k-q)) ( G_0(k-p_2)+\tilde G(k-p_2)). 
\end{align}
Taking into account only $1\delta$ cases, we have
\begin{align}p_{1\mu}\Gamma^{\lambda\mu\nu}(p_1,p_2,\beta)=&
-\int d^4k \text{Tr}[\slashed k\gamma^\lambda\gamma^5(\slashed k-\slashed q)\gamma^\nu]( G_0(k)\tilde G(k-q)+\tilde G(k) G_0(k-q))  \notag\\
&+\int d^4k \text{Tr}[\gamma^\lambda\gamma^5(\slashed k-\slashed q)\gamma^\nu(\slashed k-\slashed p_1)]( G_0(k-q)\tilde G(k-p_1)+\tilde G(k-q) G_0(k-p_1))  \notag\\
&-\int d^4k \text{Tr}[\gamma^\nu\slashed k\gamma^\lambda\gamma^5(\slashed k-\slashed p_2)] ( G_0(k)\tilde G(k-p_2)+\tilde G(k) G_0(k-p_2)) \notag\\ 
&+\int d^4k \text{Tr}[\gamma^\nu\slashed k\gamma^\lambda\gamma^5(\slashed k-\slashed q)]( G_0(k-q)\tilde G(k)+\tilde G(k-q) G_0(k)) 
\end{align}
The first and the fourth integrals cancel each other. Through a shift of the variable of integration, namely $k\to k-p_1$, we have also that the third integral cancels out the second one.  We can observe that the contributions with two $\delta$'s are trivially null when we consider the vectorial WIs. A shift is always allowed in the hot part of the AVV diagram because the integrals are not linearly divergent as in the cold part, in fact 
\begin{equation}
\Gamma^{\lambda\mu\nu}_{\text{hot}}\sim \int_0^\mu dk k^\alpha,
\end{equation}
with $\alpha > 0$.


\providecommand{\href}[2]{#2}\begingroup\raggedright\endgroup

\end{document}